\documentclass[]{elsart}

\usepackage{epsfig}
\usepackage{graphics}
\usepackage{graphicx}
\usepackage{amsmath}
\usepackage{url}
\usepackage{subfigure}

\newtheorem{finding}{Finding}

\pdfoutput=1

\begin{document}

\begin{frontmatter}

\title{On excitable $\beta$-skeletons}

\author{Andrew Adamatzky}

\address{University of the West of England, Bristol, United Kingdom\\ andrew.adamatzky@uwe.ac.uk}

\date{\today}

\begin{abstract}

\vspace{0.5cm}

\noindent
A $\beta$-skeleton, $\beta \geq 1$, is a planar proximity undirected graph of an Euclidean point set 
where nodes are connected by an edge if their lune-based neighborhood contains no 
other points of the given set. Parameter $\beta$ determines size and shape of the nodes' neighborhoods.
In an excitable $\beta$-skeleton every node takes three states --- resting, excited and refractory, and 
updates its state in discrete time depending on states of its neighbors. We design families of 
$\beta$-skeletons with absolute and relative thresholds of excitability and demonstrate that several 
distinct classes of space-time excitation dynamics can be selected using $\beta$. The classes include 
spiral and target waves of excitation, branching domains of excitation and oscillating localizations.     
 
\end{abstract}

\maketitle

\end{frontmatter}

\section{Introduction}

A planar graph consists of nodes which are points of Euclidean plane and edges which are straight segments connecting the points. A planar proximity graph is a planar graph where two points are connected by an edge if they are close in some sense. Usually a pair of points is assigned certain neighborhood, and points of the pair are connected by an edge if their neighborhood is empty.  Delaunay triangulation~\cite{delauanay}, relative neighborhood graph~\cite{jaromczyk} and Gabriel graph~\cite{matula_1980}, and indeed spanning tree, are most known examples of proximity graphs.  $\beta$-skeletons, proposed in \cite{kirkpatrick}, form a unique family of proximity graphs monotonously parameterised by parameter $\beta$.   

Proximity graphs found their applications in fields of science and engineerings:  
geographical variational analysis~\cite{gabriel_1969,matula_1980,sokal_2008}, 
evolutionary biology~\cite{magwene_2008}, 
spatial analysis in biology~\cite{legendre_1989,dale_2000,dale_2002,jombart_2008}, 
simulation of epidemics~\cite{toroczkai_2008}. Proximity graphs are used in physics 
to study percolation~\cite{billiot_2010} and analysis of magnetic field~\cite{sridharan_2010}. 

Engineering applications of proximity graphs are in message routing in ad hoc wireless networks, see e.g.~\cite{li_2004,song_2004,santi_2005,muhammad_2007,wan_2007}, and visualisation~\cite{runions_2005}. 
Road network analysis is yet another field where proximity graphs are invaluable. Road networks are well 
matched by relative neighborhbood graphs, see e.g. study of Tsukuba central 
district~\cite{watanabe_2005, watanabe_2008}. Biological transport networks also bear remarkable 
similarity to certain proximity graphs. Foraging trails of ants~\cite{adamatzky_2002}
and protoplasmic networks of slime mold \emph{Physarum polycephalum}~\cite{adamatzky_ppl_2008,adamatzky_jones_2009}
are most striking examples. 

Structure of proximity graphs represents so wide range of natural systems that it is important to uncover 
basic mechanism of activity propagation on the graphs, which could be applied in future studies of particular
natural systems. This is why we undertook computational experiments with excitable $\beta$-skeletons
to check how space-time dynamics of excitation on $\beta$-skeletons depends on $\beta$. The paper is structured 
as follows. In Sect.~\ref{themodel} we introduce models of excitable $\beta$-skeletons.
Phenomenology of skeletons with absolute threshold of excitation (a node excites depending on 
an absolute number of its excited neighbors) is provided in Sect.~\ref{absoluteexcitation}. Space-time
dynamics of skeletons with relative threshold of excitation (a node excites depending on a ratio of excited neighbors) 
is studied in Sect.~\ref{relativeexcitation}. Results of computational experiments are summarized in Section~\ref{discussion}.

\section{The model}
\label{themodel}

\begin{figure}
\centering
\subfigure[$\beta=1$]{\includegraphics[width=0.3\textwidth]{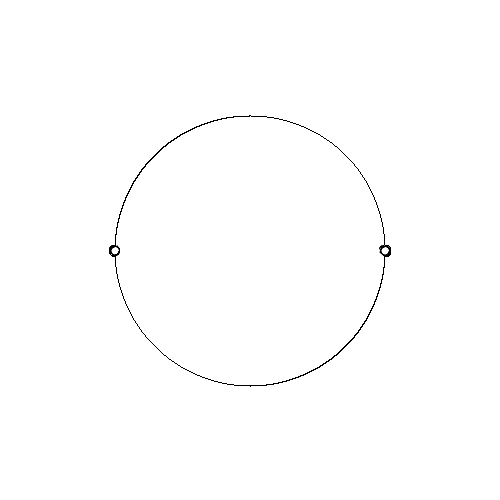}}
\subfigure[$\beta=1.2$]{\includegraphics[width=0.3\textwidth]{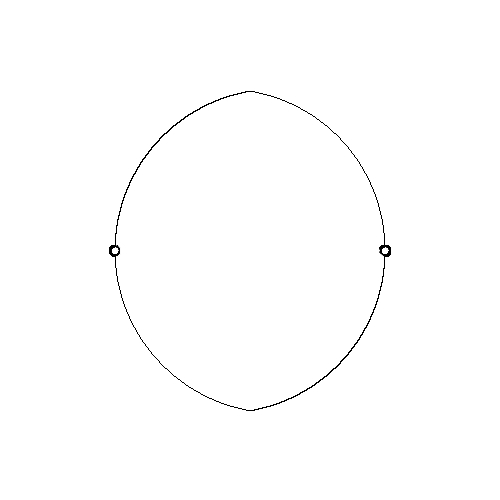}}
\subfigure[$\beta=1.5$]{\includegraphics[width=0.3\textwidth]{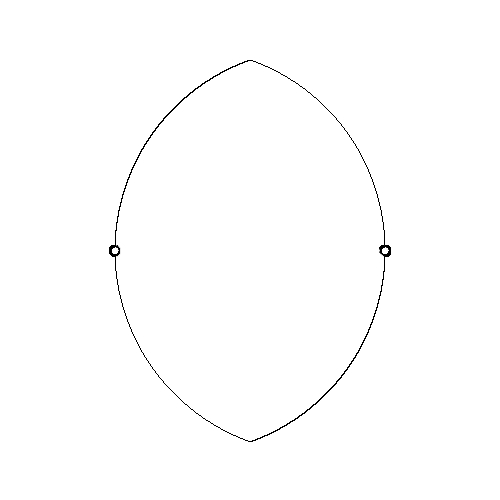}}
\subfigure[$\beta=1.7$]{\includegraphics[width=0.3\textwidth]{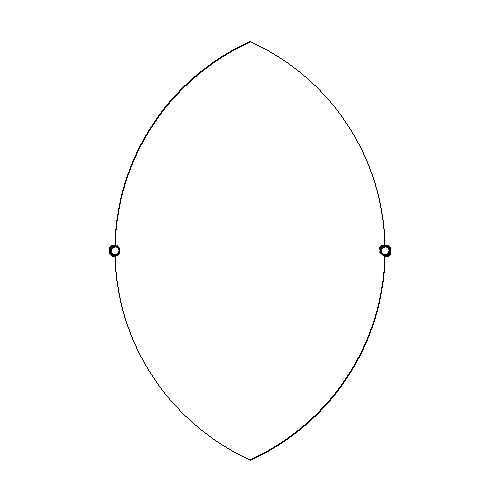}}
\subfigure[$\beta=2$]{\includegraphics[width=0.3\textwidth]{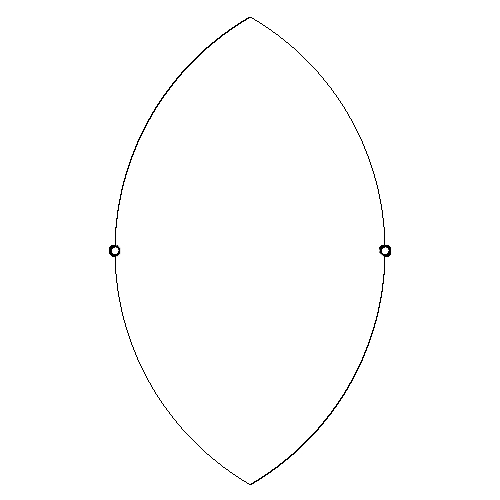}}
\caption{Examples of the lunes between two points (small circles) for various values of $\beta$.}
\label{lunes}

\vspace{0.5cm}

\end{figure}

Given a set $\mathbf V$ of planar points, for any two points $p$ and $q$ we define 
$\beta$-neighborhood $U_\beta(p,q)$ as an intersection of two discs 
with radius $\beta |p-q| / 2$ centered at points $((1-\frac{\beta}{2})p,\frac{\beta}{2}q)$ and $(\frac{\beta}{2}p, (1-\frac{\beta}{2})q)$, $\beta \geq 1$~\cite{kirkpatrick,jaromczyk}, see examples of the lunes in Fig.~\ref{lunes}.
Points $p$ and $q$ are connected by an edge in $\beta$-skeleton if the pair's $\beta$-neighborhood contains no 
other points from $\mathbf V$.

A $\beta$-skeleton is a graph $G_\beta({\mathbf V})= \langle {\mathbf V}, {\mathbf E}, \beta \rangle$, 
where nodes ${\mathbf V} \subset {\mathbf R}^2$, edges $\mathbf E$, and for $p, q \in {\mathbf V}$ 
edge $(pq) \in \mathbf E$ if $U_\beta(p,q) \cap {\mathbf V}/\{p, q\} = \emptyset$. Parameterization 
$\beta$ is monotonous: if $\beta_1 > \beta_2$ then 
$G_{\beta_1}({\mathbf V}) \subset  G_{\beta_2}({\mathbf V})$~\cite{kirkpatrick,jaromczyk}. A $\beta$-skeleton
is Gabriel graph~\cite{matula_1980} for $\beta=1$ and the skeleton is relative neighbourhood graph for $\beta=2$.  
We consider only skeletons with $1 \leq \beta \leq 2$ because $\beta$-skeletons are non-planar for $\beta<1$
and they are disconnected for $\beta>2$.

\begin{figure}[!tbp]
\centering
\subfigure[$\beta=1$]{\includegraphics[width=0.49\textwidth]{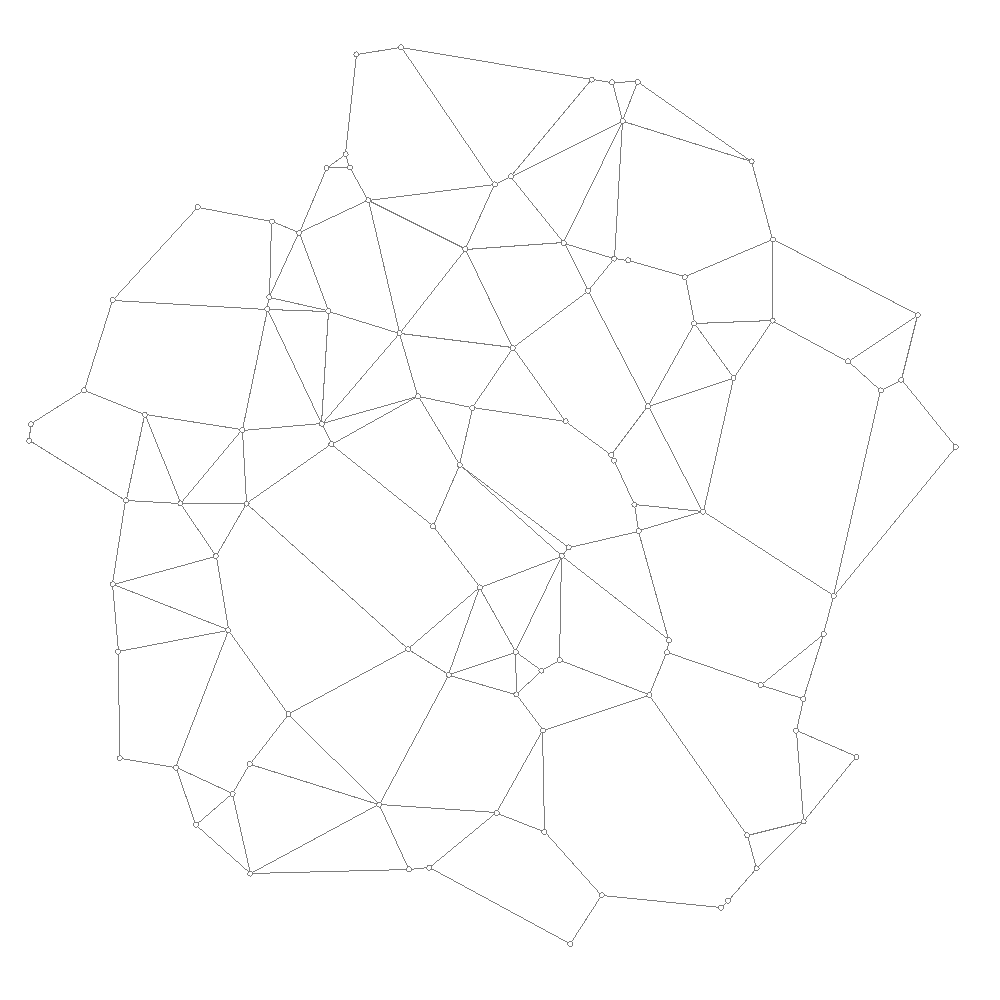}}
\subfigure[$\beta=1.5$]{\includegraphics[width=0.49\textwidth]{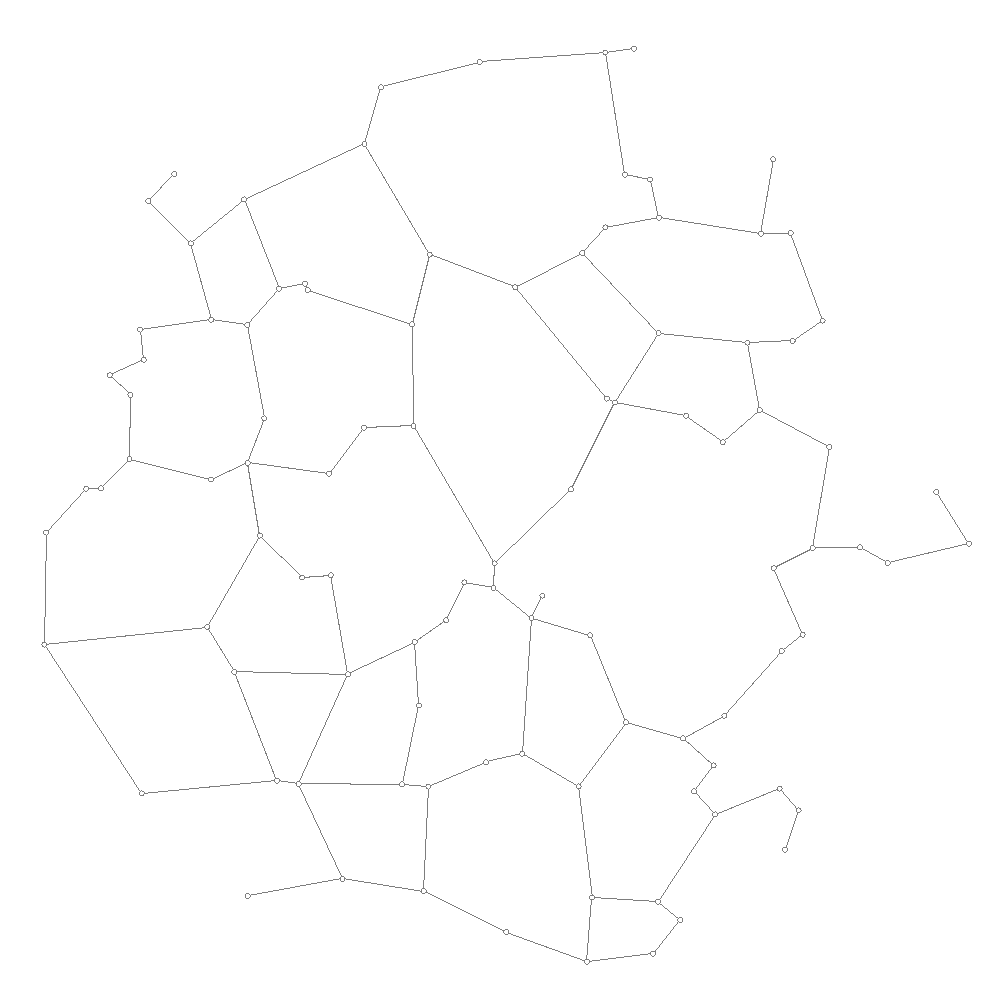}}
\subfigure[$\beta=1$]{\includegraphics[width=0.49\textwidth]{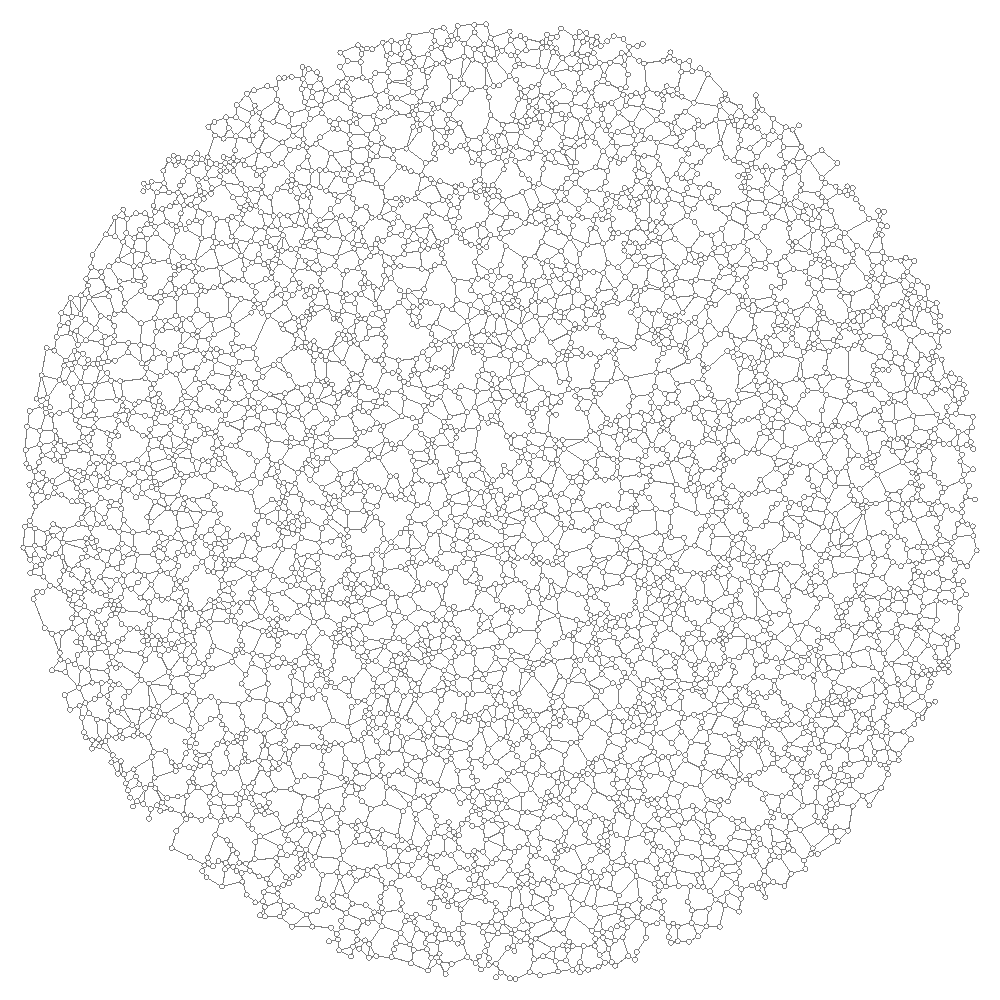}}
\subfigure[$\beta=1.5$]{\includegraphics[width=0.49\textwidth]{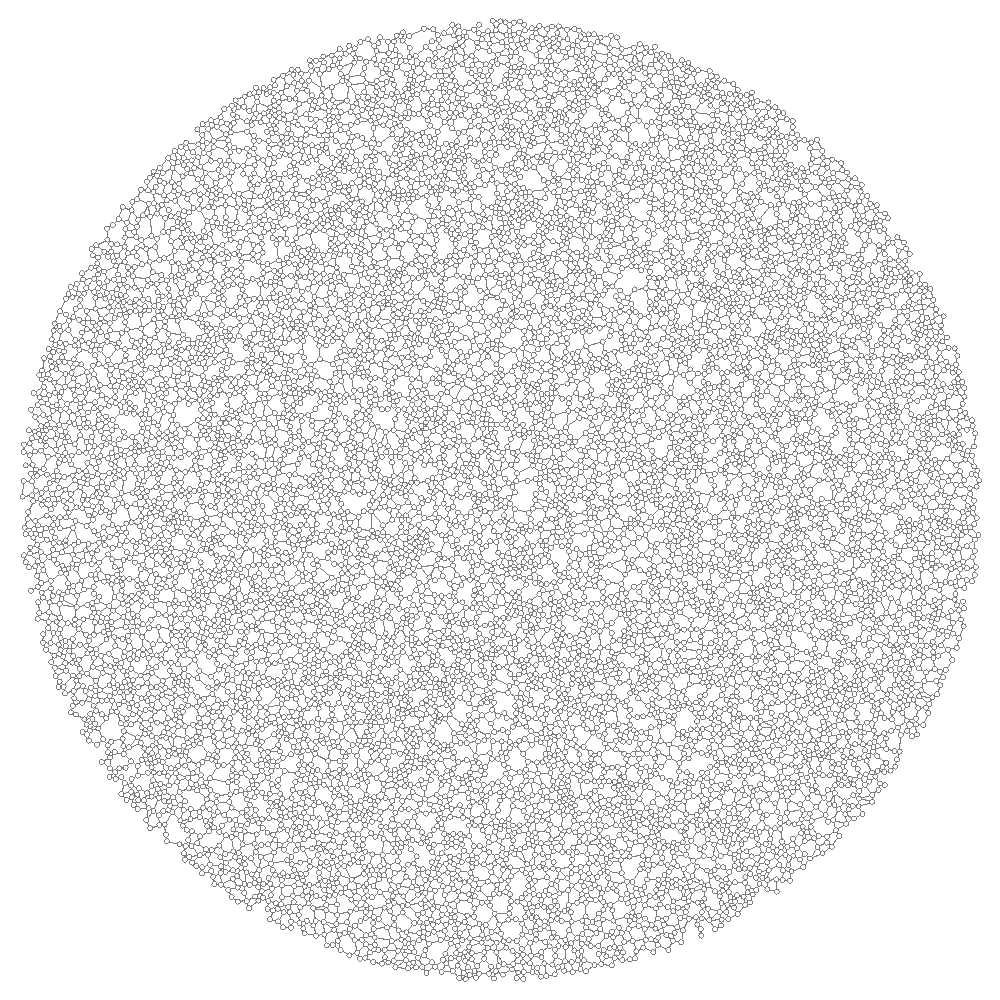}}
\caption{Examples of $\beta$-skeletons: 
(a)~$n=100$, $\varphi=0.0027$, and $\beta=1$, 
(b)~$n=100$, $\varphi=0.0027$, and $\beta=2$,
(c)~$n=5000$, $\varphi=0.136$, and $\beta=1.5$, 
(d)~$n=10000$, $\varphi=0.271$, and $\beta=1.5$.}
\label{examples}

\vspace{0.5cm}

\end{figure}  
  
We study $\beta$-skeletons of planar set of  $n$ discs.  Centers of the discs form set $\mathbf V$.
Each disc has a radius 2.5 units. The discs are randomly distributed in a large disc with radius 480 units. 
We undertake computational experiments with $n$ varying from 1000 to 15000 (Fig.~\ref{examples}). Number of 
nodes \emph{per se} is not as important as density of their packing therefore we refer to any particular $\mathbf V$ 
via density of node packing  $\varphi=0.027, \cdots, 0.407$. 

An excitable $\beta$-skeleton is defined as follows. A node $p \in \bf V$ is a finite state machine. 
Every node takes three states: resting ($\circ$), excited ($+$) and refractory ($-$). A node updates its 
states in discrete time depending on states of its neighbors. All nodes update  their states simultaneously.    

We assume that a resting node excites depending on a number of excited neighbors. If a node is 
excited at time $t$ the node takes refractory state at time step $t+1$, independently on states of 
its neighbors. Transition from refractory to resting state is also unconditional.

Let $\nu(p)=\{ q \in {\bf V}: (pq) \in {\bf E} \}$ be a neighborhood, or set of neighbors, of 
node $p$, $p^t$ a state of node $p$ at time step $t$, $\sigma^t(p)$  a number 
of excited neighbors of $p$ at step $t$, and $d(p)$ a degree, or a number of neighbours $|\nu(p)|$, 
of node $p$. Then the node-state transition function can be defined as follows:
\begin{equation}
p^{t+1}=
\begin{cases}
+, \text{ if } C(\nu(p)^t)\\
-, \text{ if } p^t=+\\
\circ,  \text{ if } p^t=- 
\end{cases}
\label{eq1}
\end{equation}
We consider two versions of the excitation condition $C(\nu(p)^t) \in $ \{ {\sc True, False} \}: 
\begin{itemize}
\item \emph{Absolute excitability}: $C(\nu(p)^t)=\sigma^t(p) \geq \theta$, $\theta=1, \cdots$\\
\item \emph{Relative excitability}: $C(\nu(p)^t=(\frac{\sigma^t(p)}{d(p)} > \epsilon), \epsilon \in [0,1]$ .
\end{itemize}
Absolute excitability is the most common approach of defining rules in excitable discrete systems however it does not
account for diversity of node degrees in disordered systems. Thus we also explore skeletons with relative excitability.

In the paper we illustrate space-dynamics of excitation by snapshots of skeleton configurations. We analyze integral 
dynamics of skeletons using activity $\alpha$. The activity $\alpha$ is a ratio of excited nodes to a total number 
of nodes averaged fixed number time steps. The activity $\alpha$ is measured after initial transient period, when 
excitation patterns are given a chance to occupy the whole skeleton (usually a hundred time steps is sufficient).

\section{Absolutely excitable skeletons}
\label{absoluteexcitation}

In $\beta$-skeletons governed by absolute excitability rule excitation persists only for threshold $\theta=1$.

\begin{finding}
Activity $\alpha$ is proportional to density $\varphi$ and inversely proportional to $\beta$.  
\end{finding}

\begin{figure}[!tbp]
\centering
\includegraphics[width=0.7\textwidth]{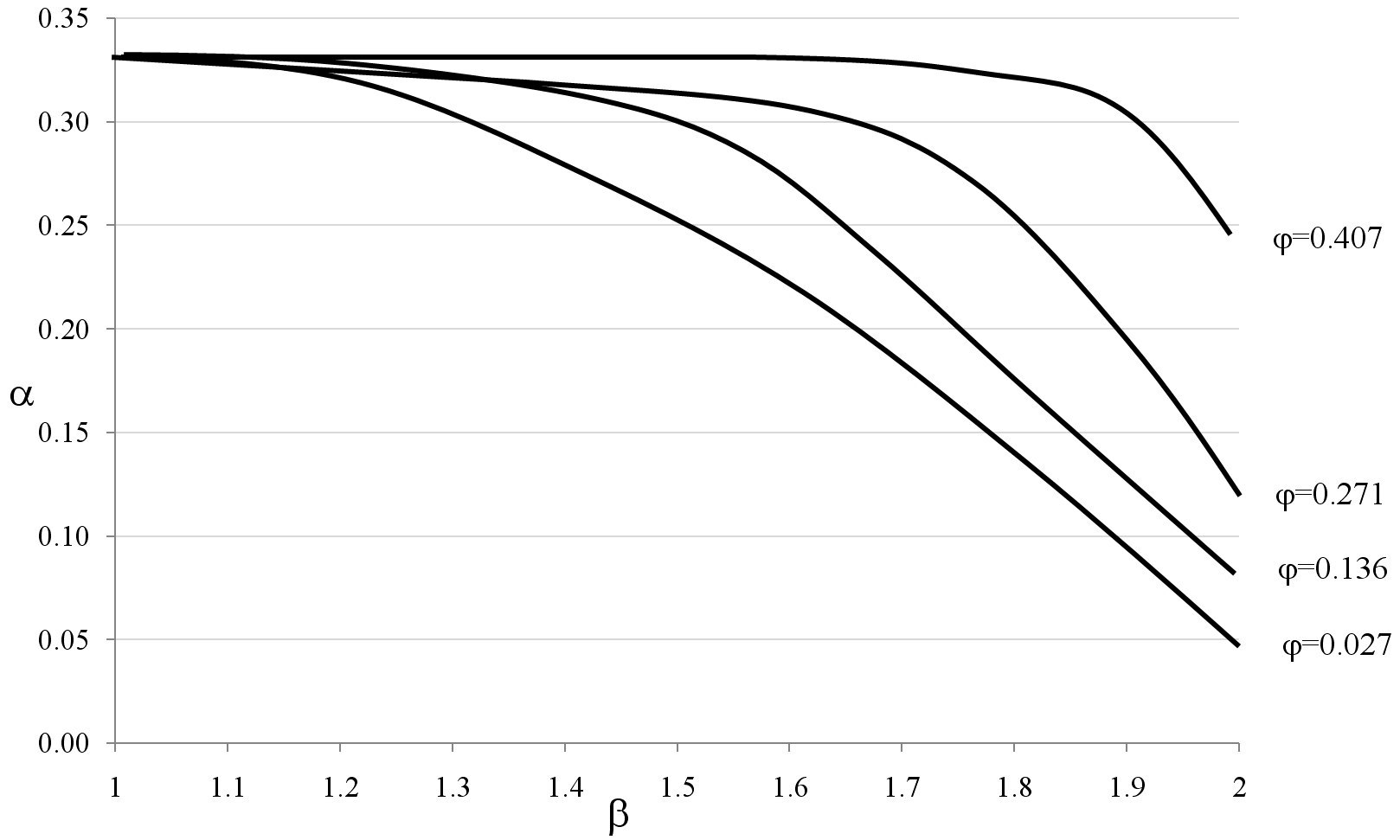} 
\caption{Activity $\alpha$ versus $\beta$ for various densities $\varphi$ of node packing in 
$\beta$-skeletons with absolute excitation threshold $\theta=1$. Skeletons are developed from initial random configuration, where a 
node is assigned excited state probability 0.1 and is resting otherwise.}
\label{excitationlevelabsoluteexcitability}

\vspace{0.5cm}

\end{figure}

Figure~\ref{excitationlevelabsoluteexcitability} shows exactly how activity $\alpha$ depends on $\beta$. Note 
that maximum possible $\alpha$ equals 0.33, for sustainable activity, because  an excited node always becomes 
refractory, and a refractory node always becomes resting. Thus even if the whole skeleton is active, only 
third of its nodes are excited at any given time step. The activity decreases polynomially with increase of 
$\beta$, degree of the polynomial is proportional to density $\varphi$ of nodes in the skeleton.

\begin{figure}[!tbp]
\centering 
$$
 \begin{array}{c|cccc}
	\beta \backslash \varphi  & 0.027 & 0.136 & 0.271 & 0.407 \\ \hline
1.0		&   \includegraphics[scale=0.08]{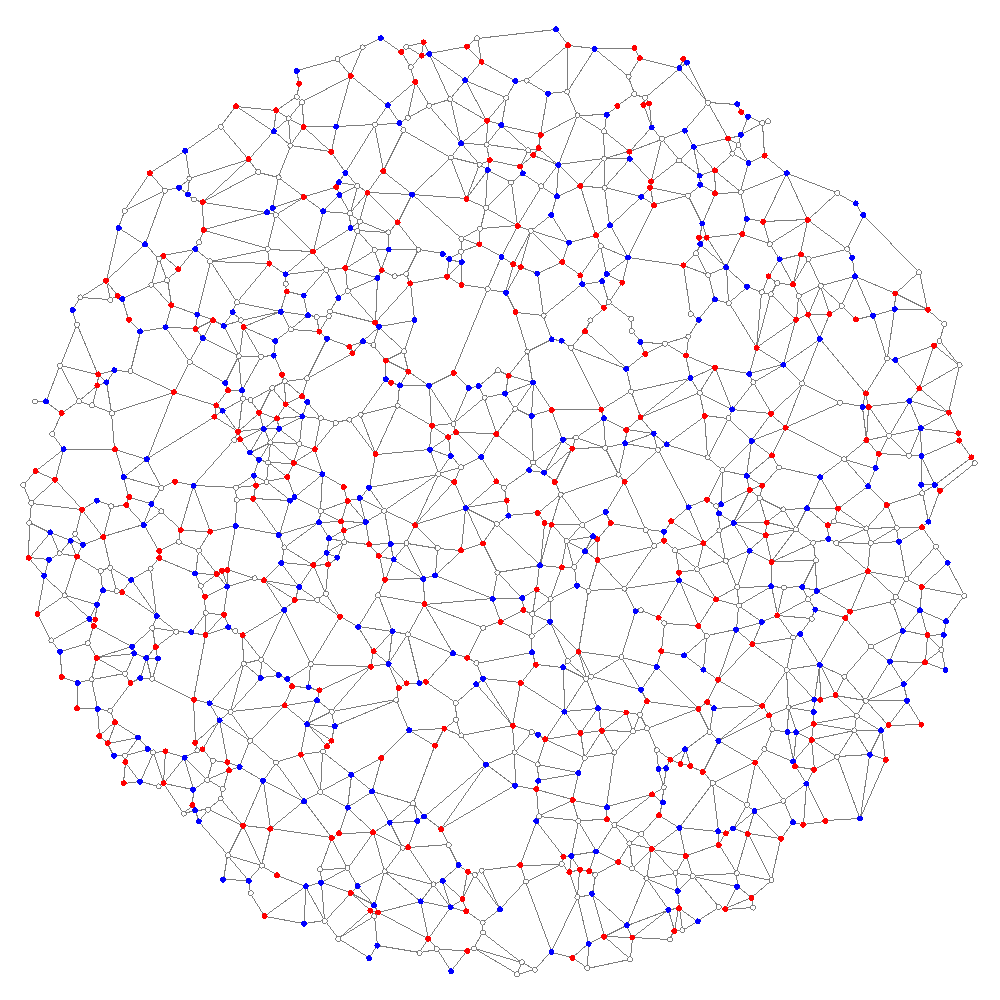}   &  \includegraphics[scale=0.08]{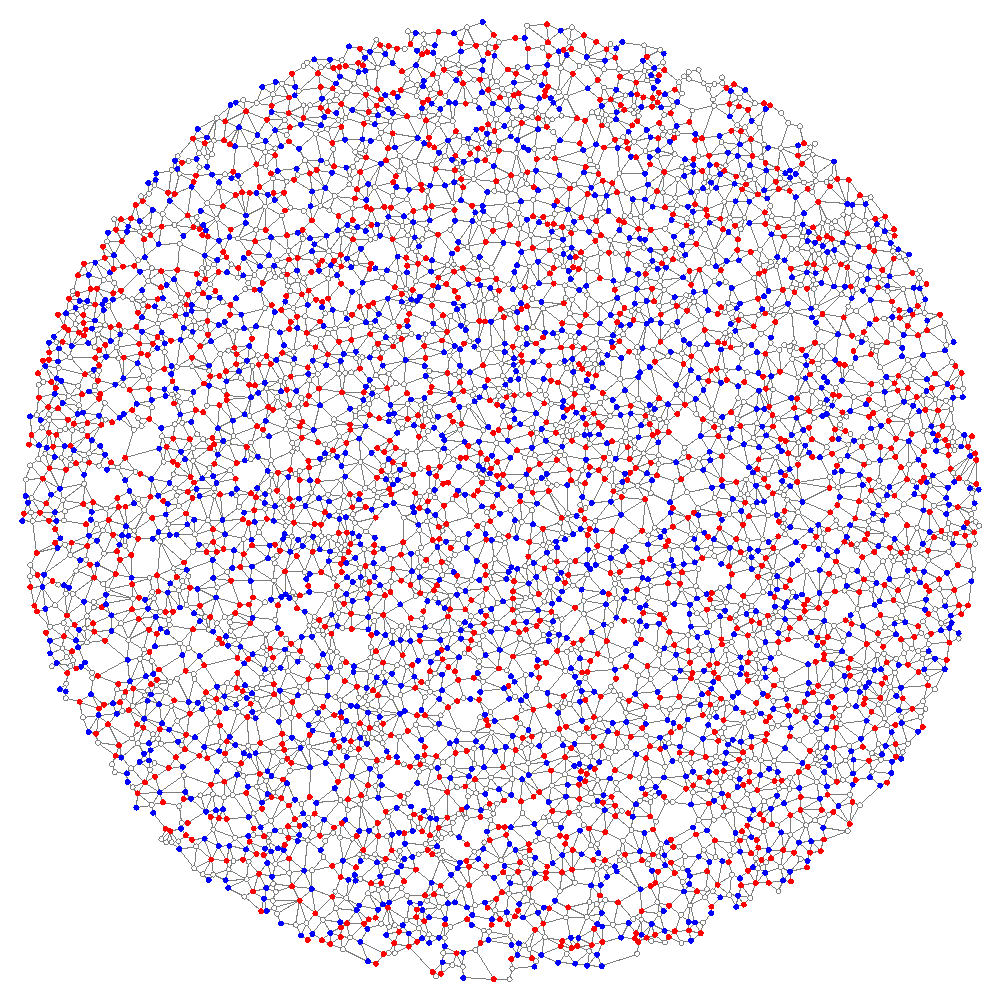}    &\includegraphics[scale=0.08]{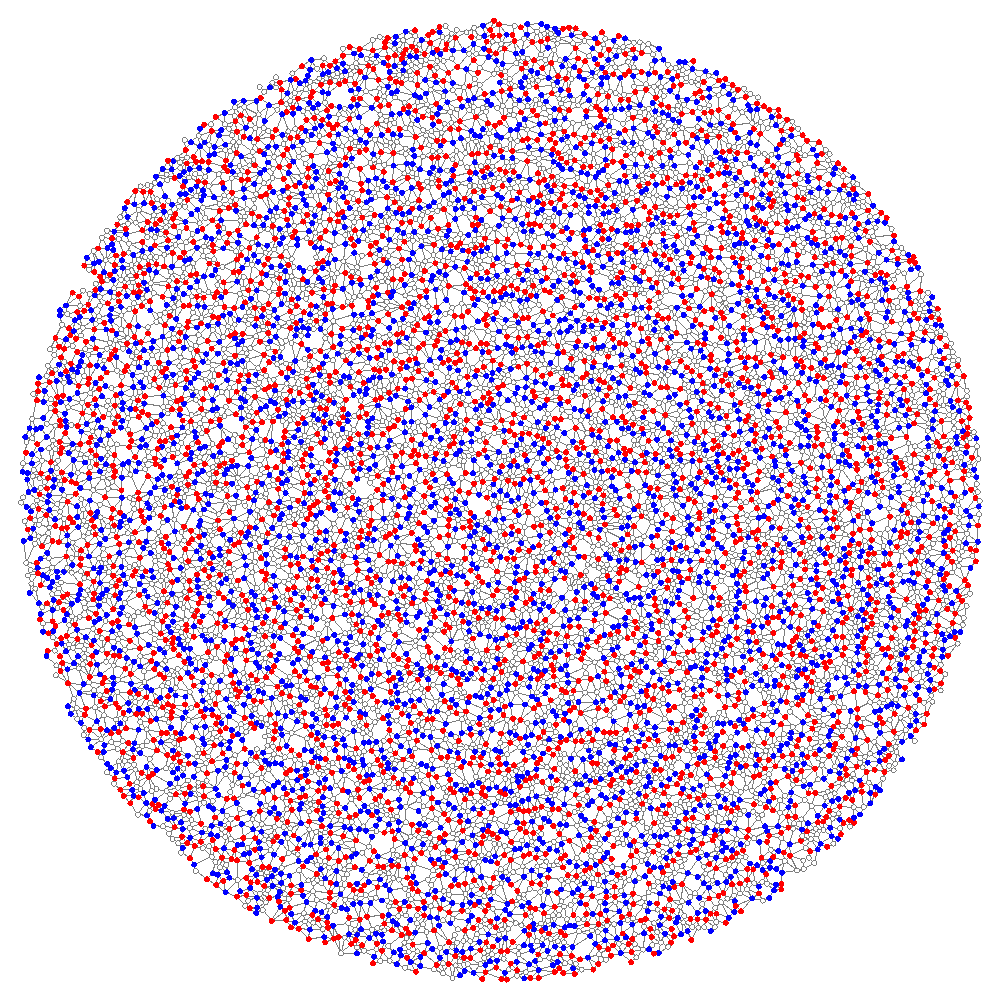}       & \includegraphics[scale=0.08]{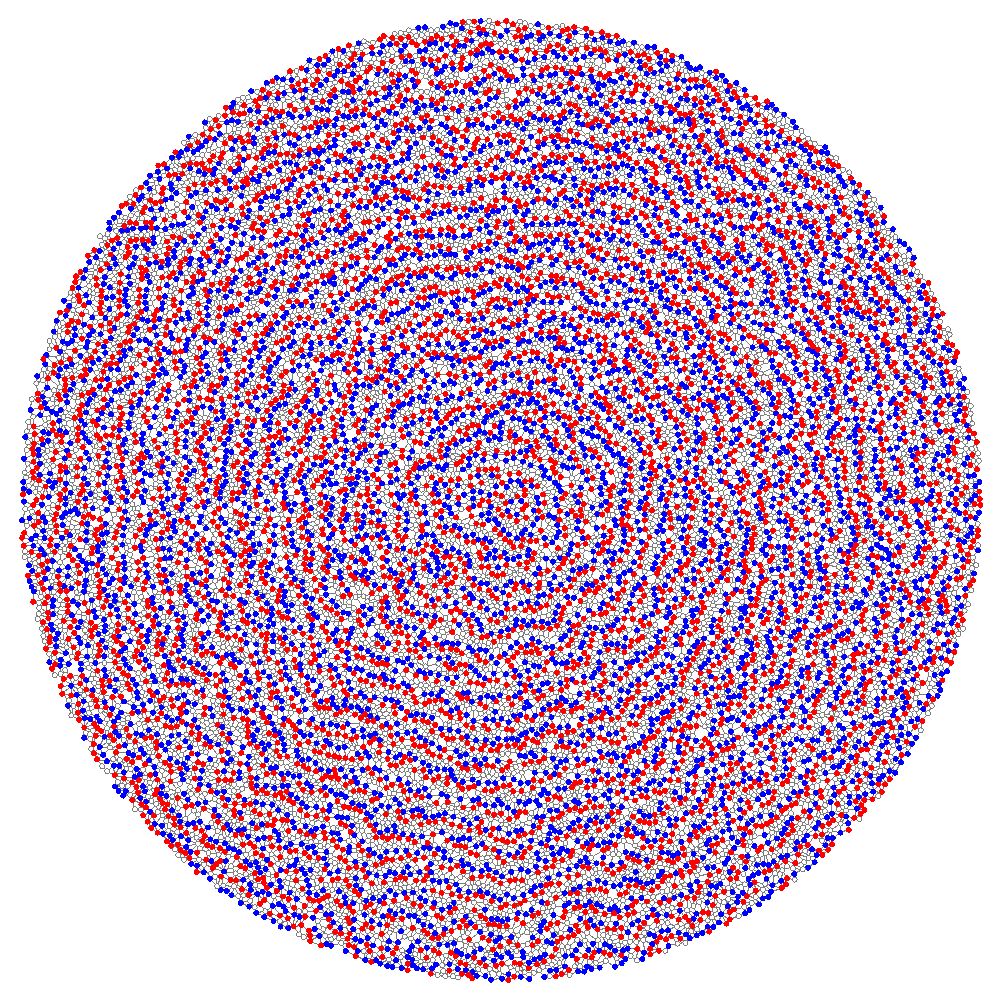}       \\
1.2		&   \includegraphics[scale=0.08]{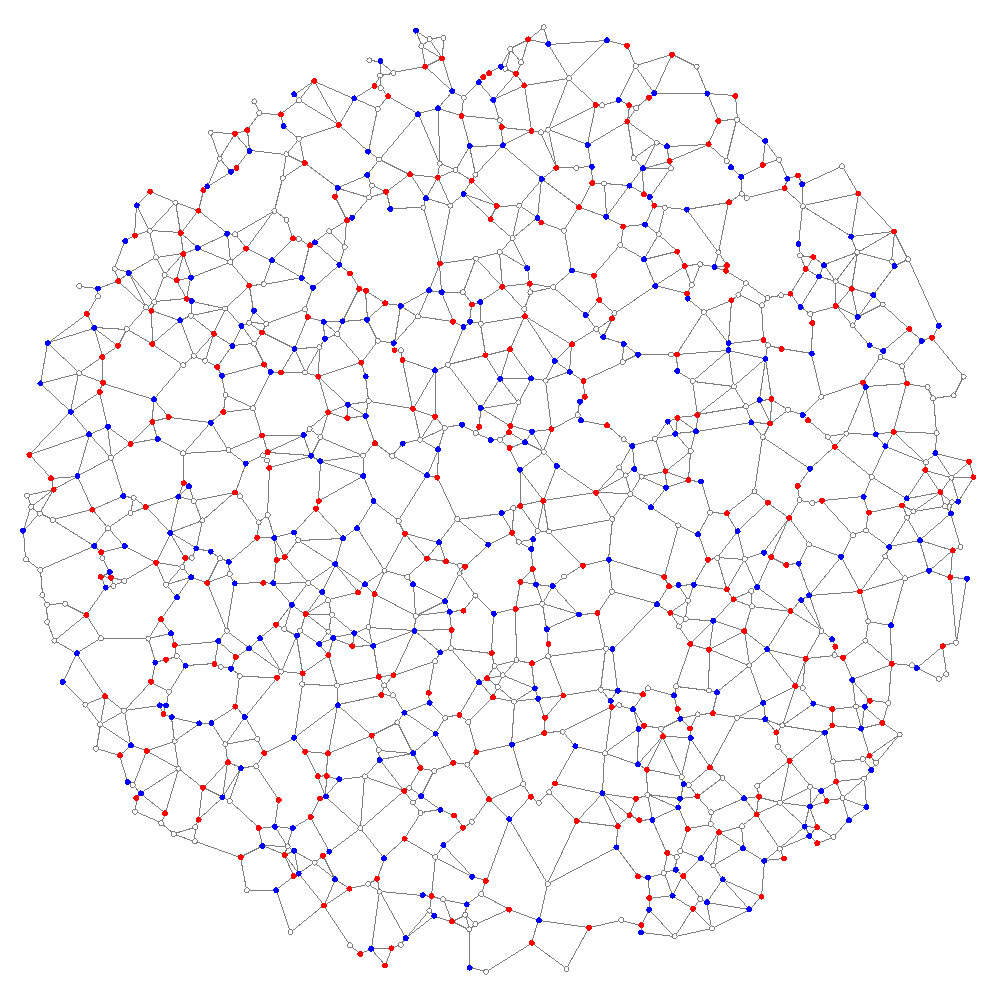}   &  \includegraphics[scale=0.08]{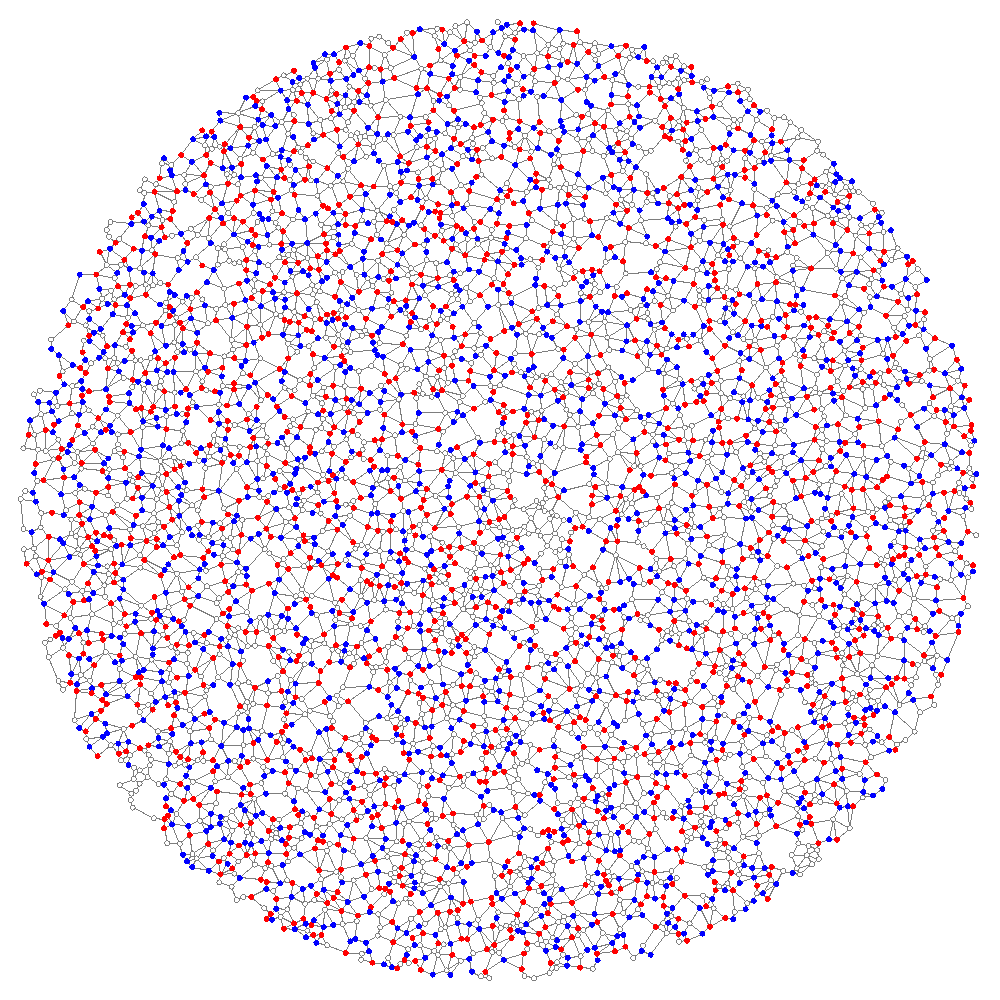}    &\includegraphics[scale=0.08]{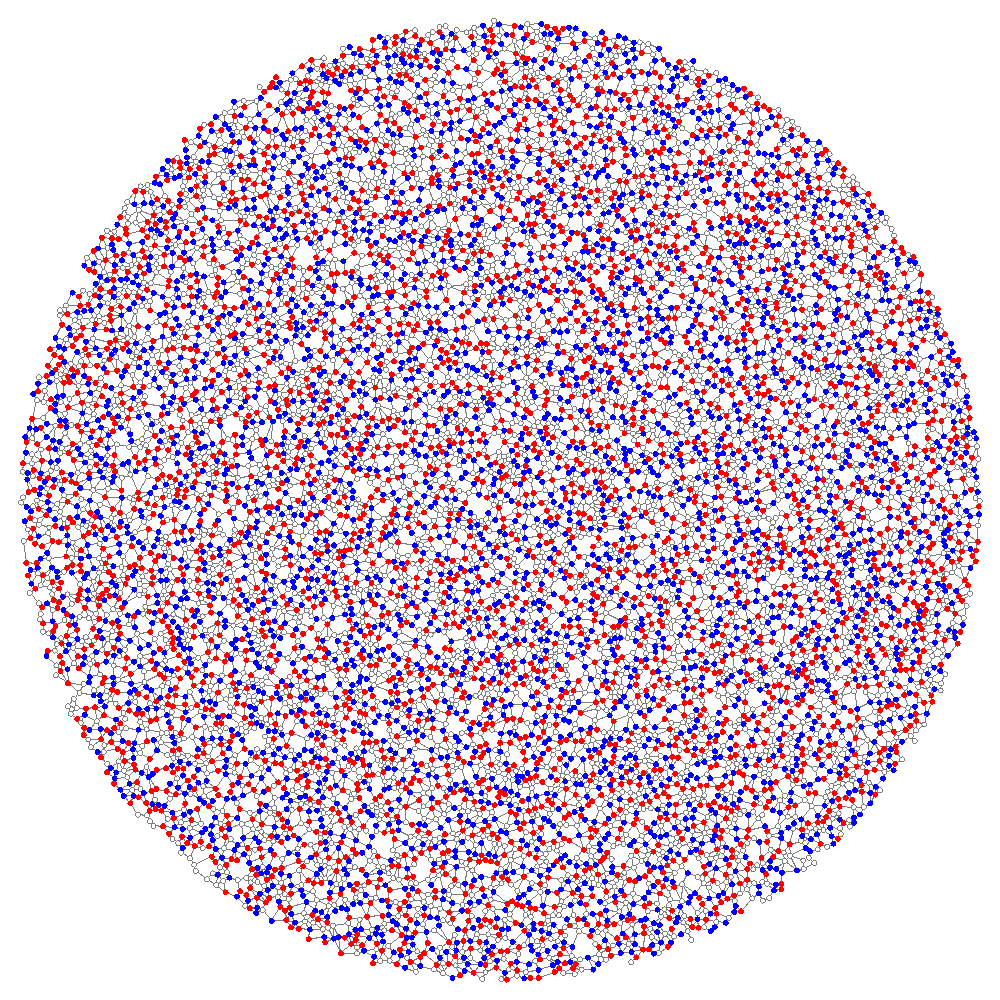}       & \includegraphics[scale=0.08]{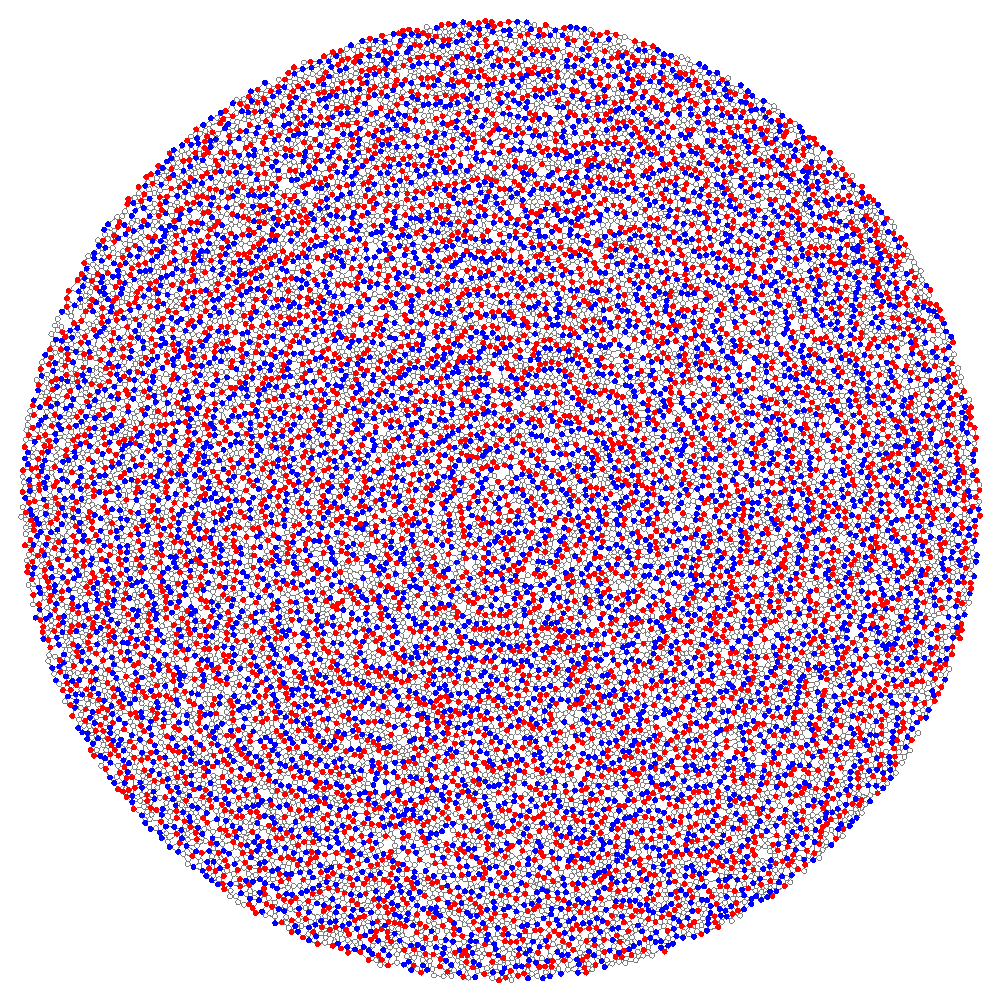}       \\
1.4		&   \includegraphics[scale=0.08]{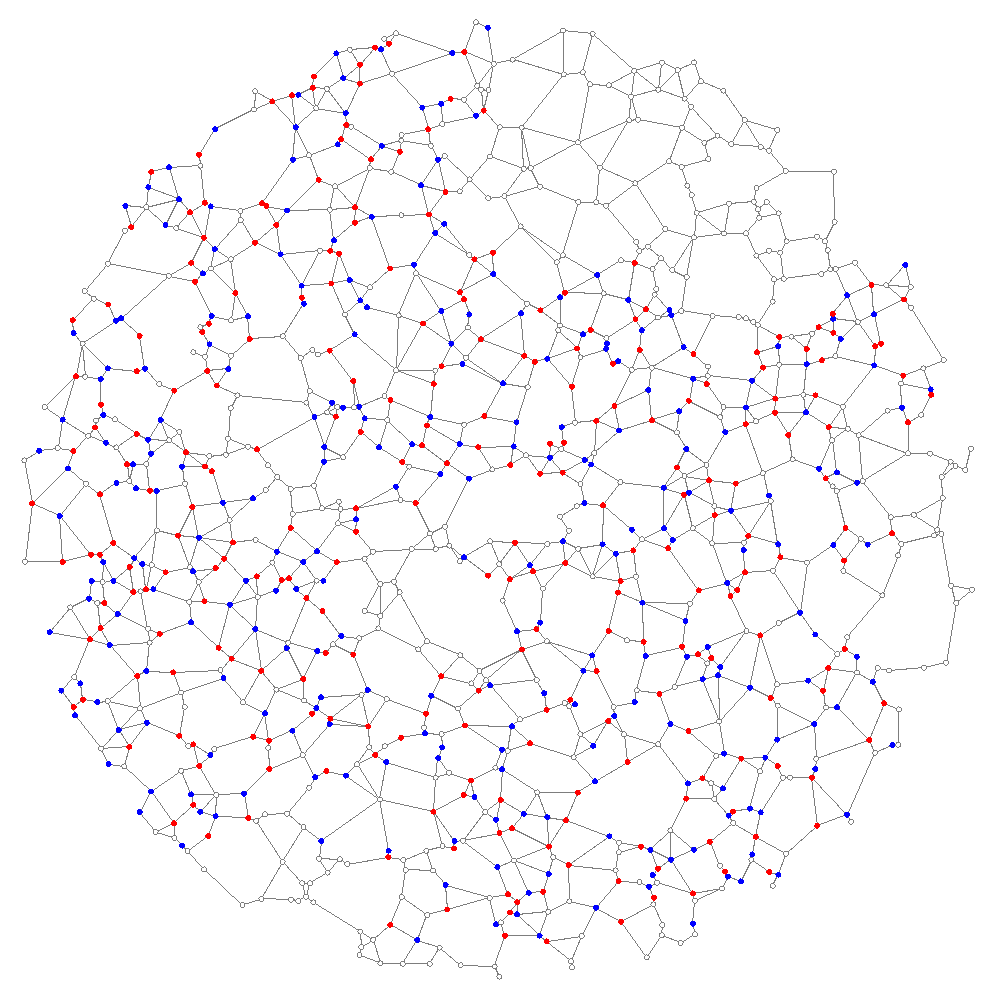}   &  \includegraphics[scale=0.08]{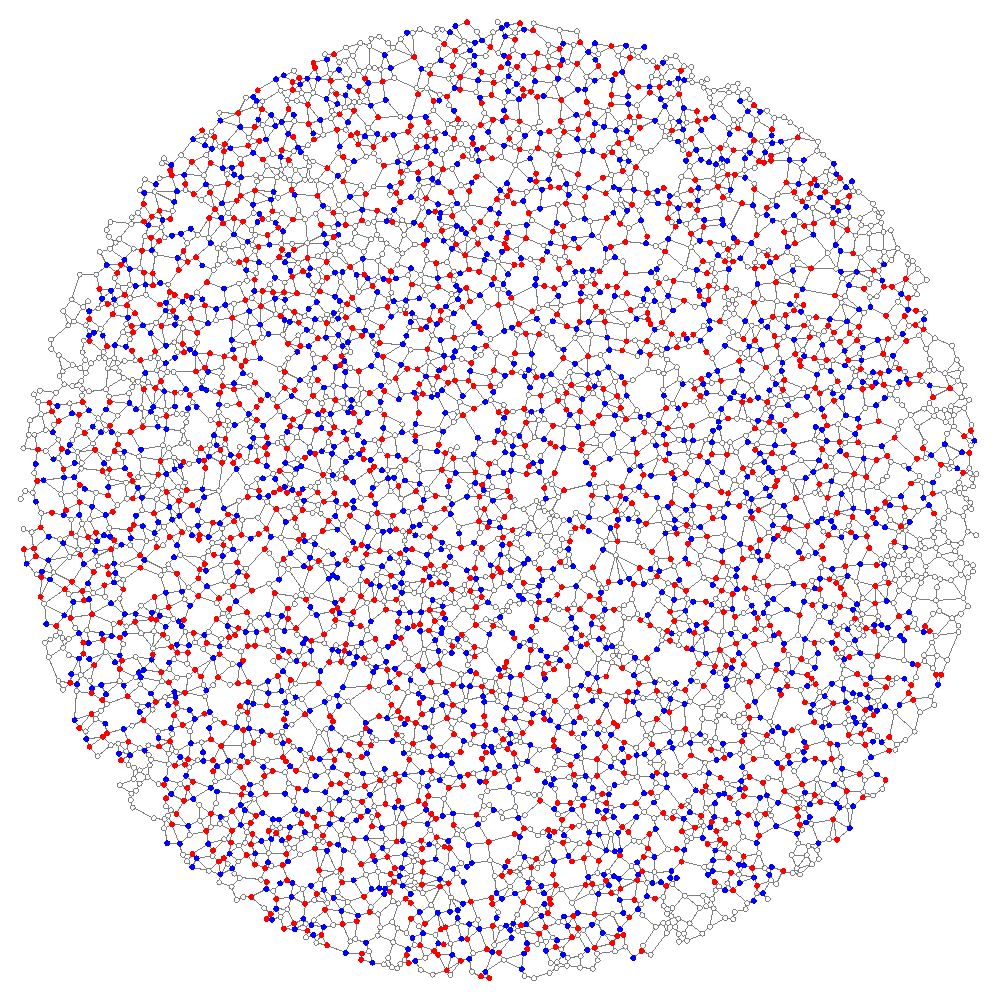}    &\includegraphics[scale=0.08]{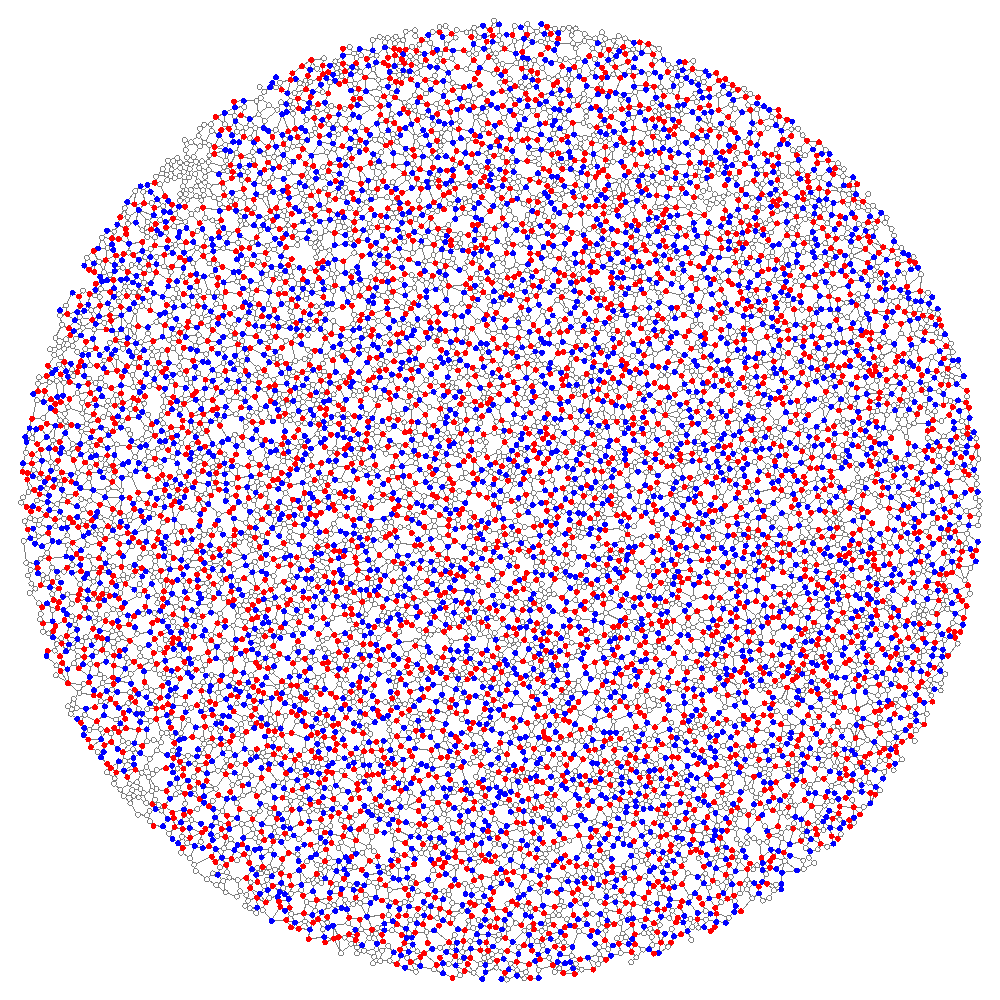}       & \includegraphics[scale=0.08]{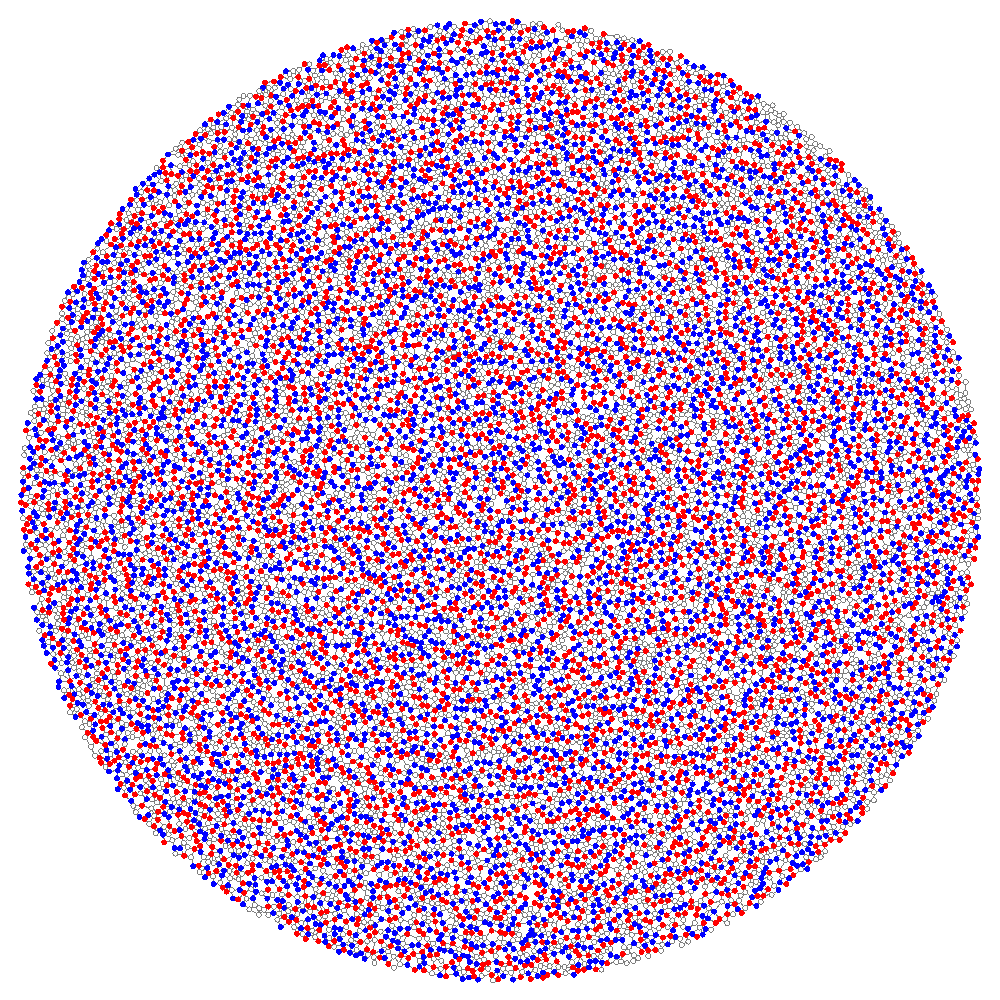}       \\
1.6		&   \includegraphics[scale=0.08]{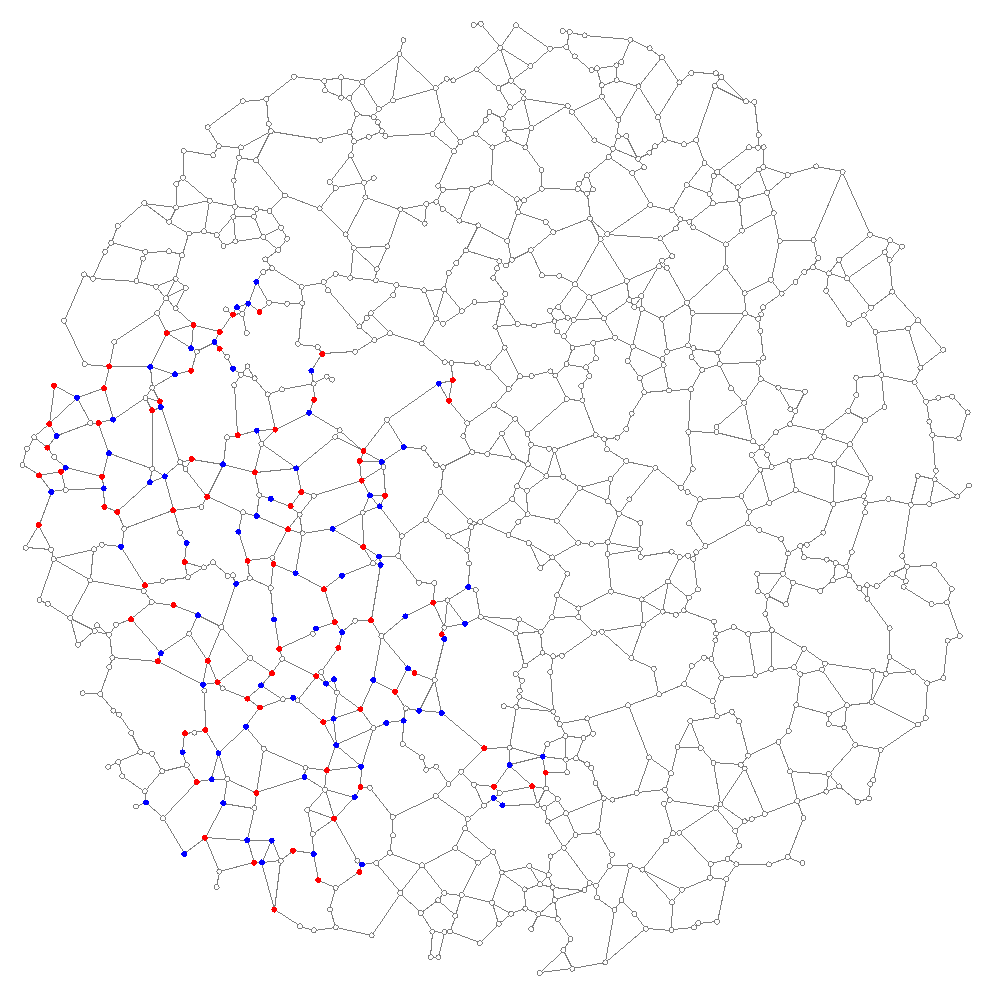}   &  \includegraphics[scale=0.08]{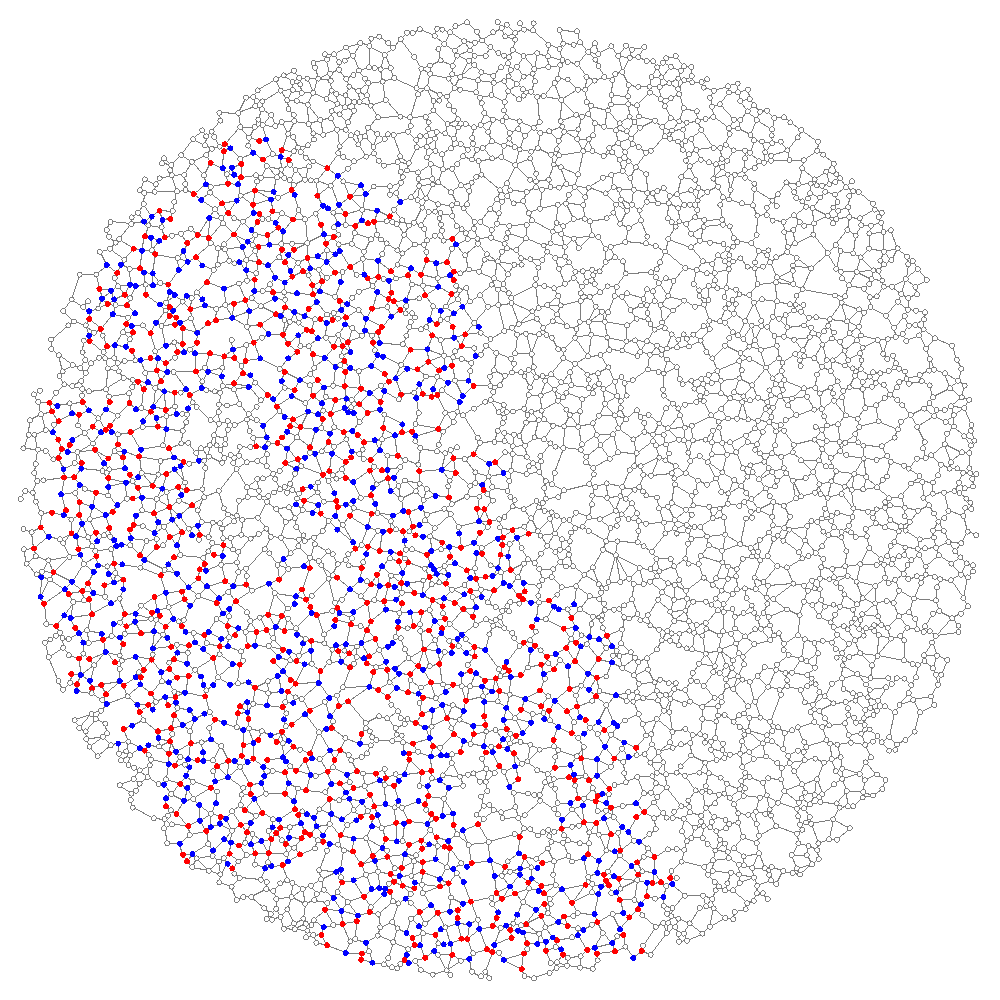}    &\includegraphics[scale=0.08]{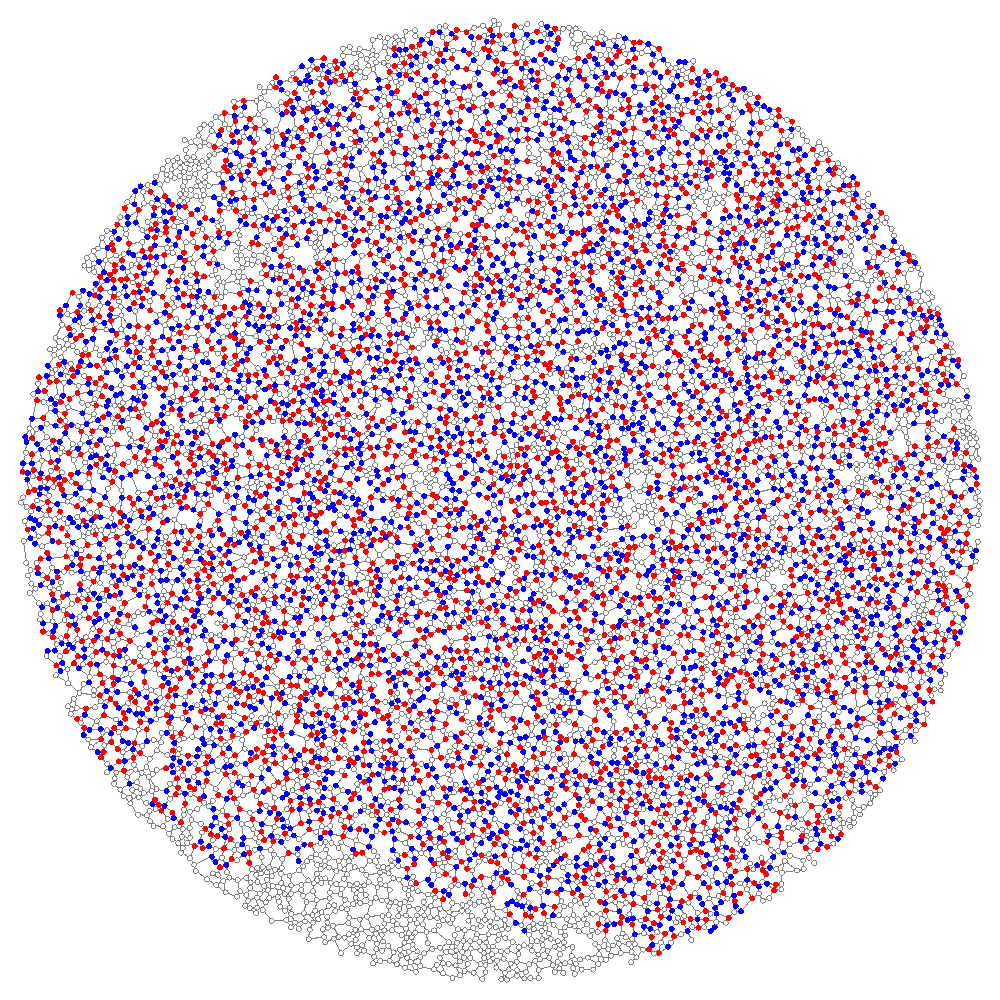}       & \includegraphics[scale=0.08]{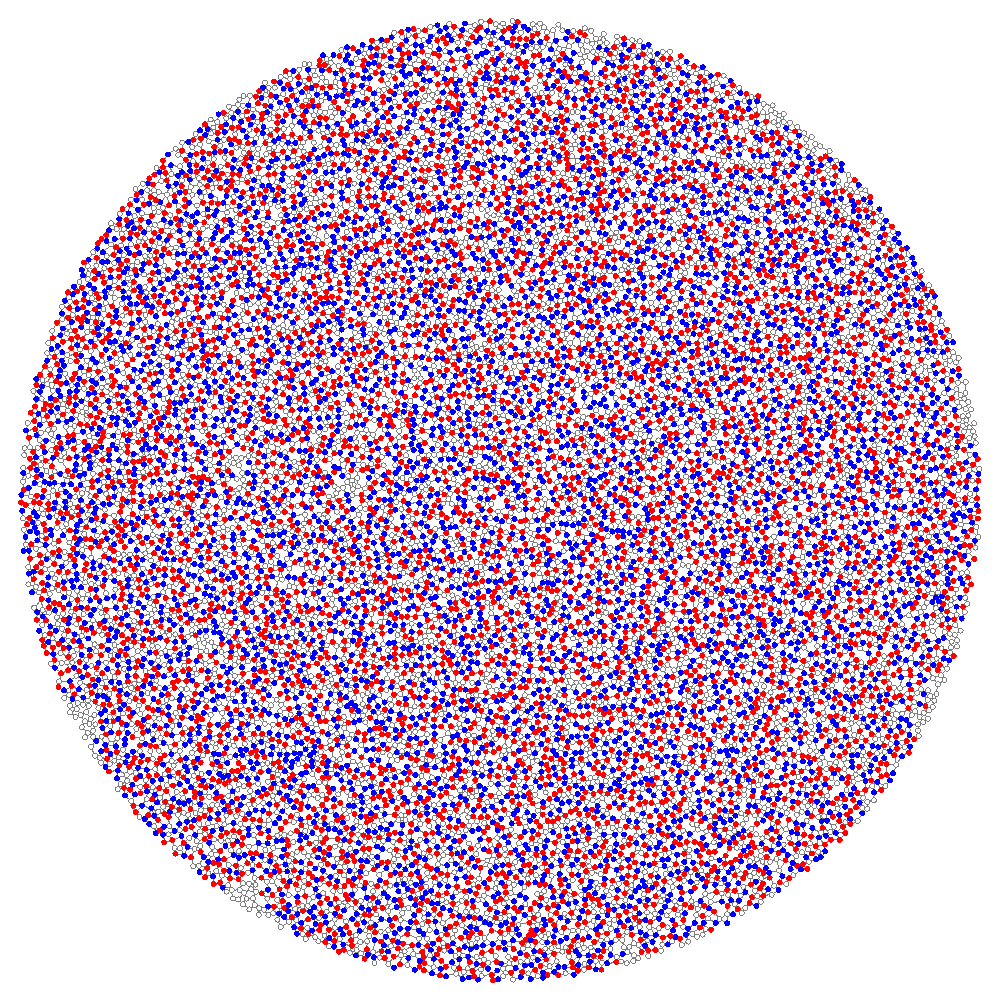}       \\
1.8		&   \includegraphics[scale=0.08]{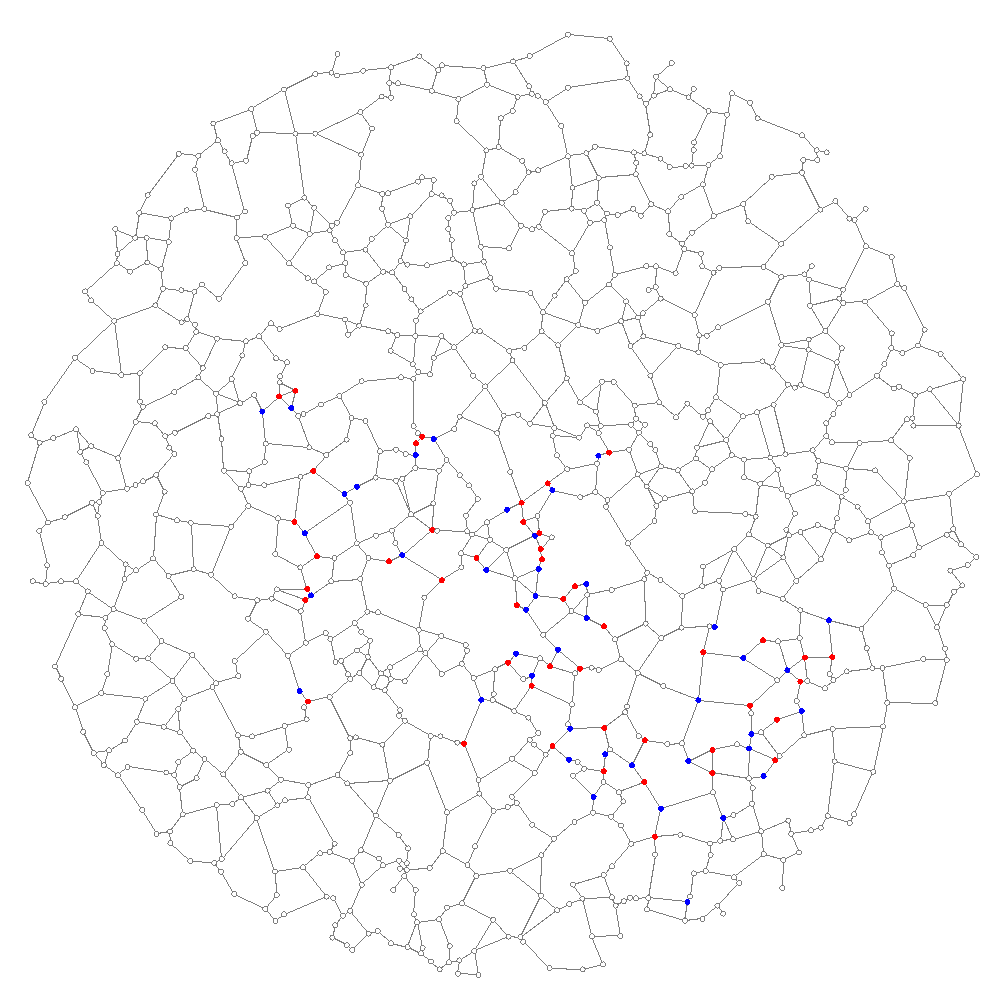}   &  \includegraphics[scale=0.08]{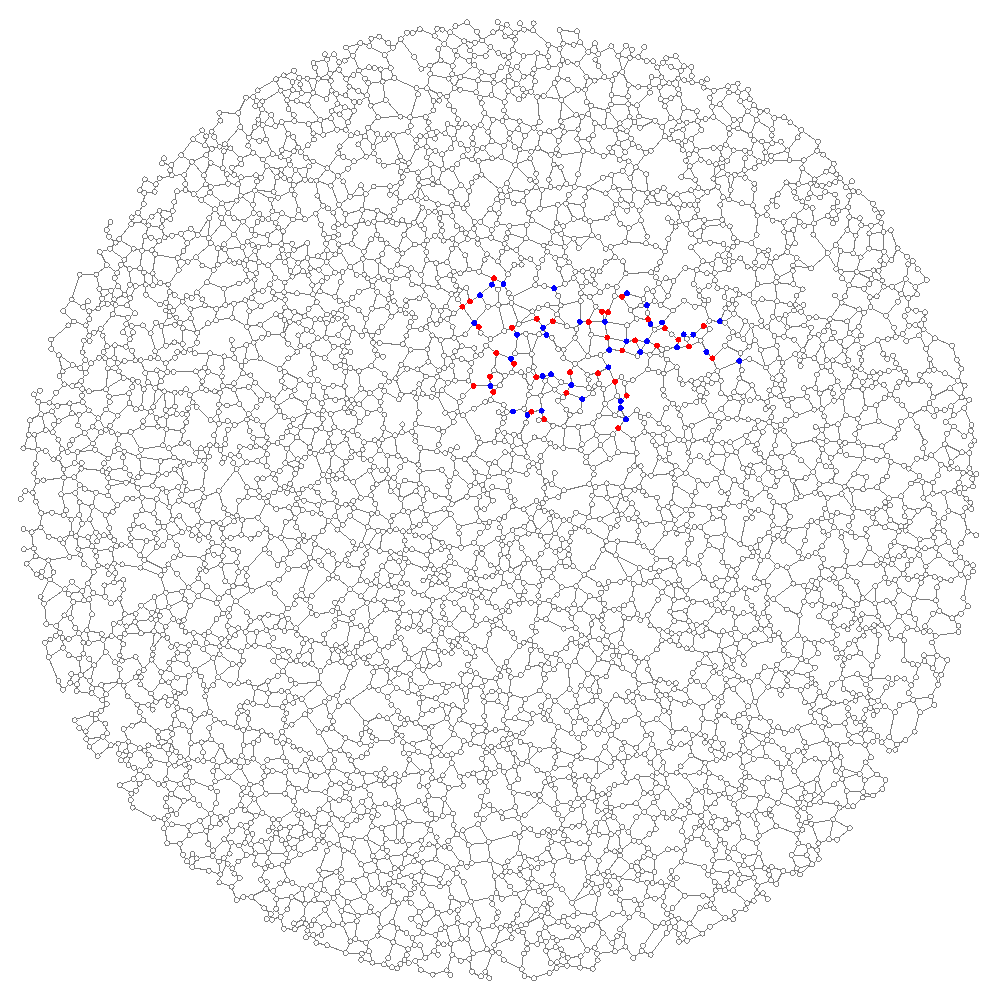}    &\includegraphics[scale=0.08]{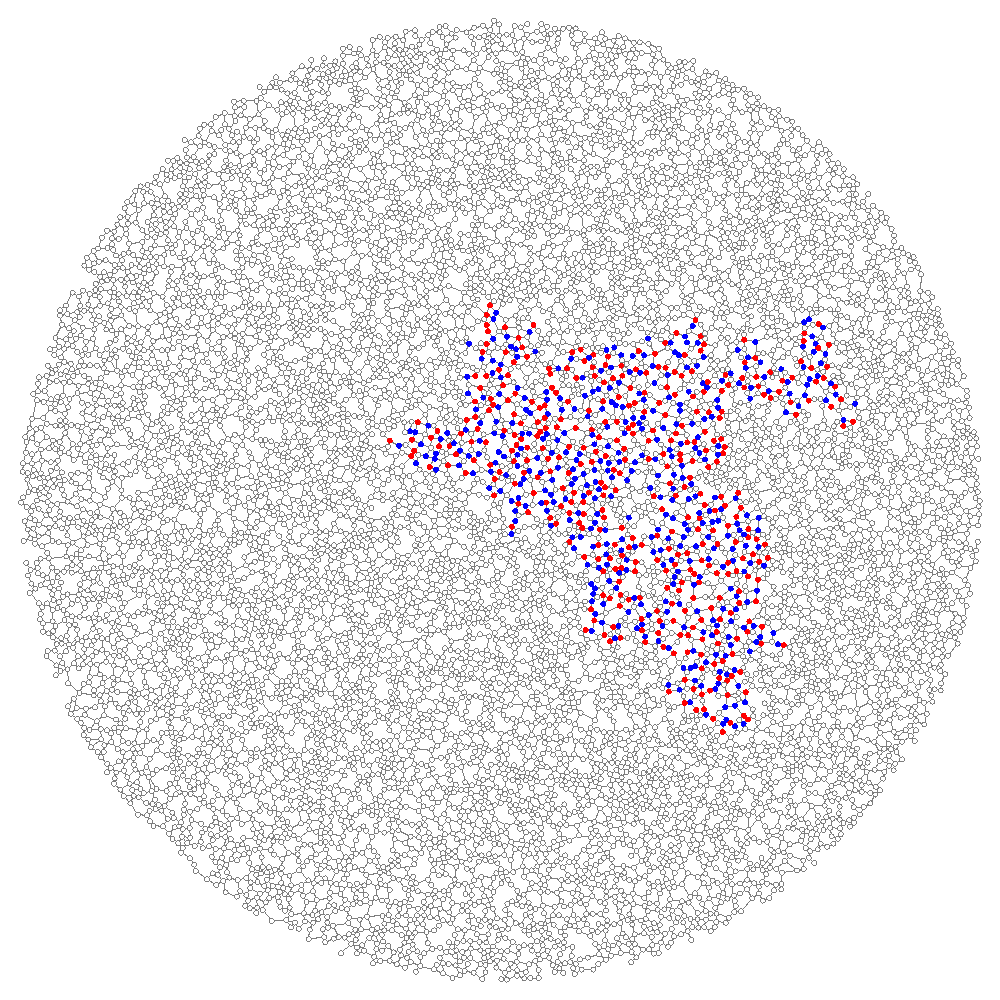}       & \includegraphics[scale=0.08]{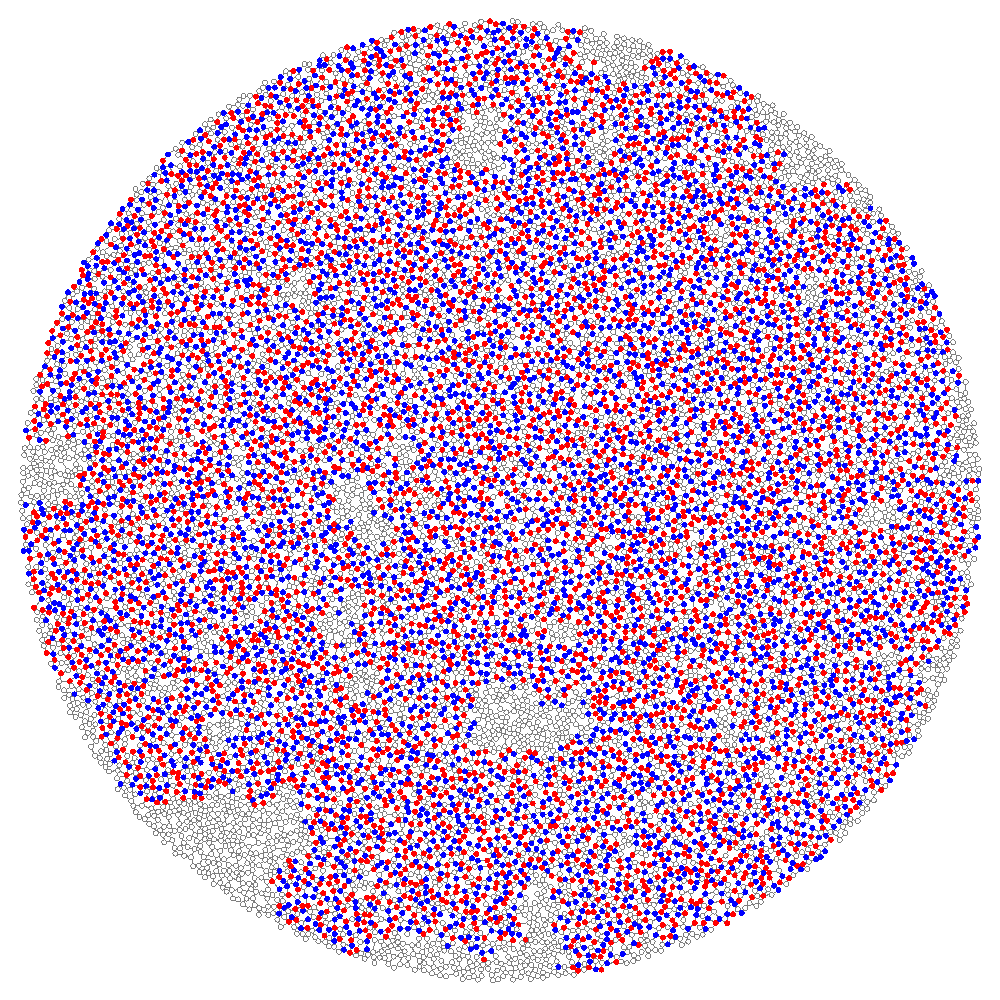}       \\
2.0	&   \includegraphics[scale=0.08]{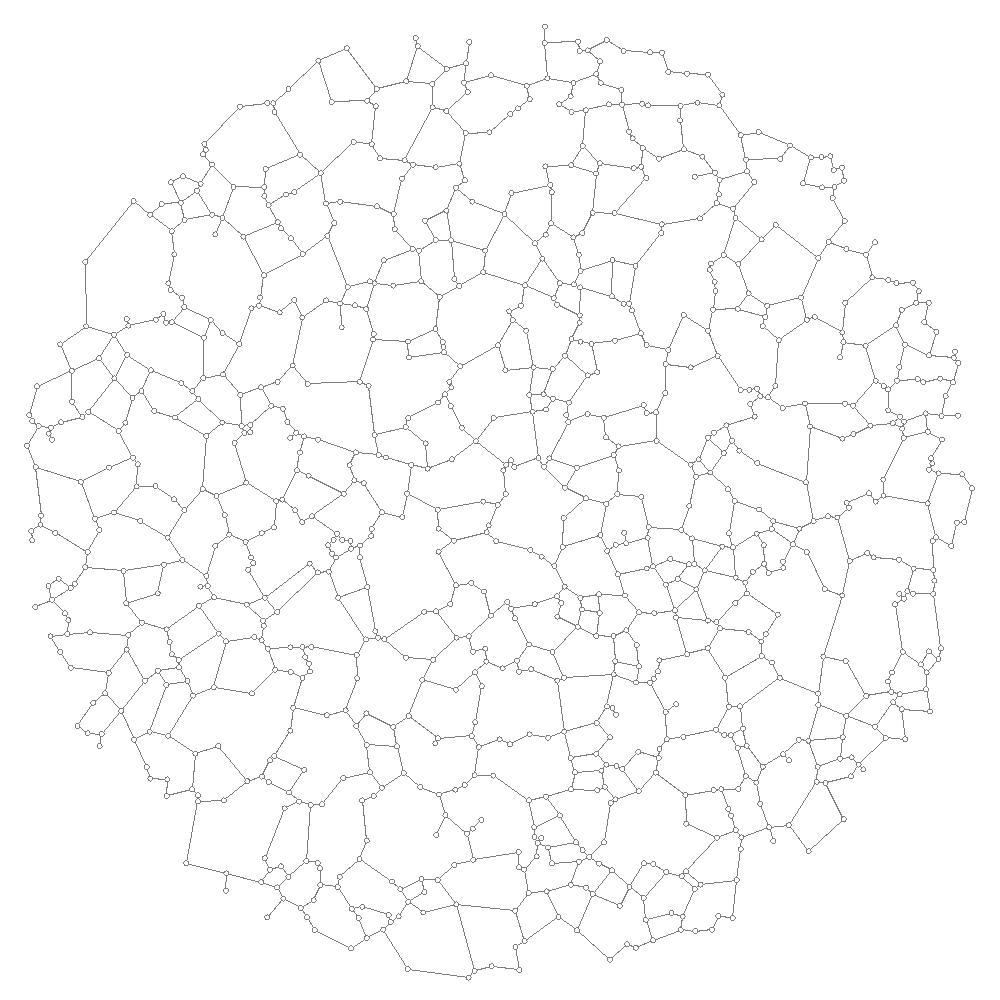}   &  \includegraphics[scale=0.08]{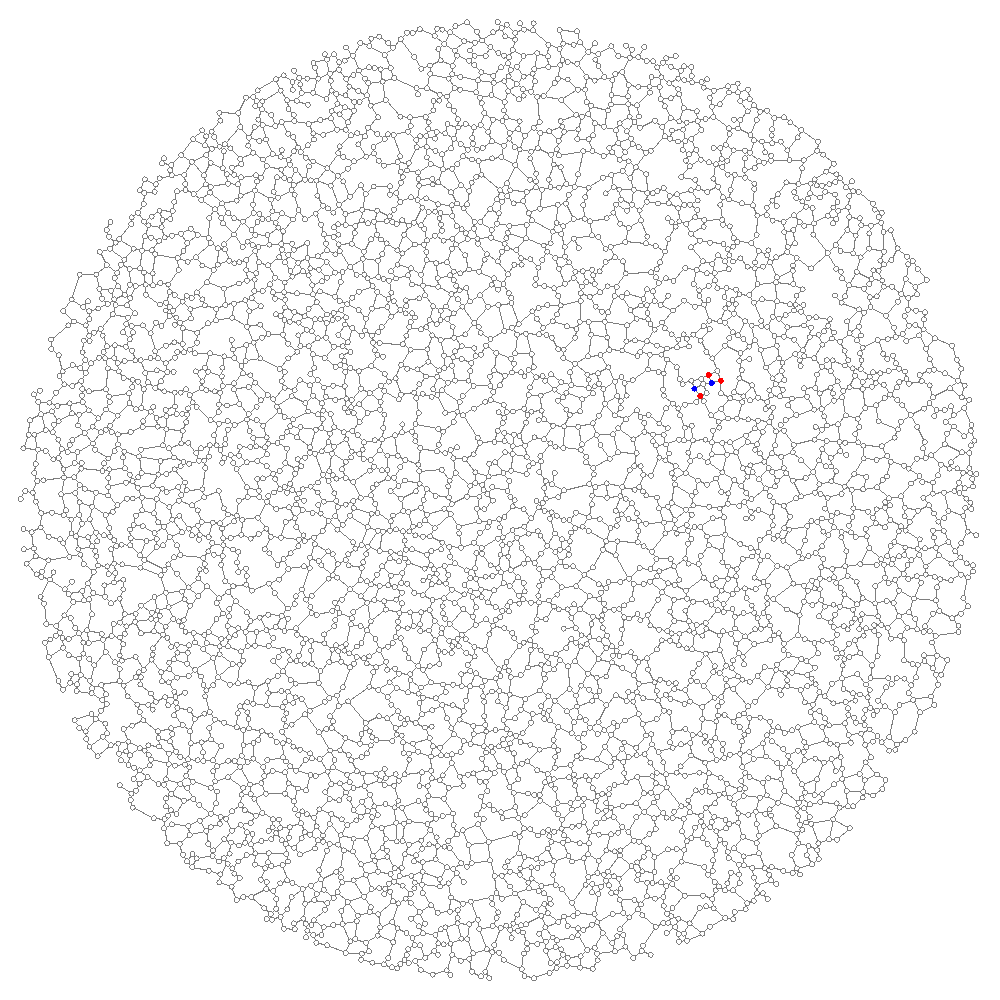}    &\includegraphics[scale=0.08]{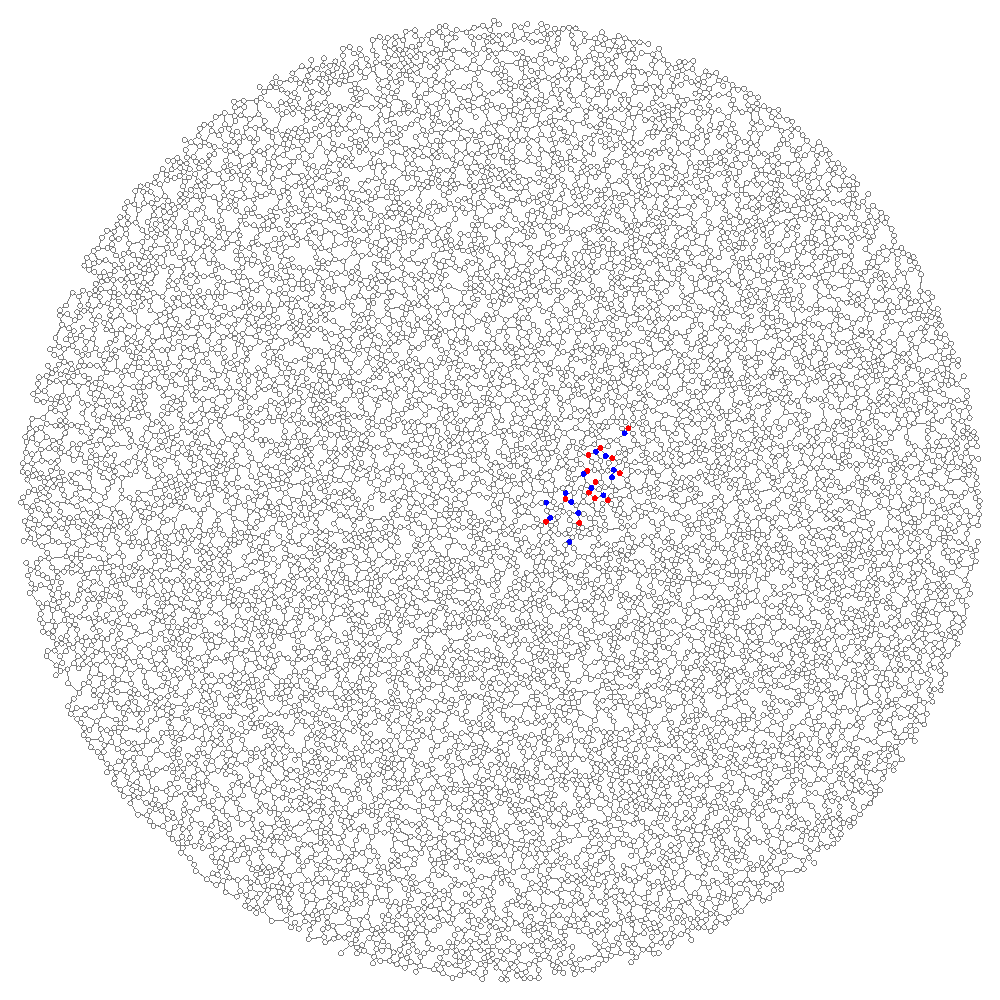}       & \includegraphics[scale=0.08]{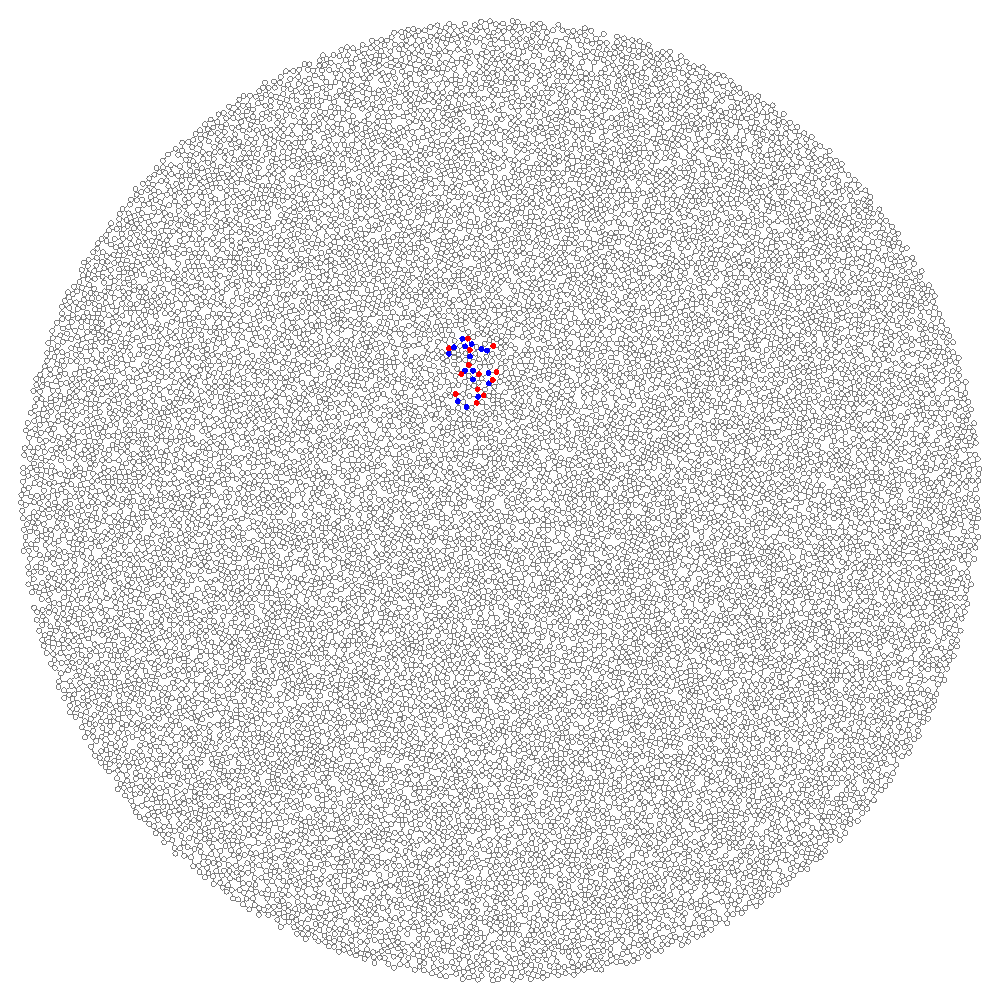}       \\
 \end{array}
 $$
 \caption{Configurations of excitation on $\beta$-skeletons with node density $\varphi=0.027, 0.136, 0.271, 0.407$
 and absolute threshold of excitation $\theta=1$.  All configurations are developed from a localized initial excitation: all nodes
 but are resting, one node is excited. The configurations are  snapped after 50 steps of development. Red (light-gray) coloured
 node are excited, blue (dark-gray) coloured nodes are refractory.}
\label{absolutethreshold}

\vspace{0.5cm}

\end{figure}

Spatial counterparts of integral activity are shown in Fig.~\ref{absolutethreshold}. These are examples 
of excitation patterns emerged after a single node of the resting skeleton was excited. In $\beta$-skeletons with 
high density of nodes packing $\varphi=0.271$ and $0.407$ excitation of a single node of a resting skeleton leads 
to formation of a generator of spiral and target waves if $1 \leq \beta < 1.4$. The most noticeable pattern of the 
waves is observed for $\varphi = 0.407$ and $\beta=1$. Increase of $\beta$ leads to break up wave-fronts and only 
localized compact clusters of activity remain on the skeleton.

What values of $\varphi$ and $\beta$ cause excitation fail to span the whole graph? For what values of $\varphi$ and 
$\beta$ only tiny localized clusters of oscillating activity remain in a skeleton? We call values $\beta$ critical if 
patterns of excitation change abruptly, e.g. excitation dynamics changes from target waves filling the whole skeleton to 
slowly growing, $\beta_d$, or even localized domains of activity, oscillators, $\beta_o$.  
  
\begin{finding}
Critical values $\beta_d$ and $\beta_o$ are linearly proportional to density $\varphi$ of nodes in skeleton.
\end{finding}

\begin{figure}[!tbp]
\centering
\includegraphics[width=0.7\textwidth]{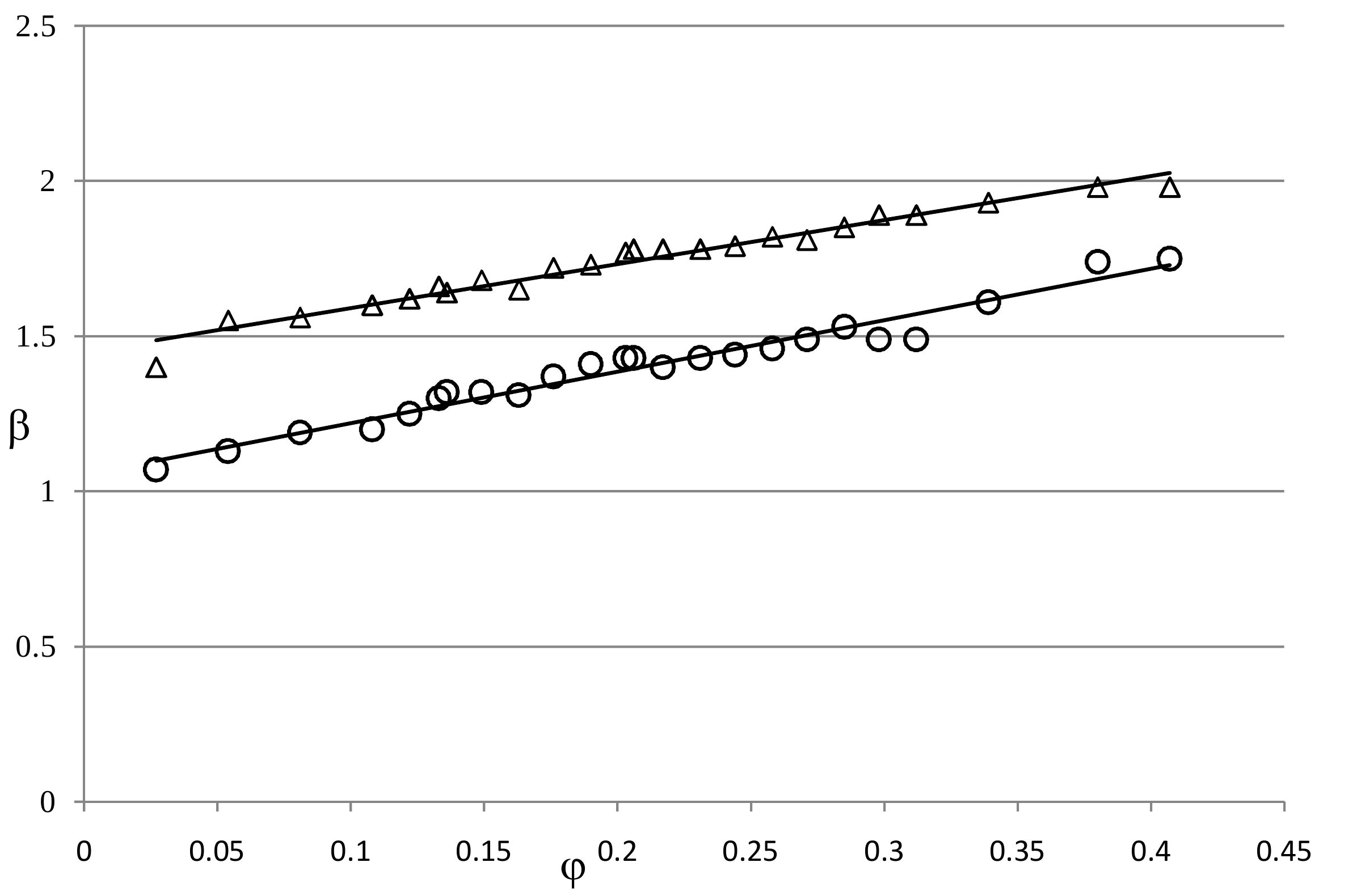}
\caption{Dependence of critical values $\beta_d$ and $\beta_o$ on $\varphi$. Data points for $\beta_d$ are shown by circles, for
$\beta_o$ by triangles. Trend lines are $\beta_d=1.65\varphi + 1.05$, $R^2=0.961$ and $\beta_a=1.42\varphi + 1.45$, $R^2=0.964$.}
\label{phibetacritical}

\vspace{0.5cm}

\end{figure}

Assuming that a skeleton fully occupied by target waves has activity level $\alpha=0.33$ even 
tiniest decrease in $\alpha$ indicates presence of a stationary resting domain. See example in 
Fig.~\ref{absolutethreshold}, $\varphi=0.271$ and $\beta=1.4$. Let us assume that excitation fails to span a skeleton
when $\alpha \leq 0.32$ and only bounded domains of excitation are present if $\alpha \leq 0.25$.  
Let $\beta_d$ and $\beta_o$ be critical values of $\beta$  thus that if for a given $\beta$-skeleton value 
$\beta$ exceeds $\beta_d$ then the skeleton is not fully occupied by excitation, and if $\beta \geq \beta_o$ 
then only localized oscillators are formed. Figure~\ref{phibetacritical} presents critical values of $\beta$ 
computed for 25 sample configurations of $\beta$-skeletons with $\varphi$ varying from 0.027 to 0.407.
What are structural correlates responsible for the transition from a wide-spread excitation to 
localized domains? 

\begin{finding}
Excitation ceases to occupy a whole skeleton when the mode of node degree distribution of the skeleton 
changes from 4 to 3. 
\end{finding}

\begin{figure}[!tbp]
\centering
\begin{scriptsize}
\input{figs/mode.tex}
(a)
\input{figs/average2.tex}
(b)
\end{scriptsize}
\caption{Mode~(a) and average~(b) node degree of $\beta$-skeletons for $\beta \in [1,2]$ and
$\varphi=0.027, \cdots, 0.407$. In (a) entries are boldfaced to increase visibility. In (b) entries corresponding 
to $\beta_o(\varphi)$ are underlined.}
\label{mode}

\vspace{0.5cm}

\end{figure}

We found this by directly comparing $\beta_d$ and integral characteristics of node degree distributions for 
a range of $\beta$ and $\varphi$ (Fig.~\ref{mode}). Mode of node degree distribution is not helpful however 
in detecting skeletons supporting localized oscillators. As shown in Fig.~\ref{mode} skeletons with 
mode 3 of degree distribution can exhibit very large and very small domains of oscillating activity.  
An average node degree of a skeleton gives us a bit more detailed picture of $\varphi$-$\beta$-induced 
structural transitions  and thus can be used to detect when localized oscillators are formed. 

\begin{finding}
A skeleton exhibits only localized oscillating domains of activity when average node degree drops below 3. 
\end{finding}
 
 \begin{figure}[!tbp]
 \centering
 \includegraphics[width=0.6\textwidth]{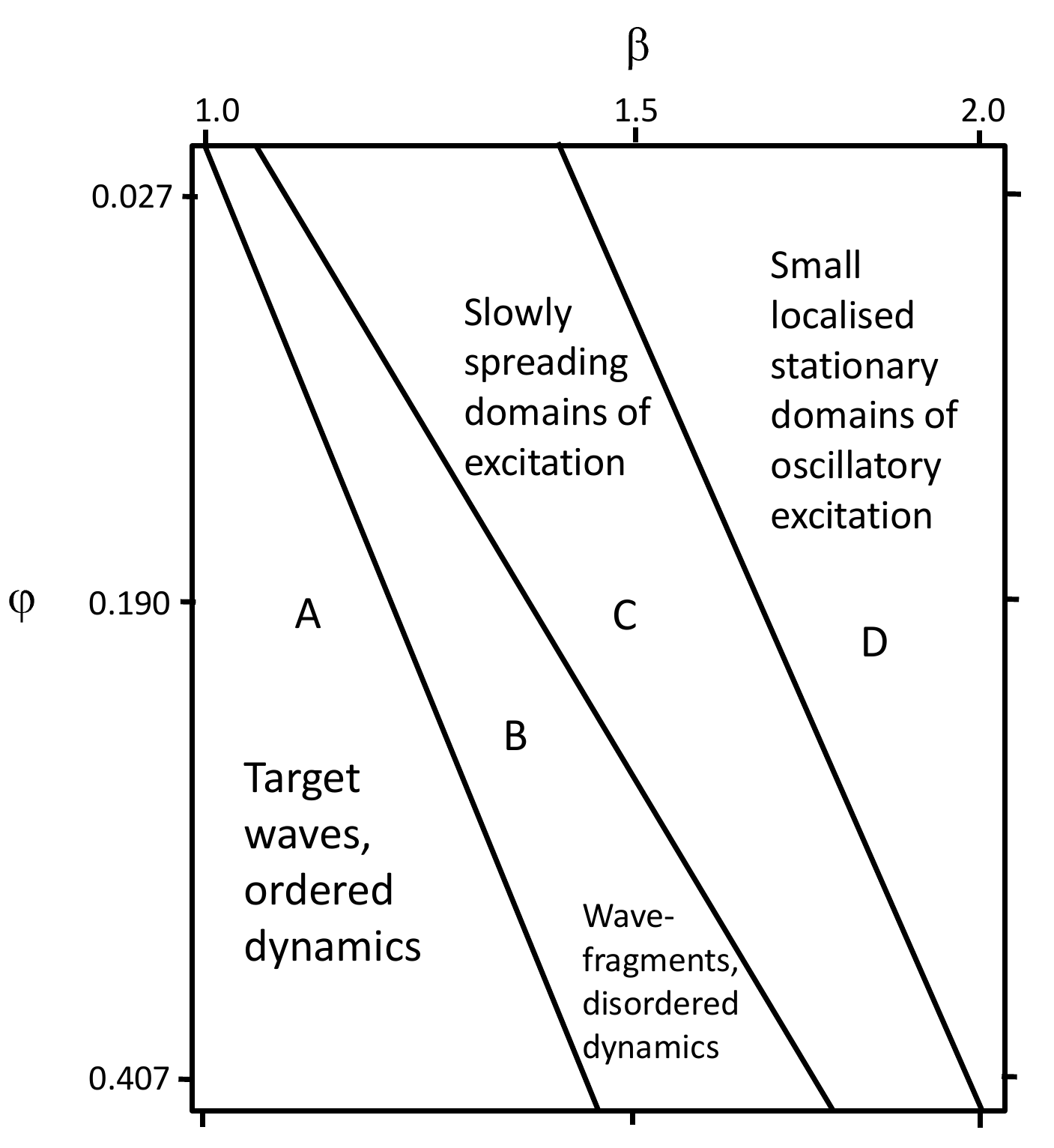}
 \caption{Parameterisaton of a space-time dynamic of $\beta$-skeleton with absolute excitation threshold
 by node density $\varphi$ and $\beta$.}
 \label{parameterisation}

\vspace{0.5cm}

 \end{figure}
 
The average node degrees corresponding to onset of localized oscillations are underlined in Fig.~\ref{mode}.
Further comparison of average node degrees (Fig.~\ref{mode}), and corresponding space-time configurations of 
activity (see e.g. Fig.~\ref{absolutethreshold}) allows us to state the following observation. Average node 
degree $\delta$ of  a skeleton determines space-time dynamics of excitation as follows: 
\begin{itemize}
\item $3.9 \leq \delta$: classical wave-like dynamics, generators of target waves, ordered dynamics (Fig.~\ref{parameterisation}A),
\item $3.7 \leq \delta < 3.9$: generators of wave-fragments, disordered excitation activity (Fig.~\ref{parameterisation}B),
\item $3.0 \leq \delta < 3.7$: formation of slowly propagating domains of excitation, some loci of a skeleton remain resting (Fig.~\ref{parameterisation}C),
\item $\delta \leq 3.0$: only small localized stationary domains of oscillating activity are formed (Fig.~\ref{parameterisation}D). \end{itemize}


\section{Relatively excitable skeletons}
\label{relativeexcitation}

 \begin{figure}[!tbp]
 \centering
 \subfigure[$\beta=1$]{ \includegraphics[width=0.7\textwidth]{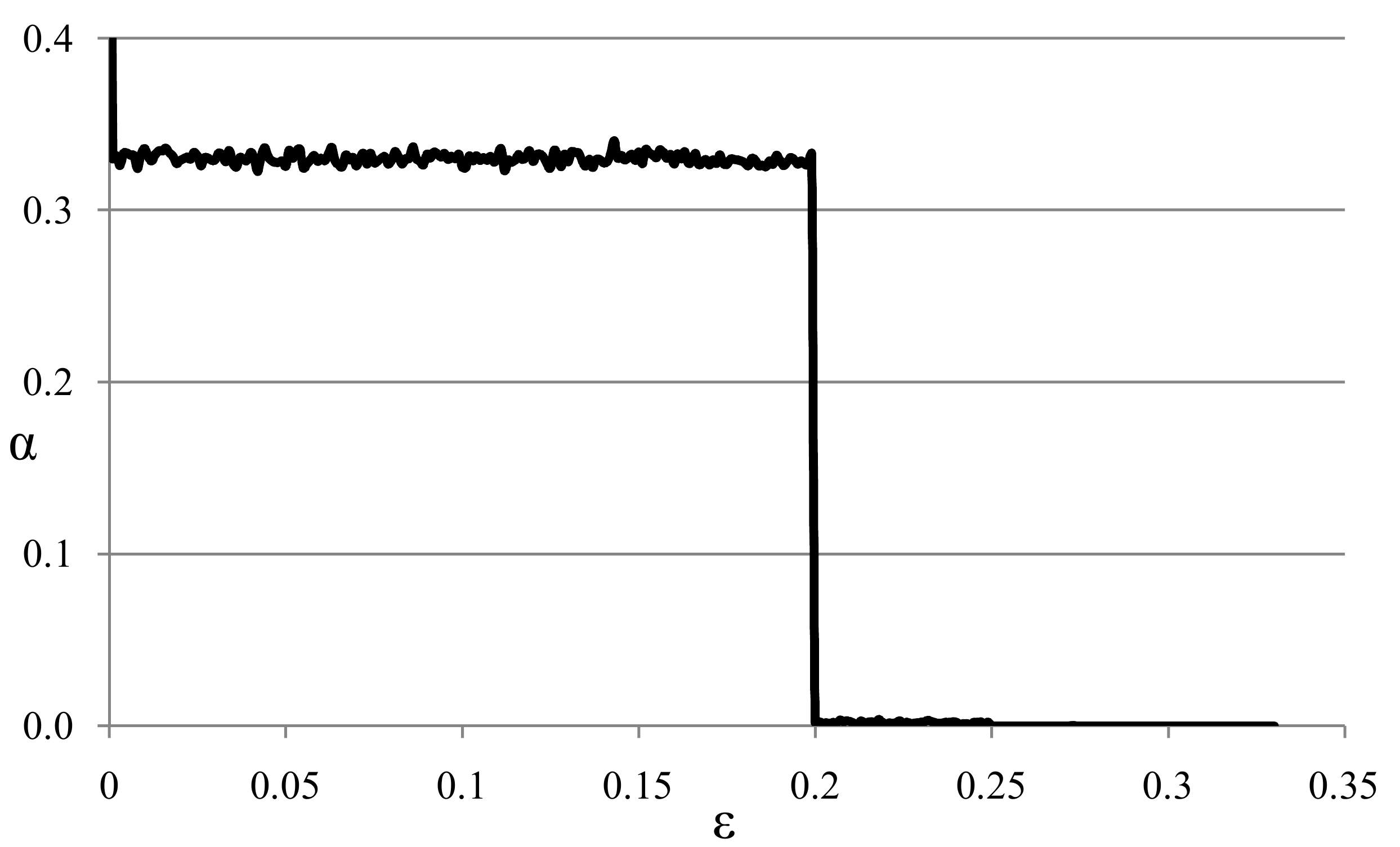}}
  \subfigure[$\beta=1.5$]{ \includegraphics[width=0.7\textwidth]{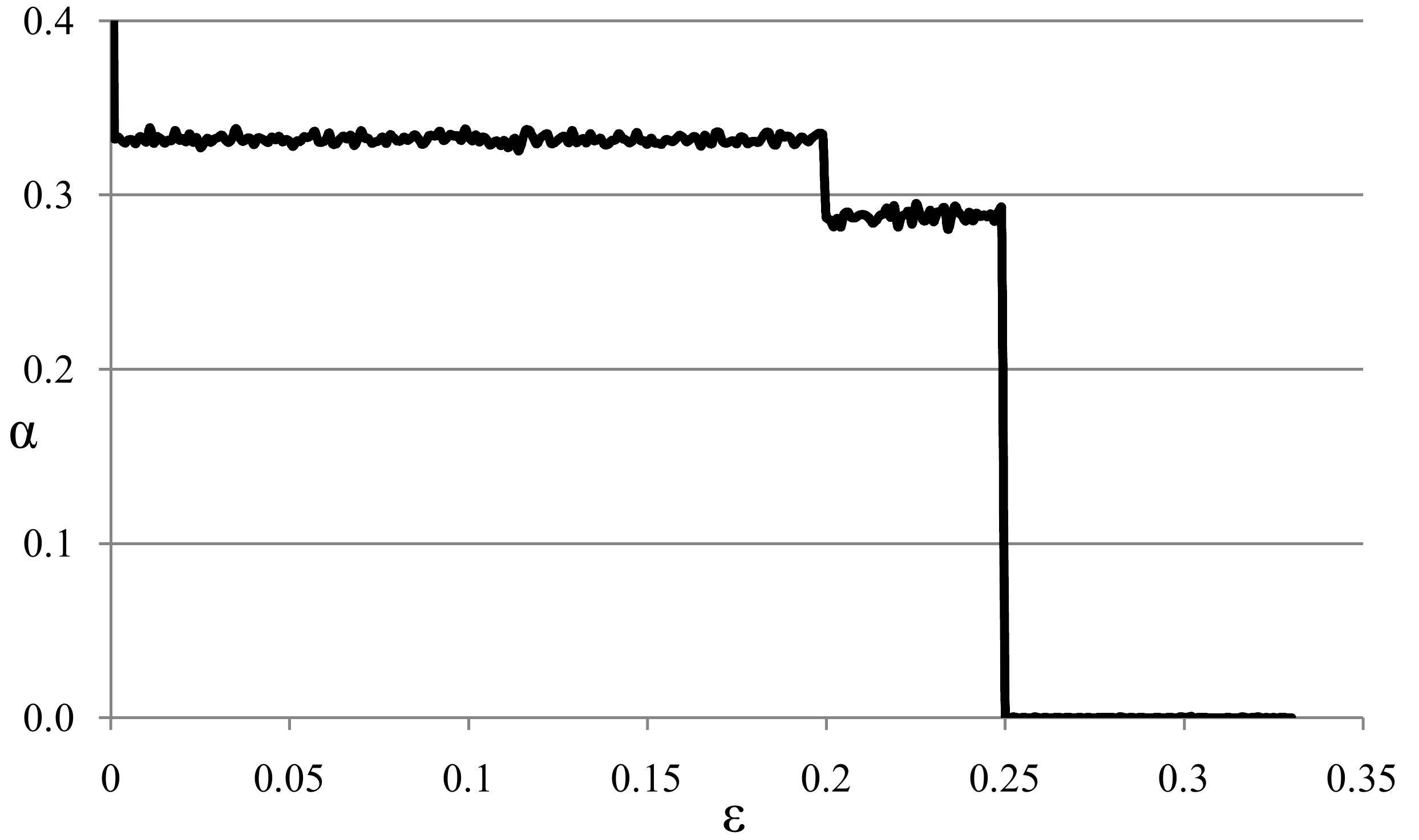}}
    \subfigure[$\beta=2$]{ \includegraphics[width=0.7\textwidth]{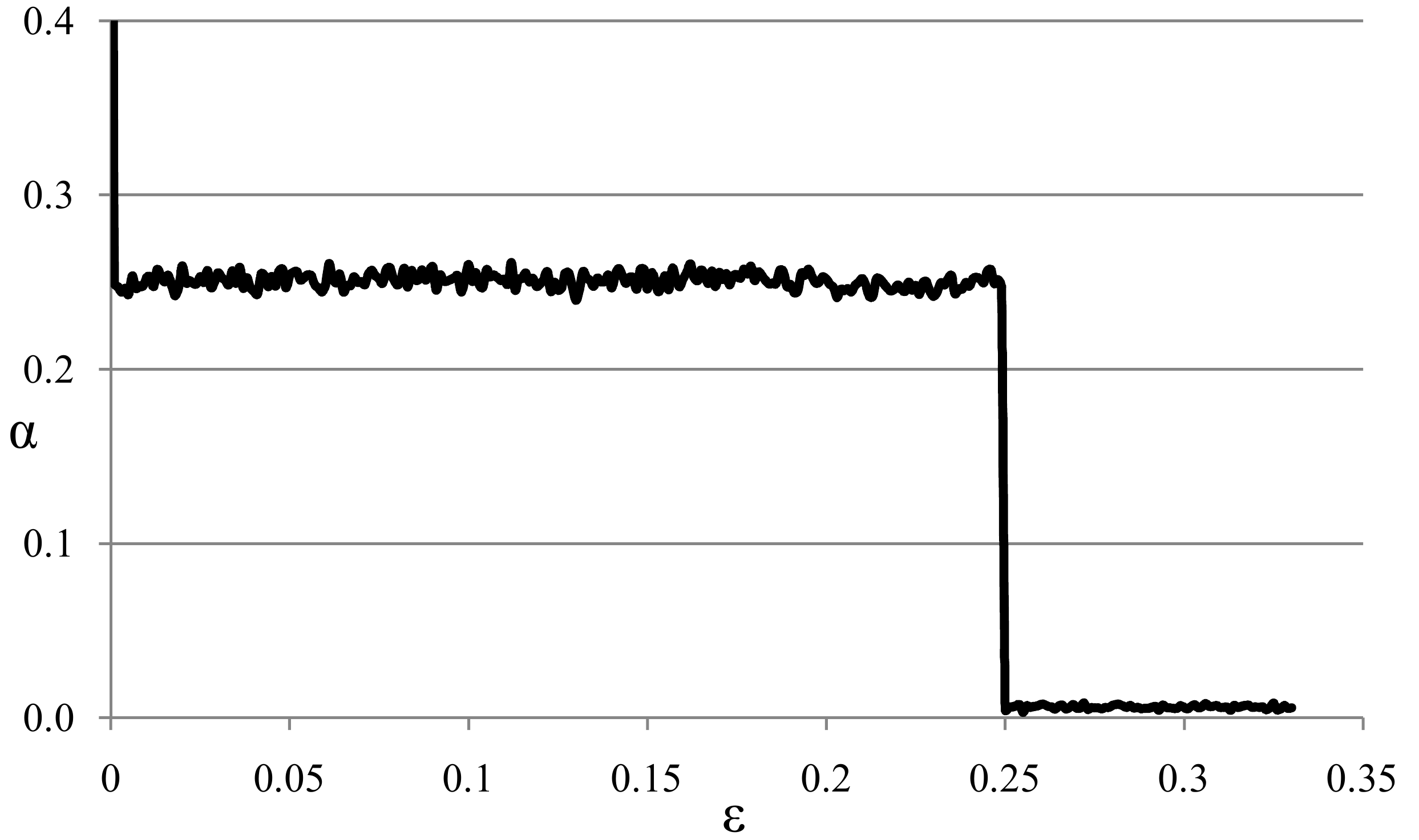}}
  \caption{Activity $\alpha$ versus relative excitation threshold $\epsilon$ of $\beta$-skeletons
  for (a)~$\beta=1$, (b)~$\beta=1.5$, (c)~$\beta=2$.}
  \label{integralactivity}  

\vspace{0.5cm}

 \end{figure}

Here we consider behavior of skeletons with node density $\varphi=0.407$. Skeletons with density $\varphi=0.407$ (15,000 disc-nodes) show a wide representation of structural properties because they exhibit widest range of modes of node degree distributions (Fig.~\ref{mode}).   In computational experiments we perturbed a  resting skeleton with small excitation: 
every node is assigned an excited state with probability 0.1. We allowed the skeleton to develop its excitation patterns for 100 iterations and recorded activity $\alpha$ then.  Examples of integral activity of skeletons for $\beta=1$, 1.5 and 2 are shown in Fig.~\ref{integralactivity}. We see that drastic changes in activity levels occur when $\epsilon$ changes from 0.199 to 2 (first drop in activity level is most clear in Fig.~\ref{integralactivity}a), and when $\epsilon$ changes from 0.249 to 0.25 (second drop in activity level, see e.g.  Fig.~\ref{integralactivity}b).

 \begin{figure}[!tbp]
 \centering
 \input{figs/relative_beta_epsilon_alpha}
 \caption{Dependence of activity level $\alpha$ of $\beta$-skeletons on $\beta$ and $\epsilon$. 
 Initially resting skeleton is set to a random configuration of excitation by exciting each node
 with probability 0.1. Values of $\alpha$ are measured 100 iterations after initial excitation 
 of skeletons. }
 \label{relativebetaepsilonalpha}

\vspace{0.5cm}

 \end{figure}

More detailed description of how $\alpha$ depends on $\beta$ is provided in Fig.~\ref{relativebetaepsilonalpha}. 
For $\epsilon \in [0, 0.2[$ activity level $\alpha$ is not changed with increase of $\beta$, $1 \leq \beta \leq 1.7$. 
However as soon as $\beta$ reaches $1.8$ activity level slightly decreases. The activity level drops down by one third when $\beta=2$ (Fig.~\ref{relativebetaepsilonalpha}). For $\epsilon \in [0.25, 0.33[$ activity monotonously increases with increase of $\beta$.  For $\epsilon \in [0.2, 0.249[$ activity $\alpha$ grows with growth of $\beta$ till $\beta=1.6$. The activity 
remains unchanged for $\beta=1.7$ and then decreases with further increase of $\beta$. Let us discuss space-time 
dynamics of these $\beta$-skeletons for selected values of $\beta$.

\subsection{$\beta=1$}
 
  \begin{figure}[!tbp]
 \centering
 \subfigure[$\beta=1$, $\epsilon=0.01$]{ \includegraphics[width=0.43\textwidth]{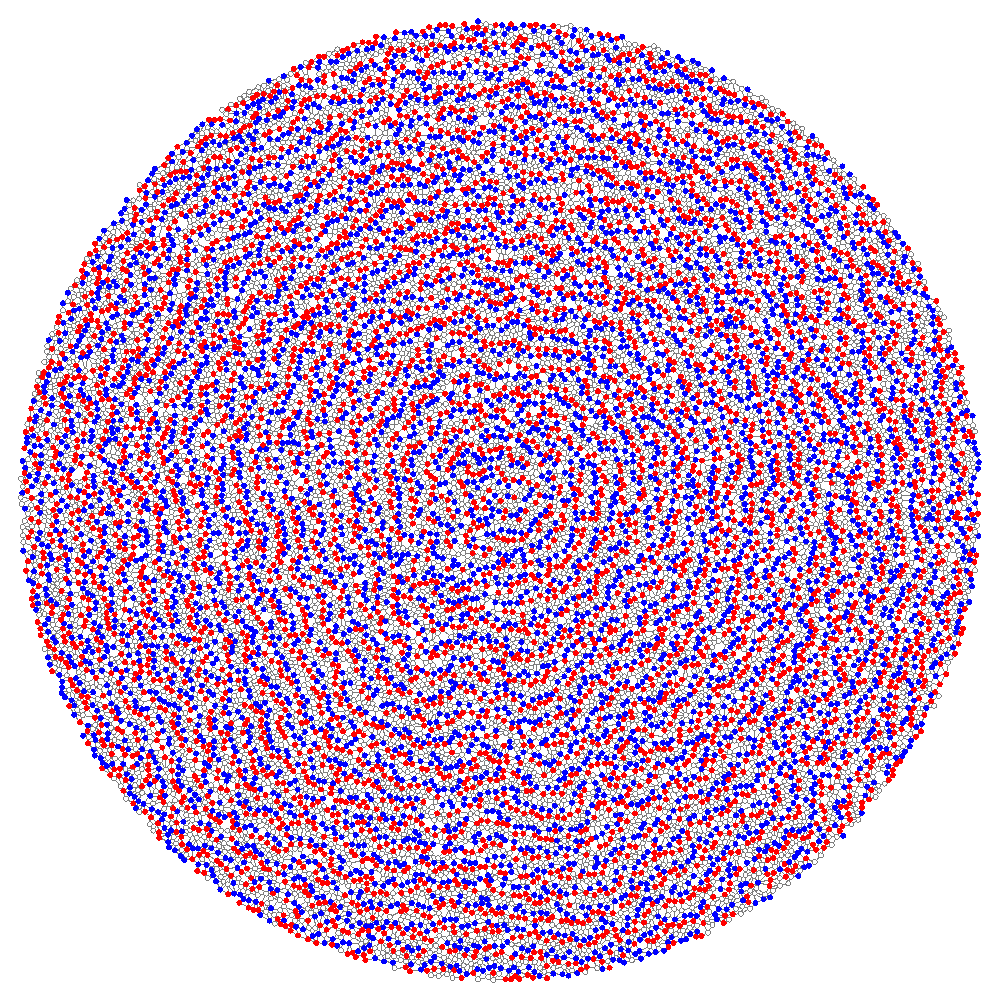}}
  \subfigure[$\beta=1$, $\epsilon=0.15$]{ \includegraphics[width=0.43\textwidth]{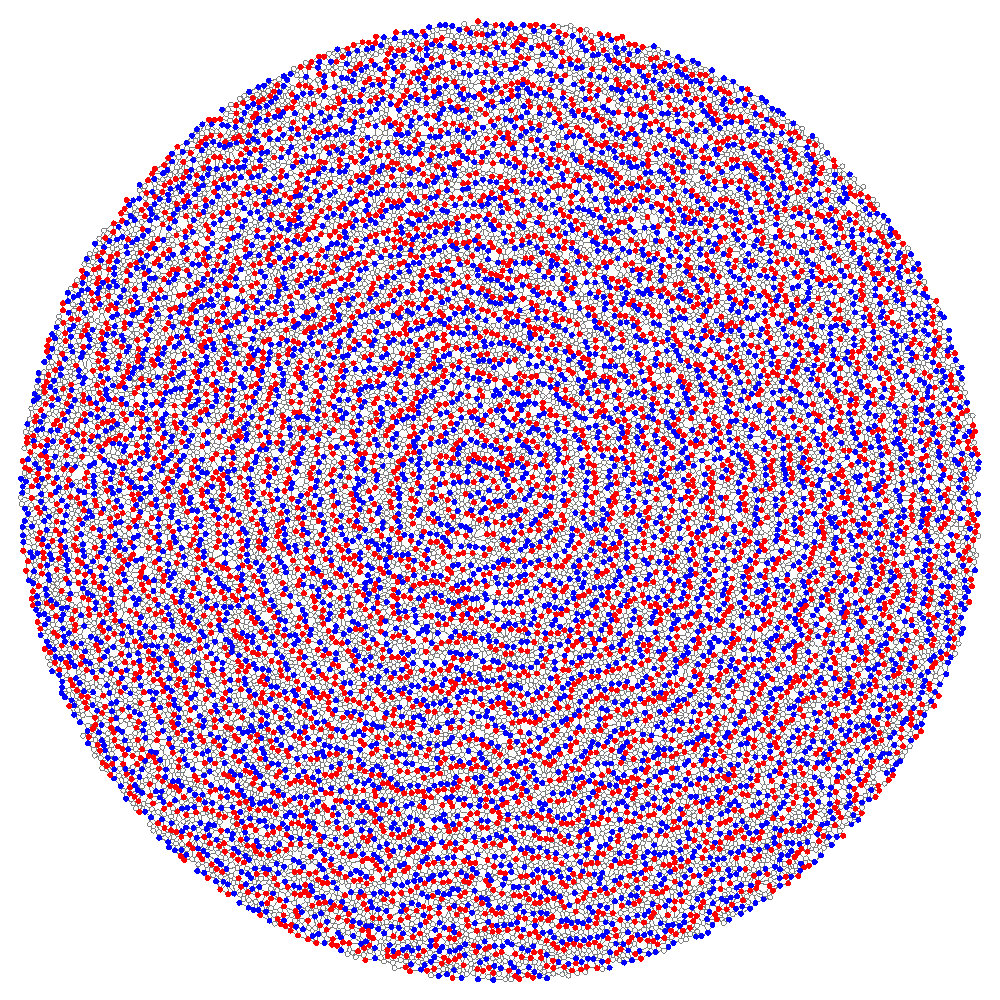}}
  \subfigure[$\beta=1$, $\epsilon=0.167$]{ \includegraphics[width=0.43\textwidth]{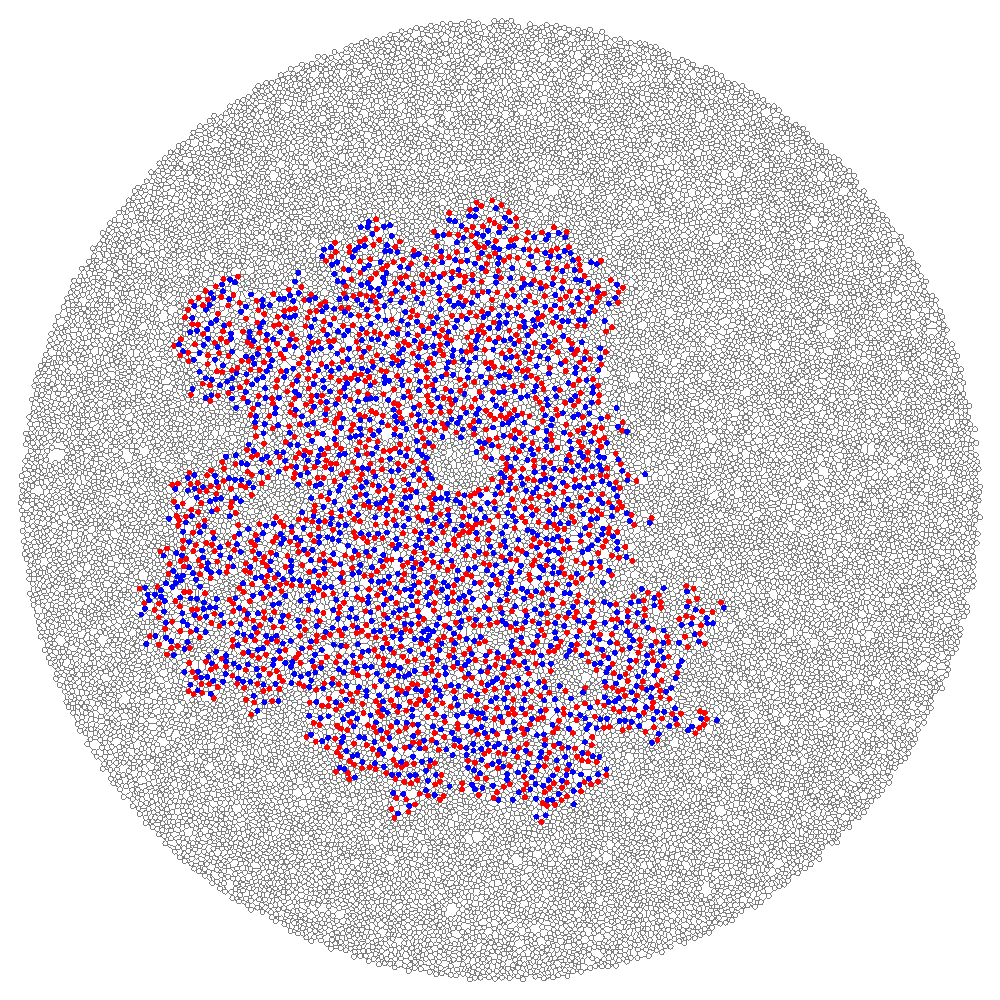}}
    \subfigure[$\beta=1$, $\epsilon=0.2$]{ \includegraphics[width=0.43\textwidth]{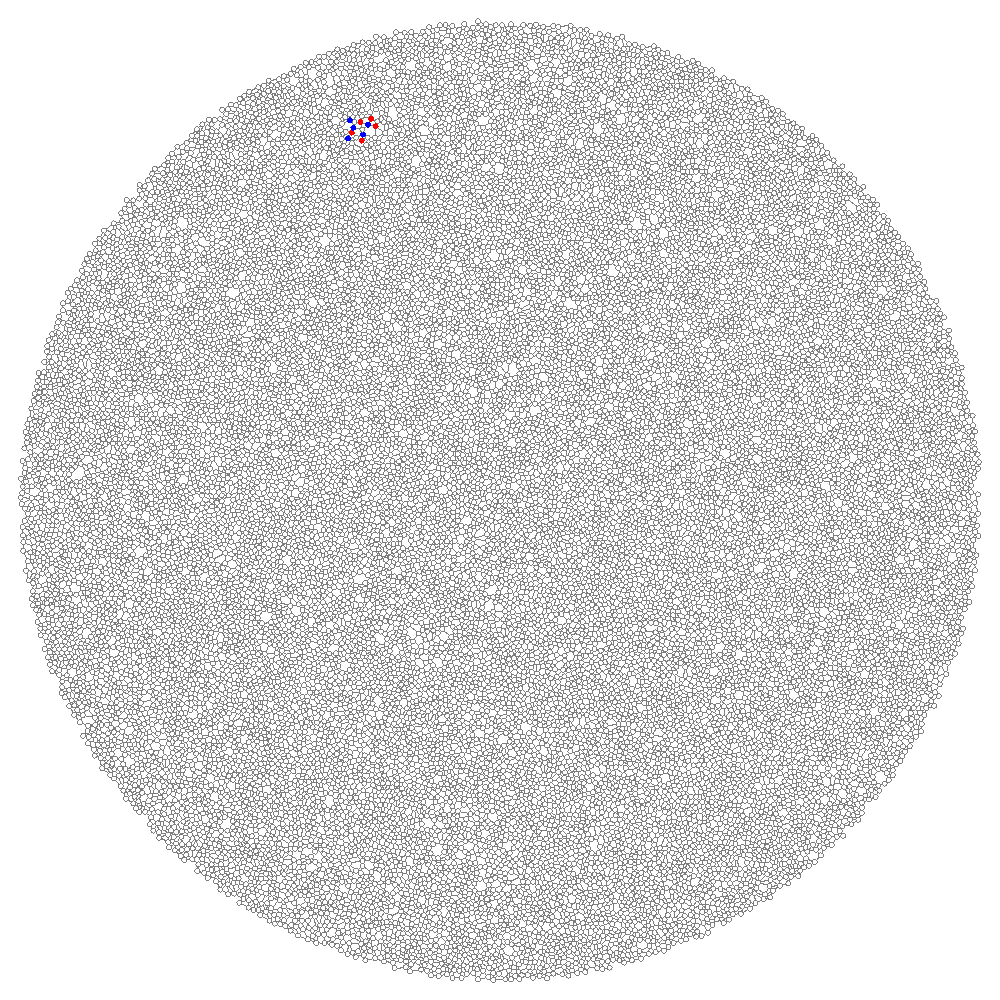}}
    \subfigure[$\beta=1$, $\epsilon=0.24$]{\includegraphics[width=0.43\textwidth]{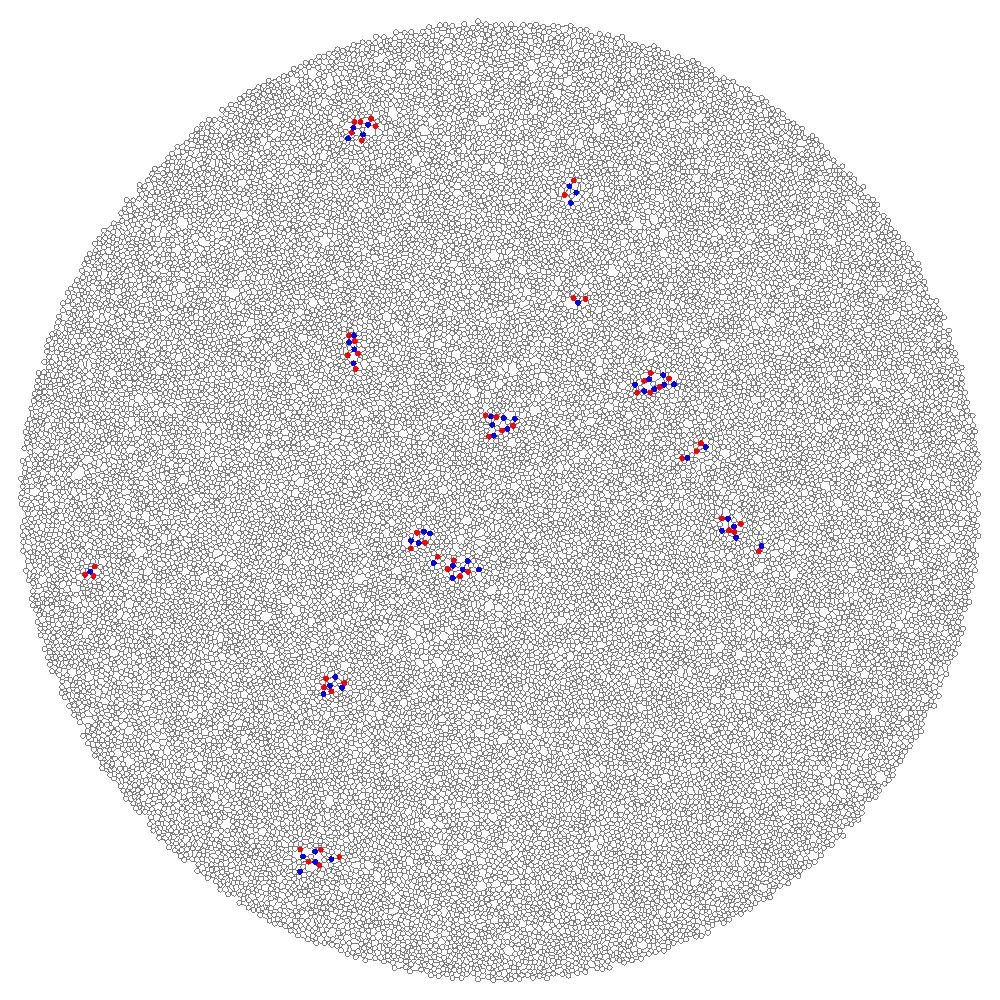}}
 \caption{Snapshots of excitation dynamics in $\beta$-skeletons ($\beta=1$) with relative excitation threshold. (a)--(d)~singular initial excitation, (e)~random initial excitation, a node is excited with probability 0.1. Red (light-gray) coloured
 node are excited, blue (dark-gray) coloured nodes are refractory.}
 \label{examplesbeta1}

\vspace{0.5cm}

 \end{figure}
 
For $\epsilon \in [0, 0.166]$ skeletons exhibit classical properties of discrete excitable media. A single excitation causes
formation of a spiral wave or a generator of spiral or target waves. Successions of  circular waves fill the whole skeleton 
(Fig.~\ref{examplesbeta1}ab). As soon as $\epsilon$ exceeds 0.166 excitation wave-fronts break up into separate wave fragments. 
Thus branching domains of excitation activity are formed (Fig.~\ref{examplesbeta1}c).

 \begin{figure}[!tbp]
 \centering
 \subfigure[$t$]{ \includegraphics[width=0.47\textwidth]{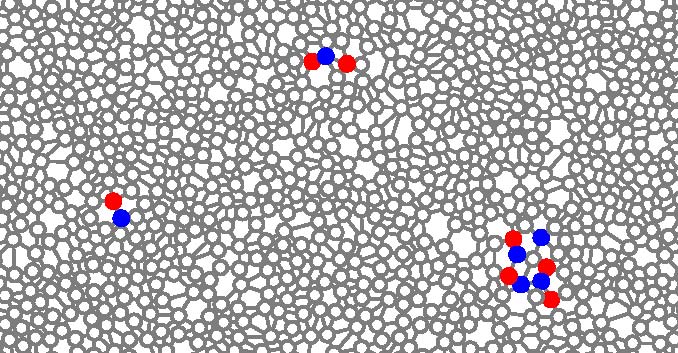}}
 \subfigure[$t+1$]{ \includegraphics[width=0.47\textwidth]{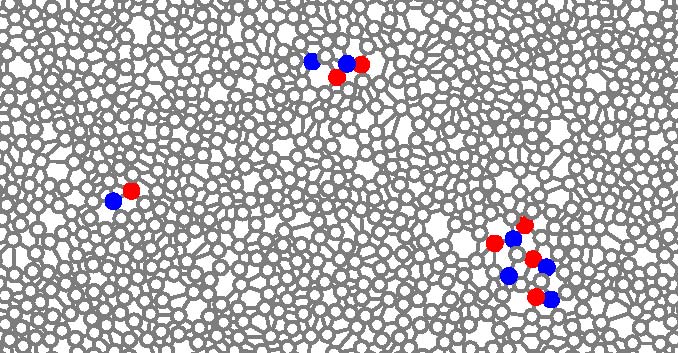}}
 \subfigure[$t+2$]{ \includegraphics[width=0.47\textwidth]{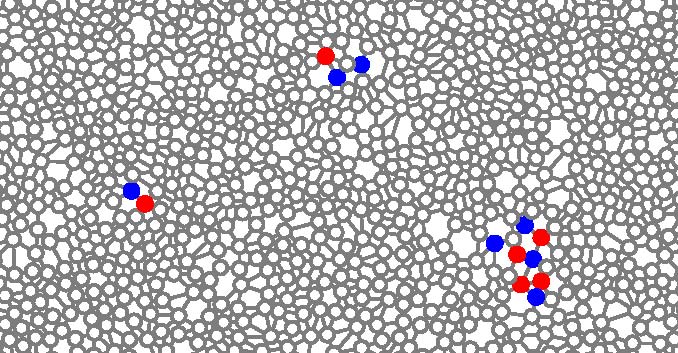}}
 \subfigure[$t+3$]{ \includegraphics[width=0.47\textwidth]{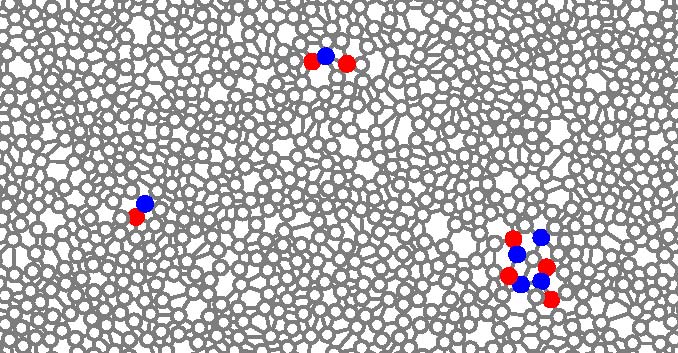}}
 \subfigure[$t+4$]{ \includegraphics[width=0.47\textwidth]{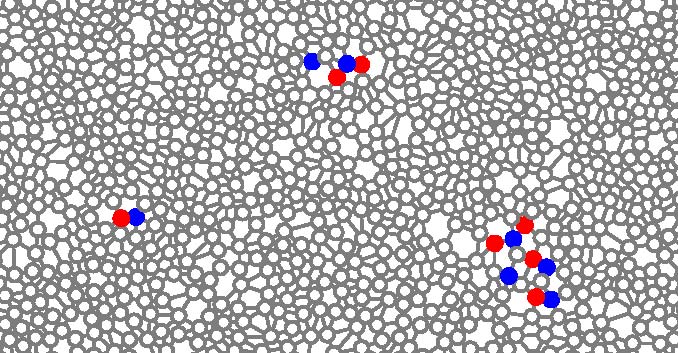}}
 \subfigure[$t+5$]{ \includegraphics[width=0.47\textwidth]{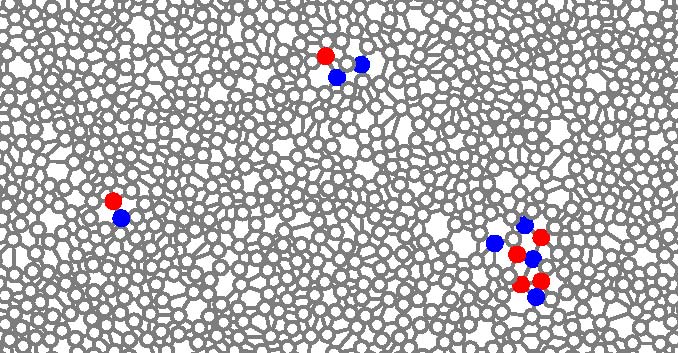}}
 \caption{Examples of oscillators in $\beta$-skeletons ($\beta=1$) with relative excitation threshold
 $\epsilon=0.24$. Red (light-gray) coloured
 node are excited, blue (dark-gray) coloured nodes are refractory.}
 \label{oscillatorbeta1epsilon024}
 
\vspace{0.5cm}

 \end{figure}

Only tiny oscillating activity domains emerge for $\epsilon \geq 0.2$ (Fig.~\ref{examplesbeta1}d). Life-cycles of three oscillators are shown in Fig.~\ref{oscillatorbeta1epsilon024}. Left oscillator in Fig.~\ref{oscillatorbeta1epsilon024} is 
a typical one. Excitation wave, consisting of one excited state and one refractory state, runs along a cycle of five nodes. 
The oscillator thus has period five (Fig.~\ref{oscillatorbeta1epsilon024}a--f). The oscillator conserves number of non-resting states.

Oscillator in the top part of snapshots in Fig.~\ref{oscillatorbeta1epsilon024} has period three. It changes 
its state between configurations with two excited and one refractory and one excited and two refractory states. 
The activation does not cycle but is exchanged between several nodes. First one node $a$ excites its two neighbors. 
These excited nodes convey excitation  to two other nodes (their neighbors), which in turn excite the original node $a$ (Fig.~\ref{oscillatorbeta1epsilon024}a--d). 

Oscillator shown in right part of snapshots in Fig.~\ref{oscillatorbeta1epsilon024} consists of four excited and four refractory states, the number of the non-resting states is conserved. The oscillator has period three (Fig.~\ref{oscillatorbeta1epsilon024}a--d). This oscillator behaves like a breather, with its compressed (Fig.~\ref{oscillatorbeta1epsilon024}c), intermediate (Fig.~\ref{oscillatorbeta1epsilon024}a)
and fully expanded (Fig.~\ref{oscillatorbeta1epsilon024}b) configuration.

Skeleton becomes non-excitable, i.e. no excitation persists, when $\epsilon$ exceeds 0.25.

\subsection{$\beta=1.5$}

  \begin{figure}[!tbp]
 \centering
 \subfigure[$\beta=1.5$, $\epsilon=0.01$]{ \includegraphics[width=0.47\textwidth]{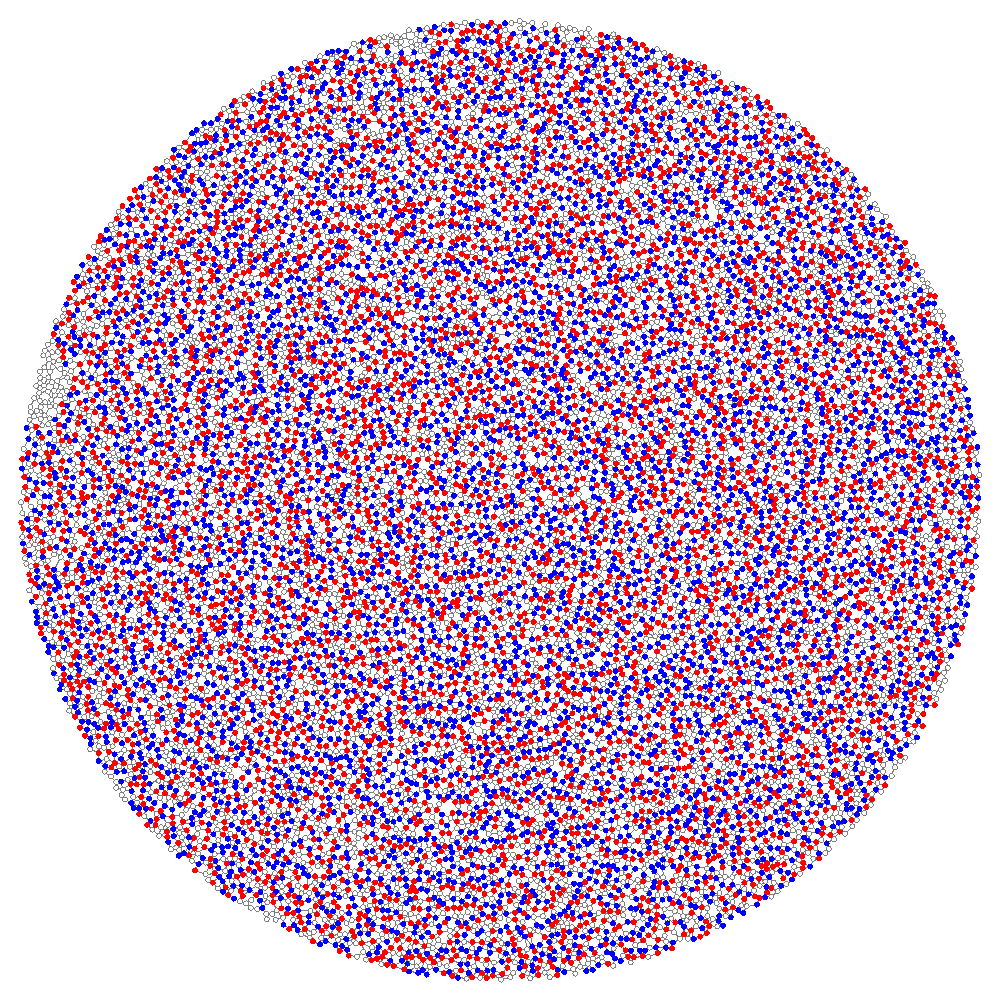}}
  \subfigure[$\beta=1.5$, $\epsilon=0.15$]{ \includegraphics[width=0.47\textwidth]{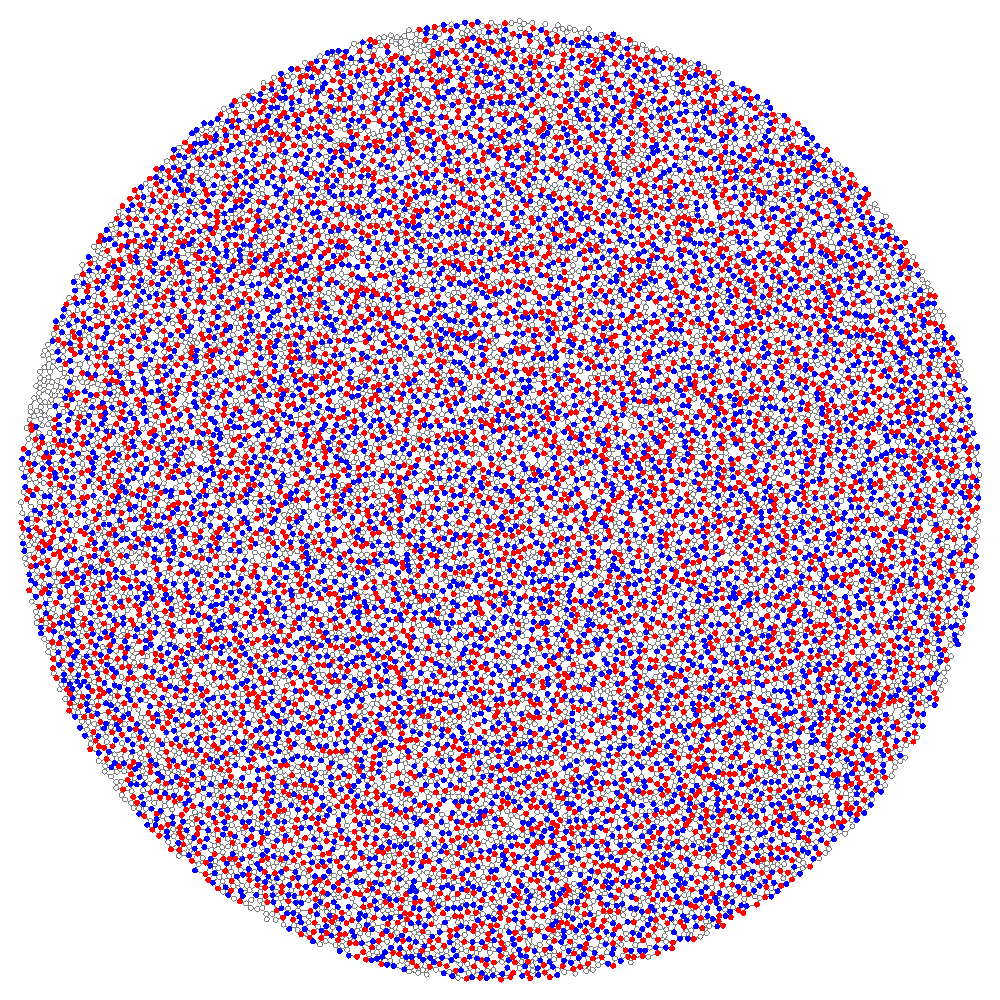}}
   \subfigure[$\beta=1.5$, $\epsilon=0.2$]{ \includegraphics[width=0.47\textwidth]{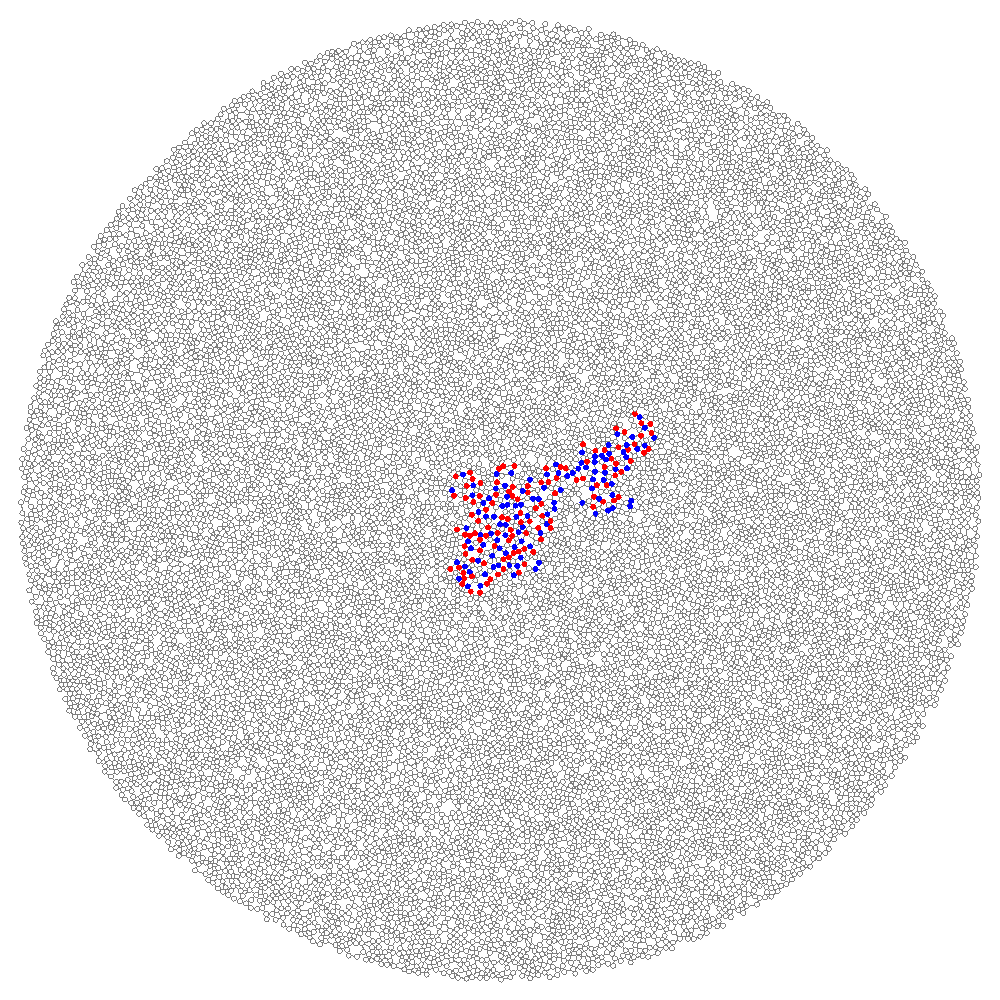}}
    \subfigure[$\beta=1.5$, $\epsilon=0.3$]{\includegraphics[width=0.47\textwidth]{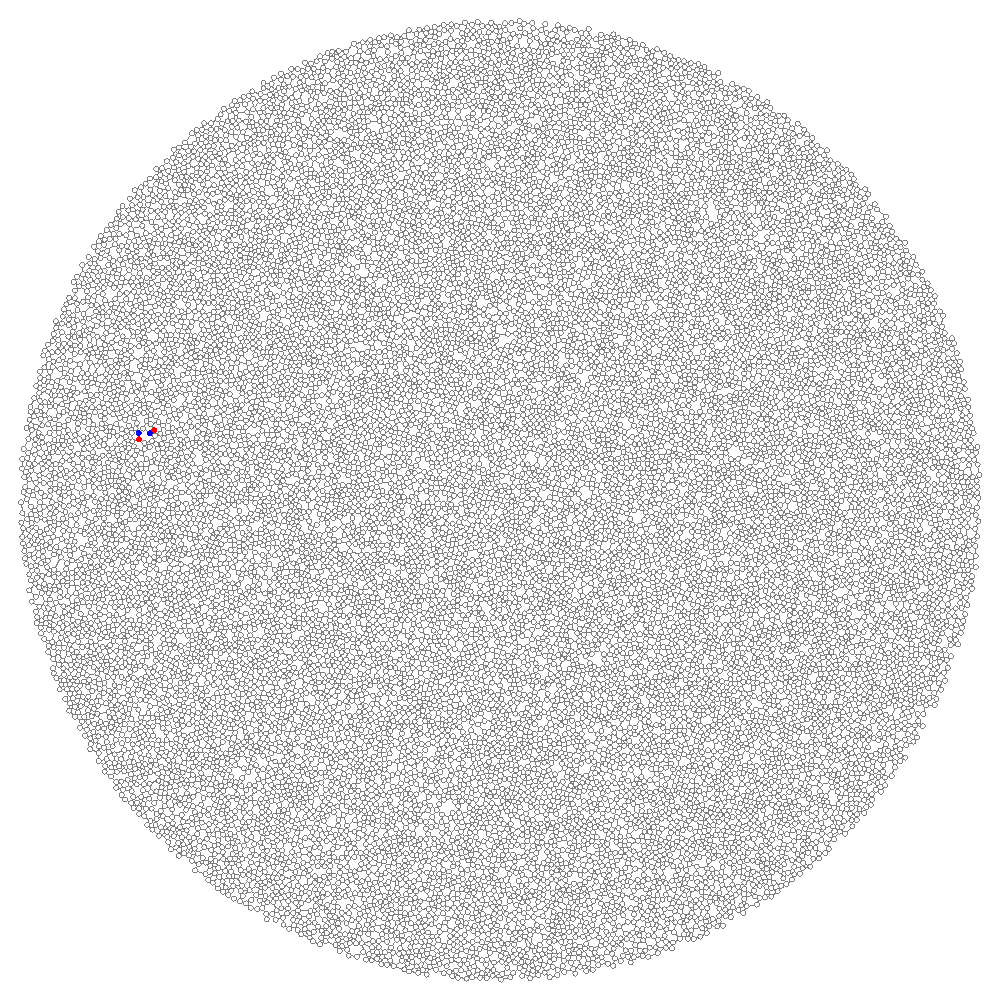}}
 \caption{Snapshots of excitation dynamics in $\beta$-skeletons ($\beta=1.5$) with relative excitation threshold. Red (light-gray) coloured
 node are excited, blue (dark-gray) coloured nodes are refractory.}
 \label{examplesbeta15}
 
\vspace{0.5cm}

 \end{figure}

Propagating domains of excitation are circularly shaped however they do not show any pronounced wave-fronts, the excitation
patterns are rather disordered for $\epsilon \in [0, 0.199]$ (Fig.~\ref{examplesbeta15}ab).  
With $\epsilon$ exceeding 0.199 activity domains change their shapes form circular to branching, tree-like propagating 
domains (Fig.~\ref{examplesbeta15}c). Size of such branching domains decreases with increase of $\epsilon$. When $\epsilon \geq 0.25$ all domains cease propagating and just tiny clusters, oscillators, of activity are formed (Fig.~\ref{examplesbeta15}d). A typical oscillator is a cycle of three or four nodes around which a quasi-one-dimensional excitation propagates. Sometimes the running excitation sends 'sparks' of activity to lateral nodes; these sparks extinguish in few steps of development.  No excitation
persists in skeletons for $\epsilon > 0.33$. 

\subsection{$\beta=2$}

   \begin{figure}[!tbp]
 \centering
 \subfigure[$\beta=2$, $\epsilon=0.01$]{ \includegraphics[width=0.47\textwidth]{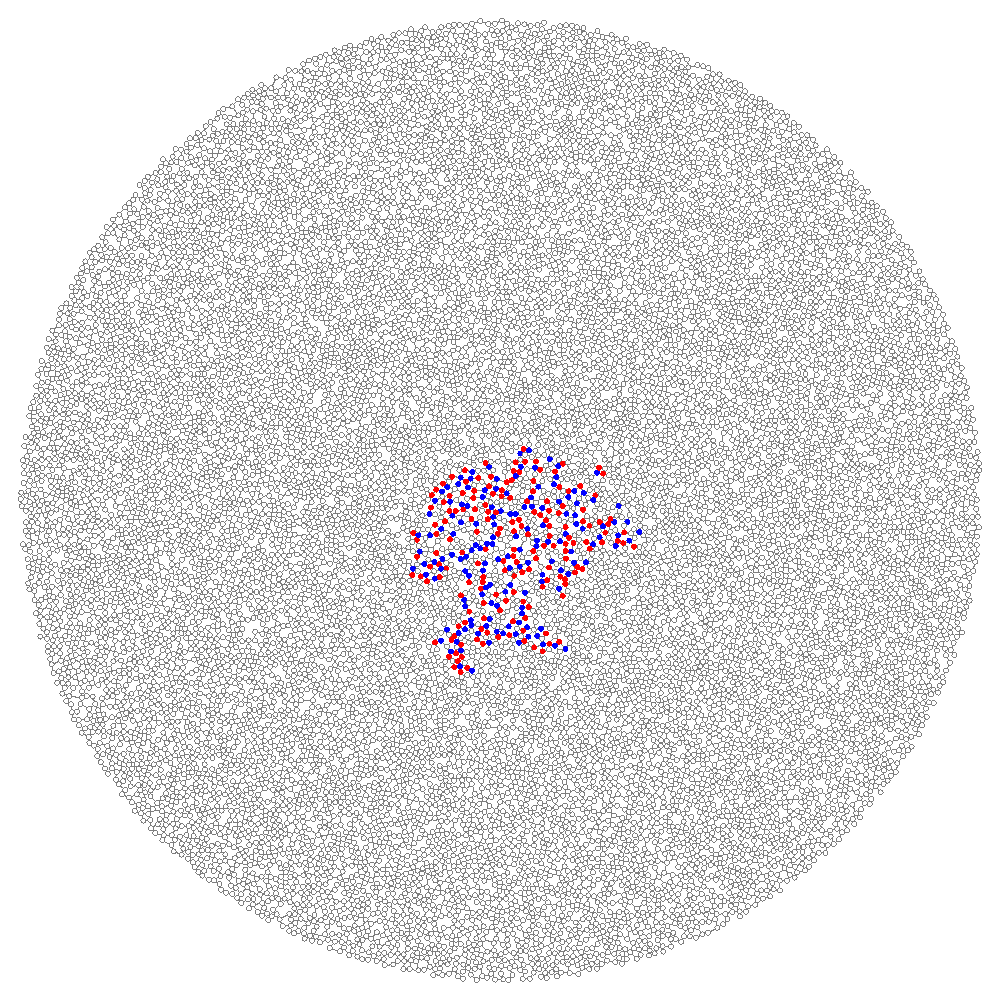}}
  \subfigure[$\beta=2$, $\epsilon=0.15$]{ \includegraphics[width=0.47\textwidth]{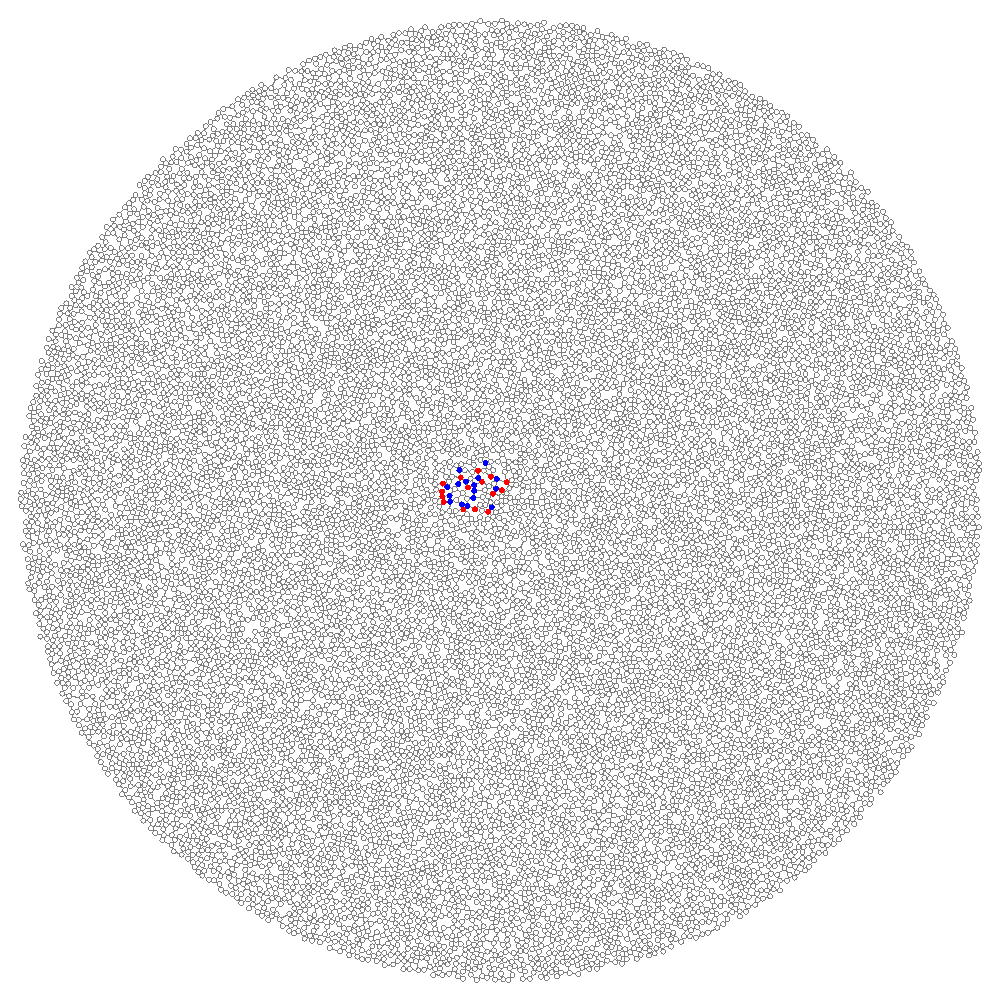}}
   \subfigure[$\beta=2$, $\epsilon=0.2$]{ \includegraphics[width=0.47\textwidth]{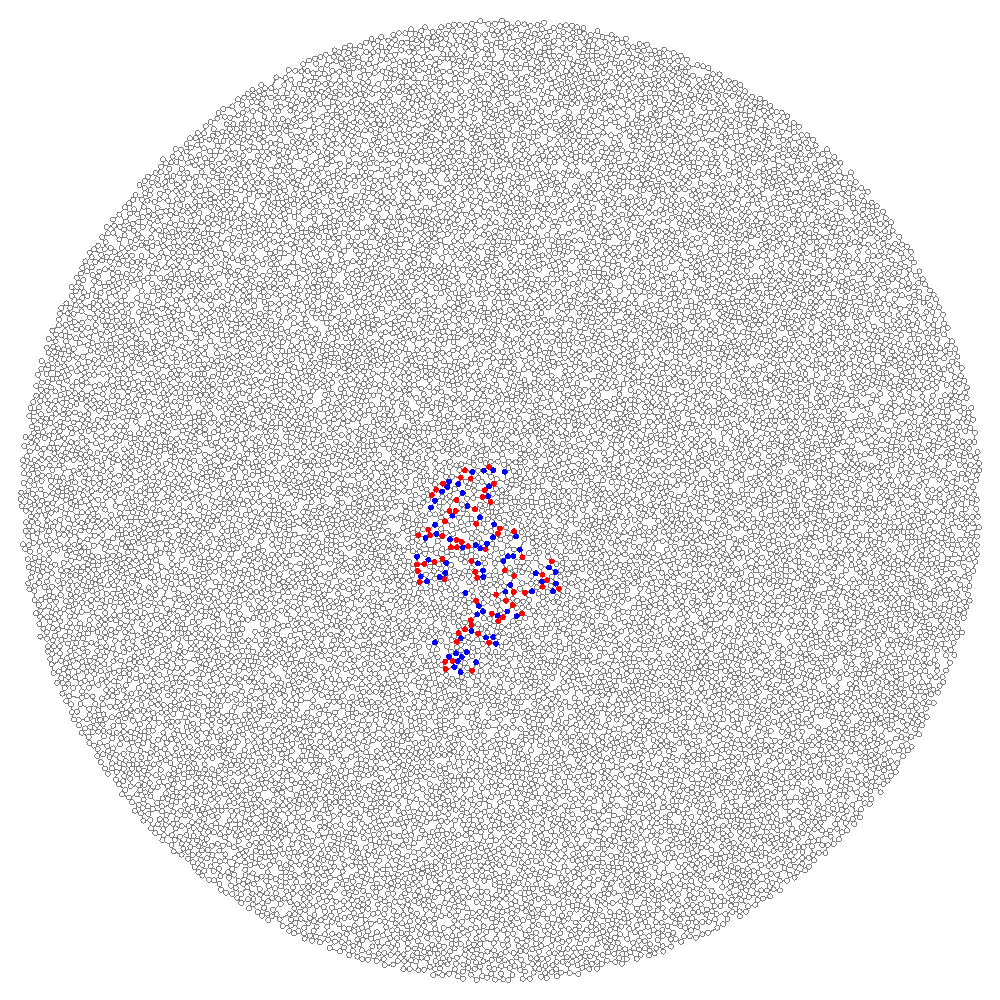}}
    \subfigure[$\beta=2$, $\epsilon=0.3$]{\includegraphics[width=0.47\textwidth]{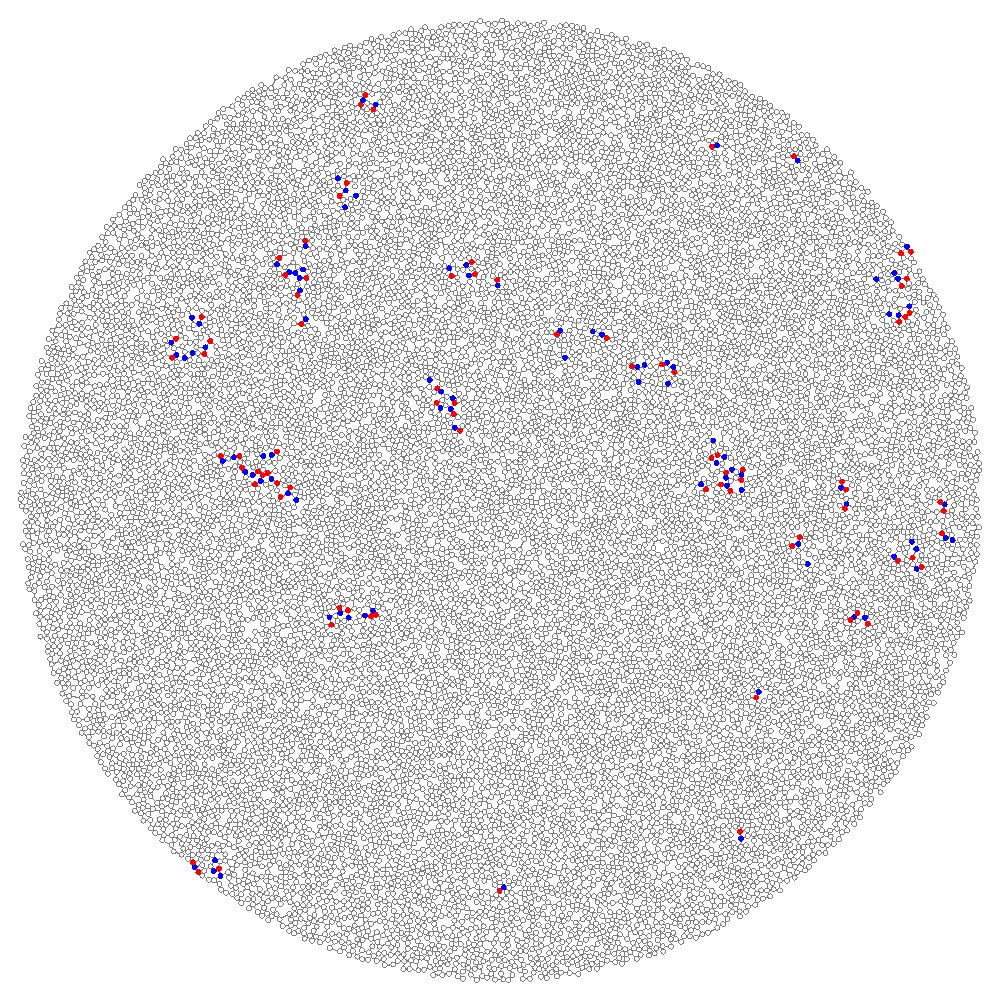}}
 \caption{Snapshots of excitation dynamics in $\beta$-skeletons ($\beta=2$) with relative excitation threshold. Red (light-gray) coloured
 node are excited, blue (dark-gray) coloured nodes are refractory.}
 \label{examplesbeta2}
 
\vspace{0.5cm}

 \end{figure}

Single excitation gives birth to  slowly propagating irregularly-shaped domains of excitation activity (Fig.~\ref{examplesbeta2}a). The domains originated form a single excitation never span the whole skeleton. Sizes of domains decrease 
with increase of $\epsilon$ (Fig.~\ref{examplesbeta2}bc).

  \begin{figure}[!tbp]
 \centering
 \subfigure[$t$]{ \includegraphics[width=0.2\textwidth]{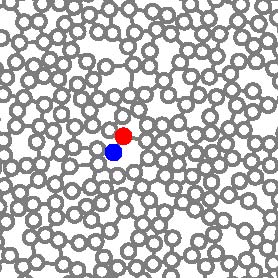}}
  \subfigure[$t+1$]{ \includegraphics[width=0.2\textwidth]{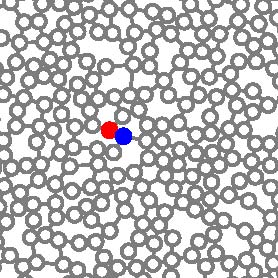}}
    \subfigure[$t+2$]{ \includegraphics[width=0.2\textwidth]{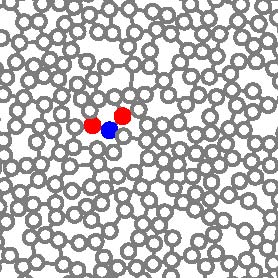}}
      \subfigure[$t+3$]{ \includegraphics[width=0.2\textwidth]{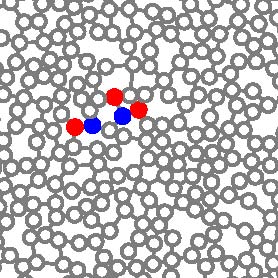}}
        \subfigure[$t+4$]{ \includegraphics[width=0.2\textwidth]{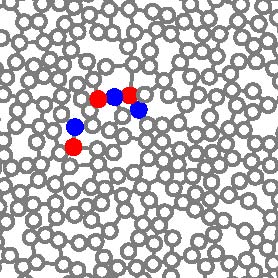}}
          \subfigure[$t+5$]{ \includegraphics[width=0.2\textwidth]{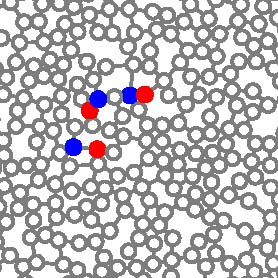}}
            \subfigure[$t+6$]{ \includegraphics[width=0.2\textwidth]{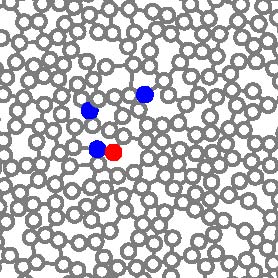}}
              \subfigure[$t+7$]{ \includegraphics[width=0.2\textwidth]{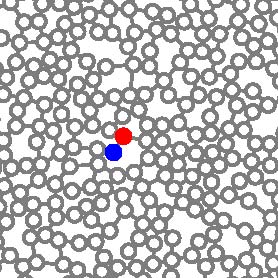}}
  \caption{Examples of oscillators in $\beta$-skeletons ($\beta=2$) with relative excitation threshold
 $\epsilon=0.29$. Red (light-gray) coloured
 node are excited, blue (dark-gray) coloured nodes are refractory.}
 \label{oscillatorbeta2epsilon29}
 
\vspace{0.5cm}

 \end{figure}

Only tiny clusters of excitation are formed for $\epsilon \in [0.25, 0.333[$ (Fig.~\ref{examplesbeta2}d). 
A life-cycle of most typical oscillator is shown in Fig.~\ref{oscillatorbeta2epsilon29}. This oscillator 
has period seven. Its core structure is a singleton-excitation (accompanied by a refractory tail) 
running around seven-node cycle, anti-clockwise. At certain moments of its life-cycle the excitation 
excites few nodes neighboring to the `core cycle'. Thus sparks of excitation are formed 
(Fig.~\ref{oscillatorbeta2epsilon29}c--f). 

No excitation persists for $\epsilon \geq 0.333$.

 \begin{figure}[!tbp]
 \centering
 \includegraphics[width=1.2\textwidth]{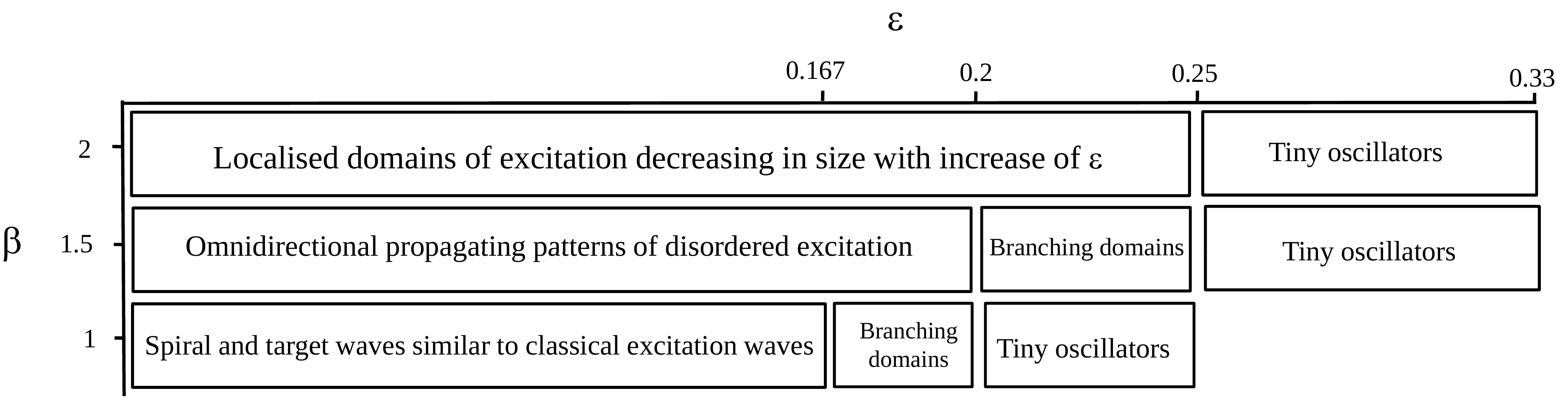}
 \caption{Parameterisaton of space-time dynamics of $\beta$-skeletons with relative excitation threshold
 by $\beta$ and $\epsilon$.}
 \label{relativethresholdparameterisaton}
 
\vspace{0.5cm}

 \end{figure}

\begin{finding}
Excitable $beta$-skeletons with relative threshold of excitation exhibit the following classes of space-time dynamics:
(1)~spiral and target waves similar to classical excitation waves, 
(2)~omnidirectional propagating patterns of disordered excitation,
(3)~localised domains of excitations, 
(4)~branching domains, and,
(5)~tiny oscillators.
\end{finding}

Position of the classes on $\beta-\epsilon$-plane is shown in Fig.~\ref{relativethresholdparameterisaton}.
There are sharp boundaries between classes, characterized by abrupt changes in morphology of excitation patterns.
Critical values of relative excitability threshold are $\epsilon=0.167, 0.2, 0.25$ and $0.33$. These thresholds 
reflect situations when a resting node excites if at least one of six ($\epsilon=0.166$), one of five ($\epsilon=0.2$), 
one of four ($\epsilon=0.25$) and one of three ($\epsilon=0.33$) neighbors is excited.  Excitability threshold 
$\epsilon=0.166$ is critical only for $\beta$-skeletons with $\beta$ very close to 1. This is because such skeletons 
have mode 5 of node degree distribution and average node degree 5.1 (Fig.~\ref{mode}, $\varphi=0.407$), i.e. nodes with 
six neighbors are second (after nodes with five neighbors) dominating nodes on the graph. For similar reasons 
$\epsilon=0.2$ is not critical threshold value for $\beta$-skeletons with $\beta$ close to 2 because over half 
of the nodes in such skeletons have three neighbors, and nodes with two or four neighbors are second dominants 
in these skeletons.

\subsection{Stability of localized oscillators}

\begin{finding}
Localized excitations, or oscillators, are stable under changes of relative 
excitability threshold.
\end{finding}

\begin{figure}[!tbp]
 \centering
 \subfigure[$\epsilon=0.204$]{\includegraphics[width=0.3\textwidth]{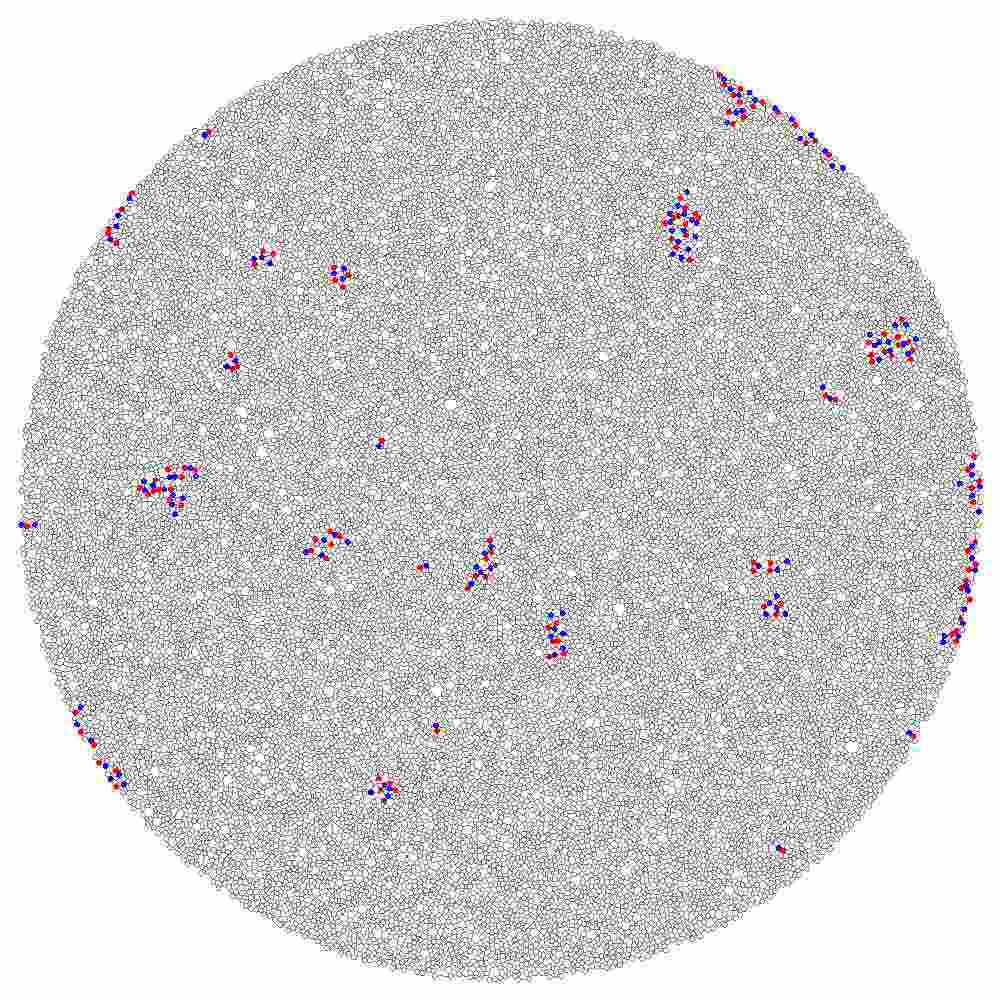}}
 \subfigure[$\epsilon=0.199$]{\includegraphics[width=0.3\textwidth]{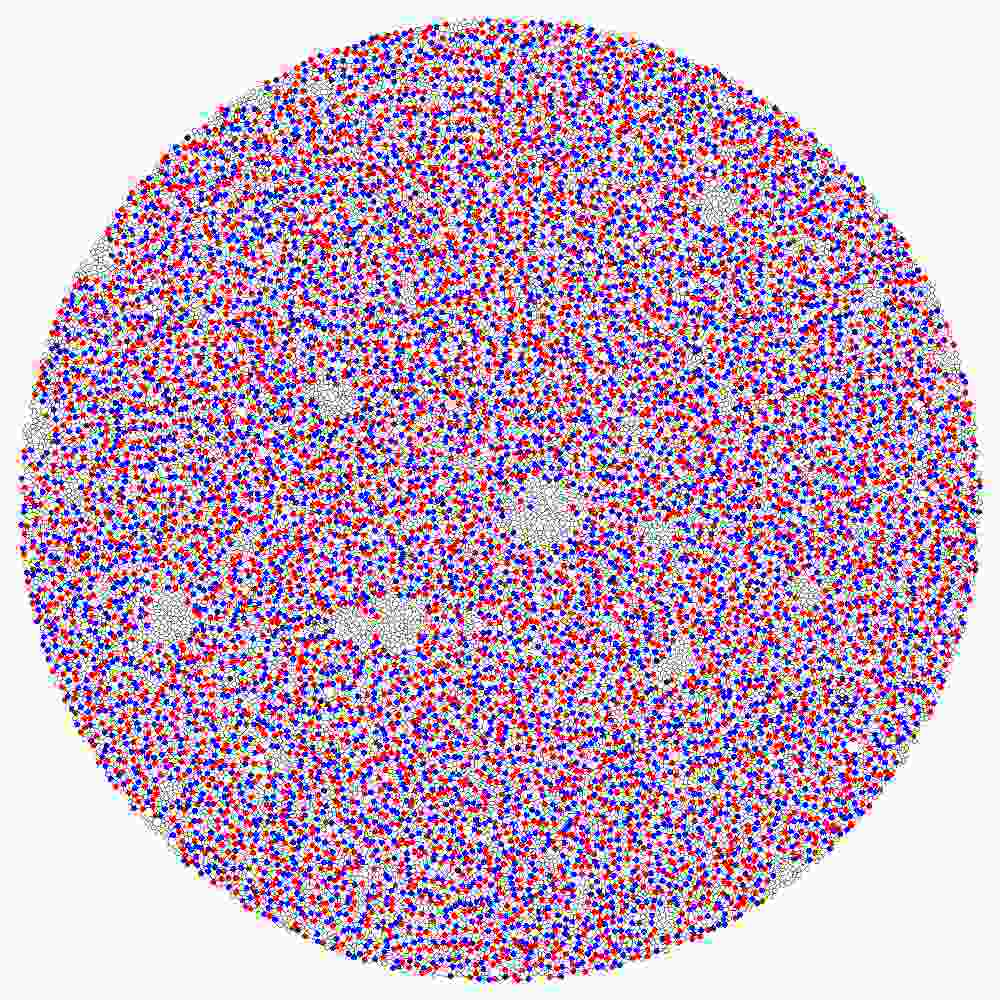}}
 \subfigure[$\epsilon=0.204$]{\includegraphics[width=0.3\textwidth]{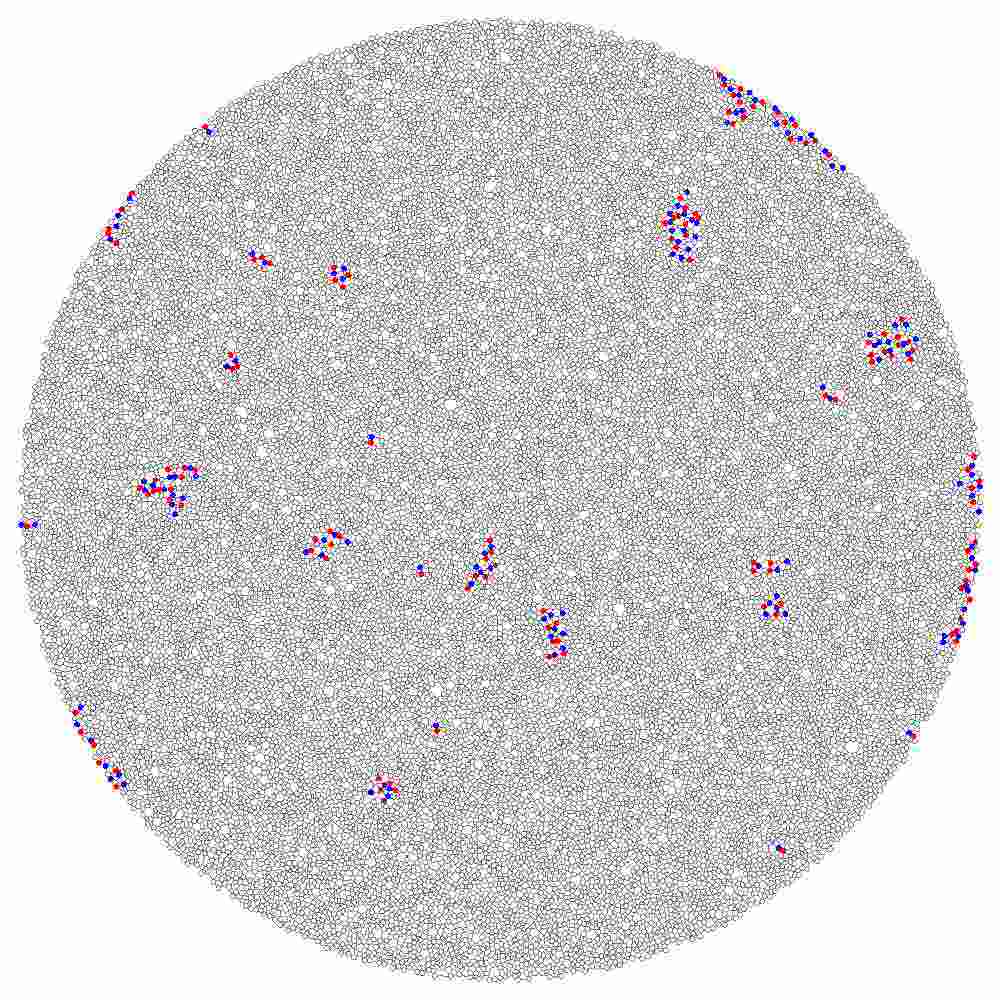}} \subfigure[$\epsilon=0.25$]{\includegraphics[width=0.3\textwidth]{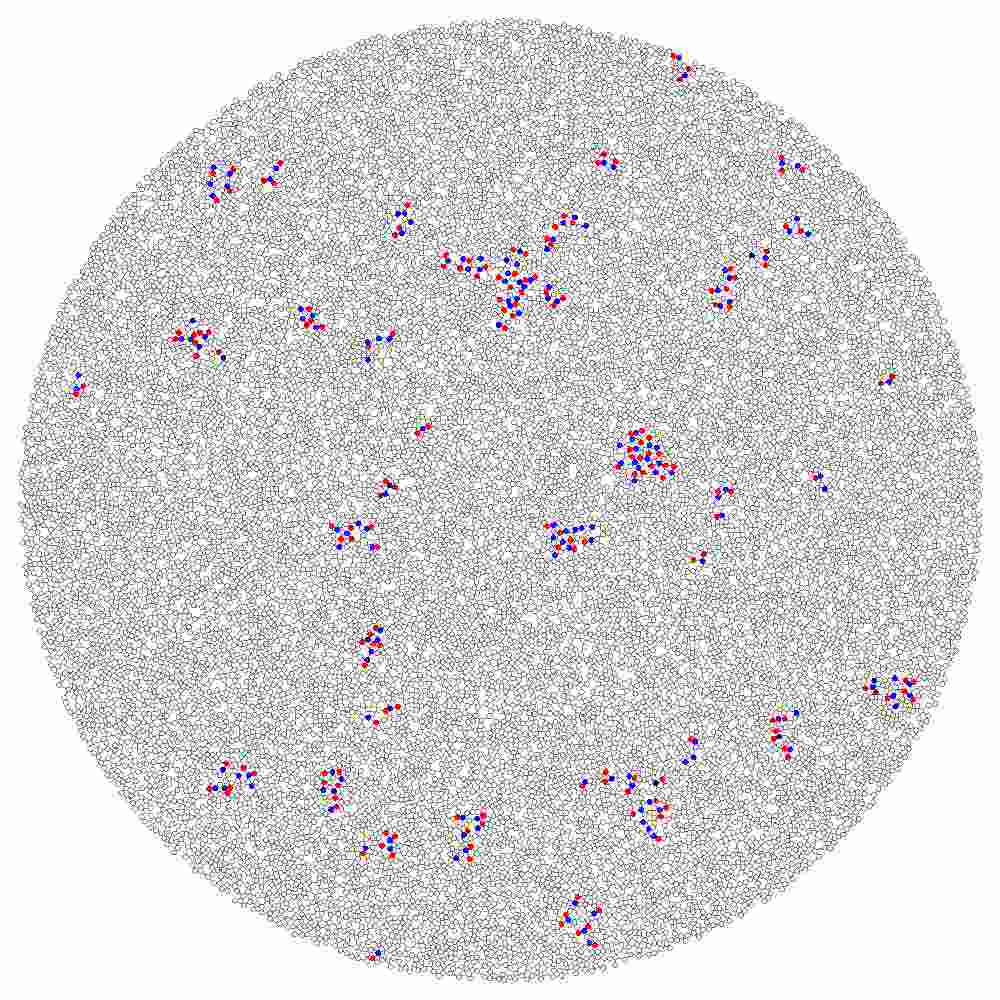}}
 \subfigure[$\epsilon=0.249$]{\includegraphics[width=0.3\textwidth]{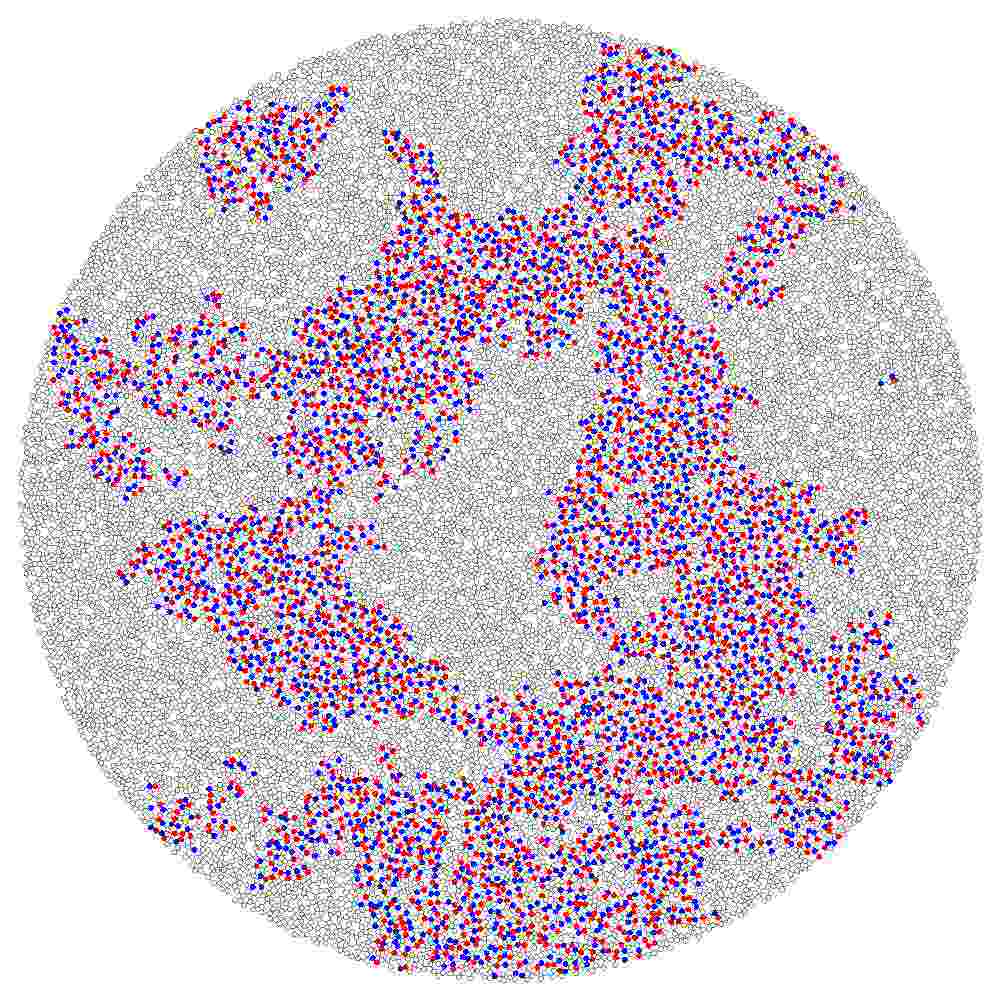}}
 \subfigure[$\epsilon=0.25$]{\includegraphics[width=0.3\textwidth]{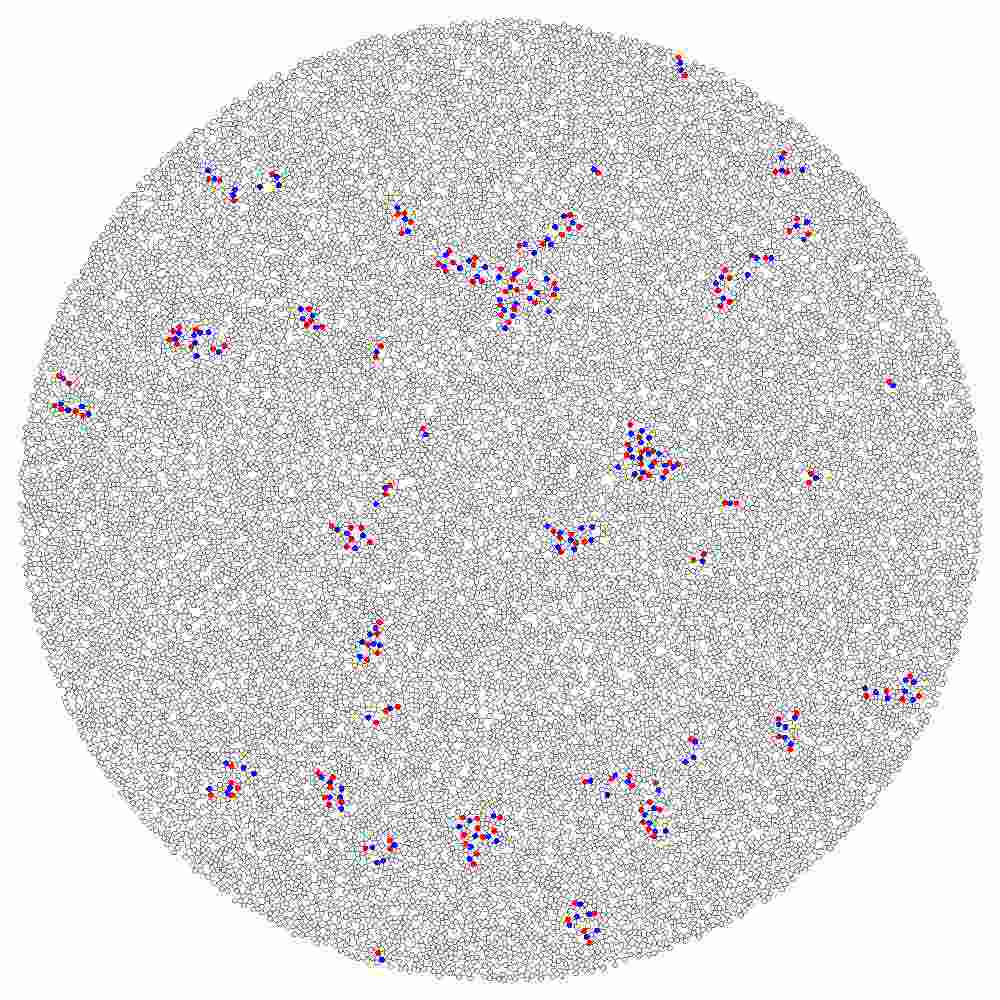}}
 \caption{Stability of oscillators under changes of $\epsilon$: (a)--(c)~$\beta=1$, (d)--(f)~$\beta=2$.}
 \label{memory}
 
\vspace{0.5cm}

 \end{figure}

The finding is illustrated  for $\beta=1$ and $\beta=2$ in Fig.~\ref{memory}. We choose 
values of threshold $\epsilon$ at the edge of excitability --- $\epsilon=0.204$ for $\beta=1$ and 
$\epsilon=0.25$ for $\beta=2$ --- and perturb $\beta$-skeleton with a random configuration of
excitations.  Few localized oscillators emerge (Fig.~\ref{memory}ad). Then we decrease
excitability thresholds --- dynamically, without resetting the skeleton to resting state  --- 
to $\epsilon=0.199$ ($\beta=1$) and $\epsilon=0.249$ ($\beta=2$). The oscillators start 
to generate excitation wave-fragments. The wave-fragments emitted by the oscillators fill 
the skeletons (Fig.~\ref{memory}be). When we reduce excitability thresholds back to 
their original levels $0.204$ and $0.25$ traveling wave-fragments disappear but oscillators  
remain untouched, they stay where they were before relative excitability threshold changed 
(Fig.~\ref{memory}cf).

\section{Discussion}
\label{discussion}

We studied discrete excitation dynamics on a smoothly parameterised family of planar proximity graphs --- $\beta$-skeletons.
The skeletons were constructed on a set of uniform discs packed into a larger disc. Thus each skeleton is characterised
by $\beta$ and density of node packing.  We considered rules of absolute excitability (a resting node excites if number of 
its neighbors exceeds certain threshold) and relative excitability (a resting node excites if a ratio of its excited 
neighbors to a total number of neighbors exceeds certain threshold). 

In computational experiments we found that overall level of activity in an absolutely excitable $\beta$-skeleton is 
proportional to node packing density and inversely proportional to $\beta$. 
We demonstrated that space-time dynamics of absolutely excitable $\beta$-skeletons can be classified as follows: spiral and target waves, disordered dynamics with irregularly traveling wave-fragments, slowly spreading domains of excitation, and small localized stationary domains of oscillatory activity. Transition of a skeleton between these classes is controlled by changing $\beta$ and density of node packing. Both $\beta$ and density affect distribution of node degrees. We provided a parameterization of the classes using just one parameter --- average node degree.

There are five classes of space-time activity of the relatively excitable $\beta$-skeletons with tightly packed nodes (density 0.407, 15000 nodes): spiral and target waves, omnidirectional propagating patterns of disordered excitation, 
localized domains of excitations, branching domains of activity and tiny oscillators. The classes are controlled by $\beta$ 
and the relative excitability threshold $\epsilon$.  Tiny oscillators usually emerge just below edge of excitability. The tiny oscillators are stable under relative excitability threshold. Decreasing threshold we can increase number of excited loci around the oscillators but we can not modify oscillators themselves. As soon as we increase the threshold all excitation activity 
but original oscillators disappear.  

 The class of branching domains is a transitional class occupying a part of $\beta-\epsilon$-space between full excitability and non-excitability. With regards to propagating patterns, waves are pronounced for $\beta=1$, and are alike classical excitation waves in discrete media.  Increase of $\beta$ to 1.5 causes wave-fronts to break up into separate wave-fragments.

We believe our results will find their applications in analysis of spreading crowd activities, e.g. riots, on a city's streets; distributed containment of traffic jams on motorways networks; study of foraging patterns of social insects and myxomycetes; control of excitation dynamics and propagation of defects in soft matter; design and simulation of conglomerates of simple excitable elements.

\section{Acknowledgment}

The work is part of the European project 248992 funded under 7th FWP (Seventh Framework Programme) FET Proactive 3: Bio-Chemistry-Based Information Technology CHEM-IT (ICT-2009.8.3).

\clearpage

\newpage



\begin{thebibliography}{99}

\bibitem{adamatzky_2002}
Adamatzky~A. and Holland~O.
Reaction-diffusion and ant-based load balancing of communication networks.
Kybernetes 31 (2002) 667--681.


 \bibitem{adamatzky_ppl_2008}
 Adamatzky A. Developing proximity graphs by Physarum Polycephalum: 
 Does the plasmodium follow Toussaint hierarchy?  Parallel Processing Letters 19 (2008) 105--127.

\bibitem{adamatzky_jones_2009}
Adamatzky~A. and Jones~J.
Road planning with slime mould: If Physarum built motorways it would route M6/M74 through Newcastle
Int J Bifurcation and Chaos (2009), in press. See also \url{arXiv:0912.3967v1}.

\bibitem{beavon}
Beavon~D.~J.~K., Brantingham~P.~L. and Brantingham~P.~J.
The influence of street networks on the patterning of property offenses. 
\url{www.popcenter.org/library/CrimePrevention/Volume_02/06beavon.pdf}



\bibitem{billiot_2010}
Billiot~J.~M., Corset~F., Fontenas~E.
Continuum percolation in the relative neighborhood graph.
\url{arXiv:1004.5292}


\bibitem{dale_2000}
Dale~M.~R.~T.
Spatial Analysis in Plant Ecology (Cambridge University Press, 2000).


\bibitem{dale_2002}
Dale~M.~R.~T., Dixon~P., Fortin~M.-J., Legendre~P., Myers~D.~E. and Rosenberg~M.~S.
Conceptual and mathematical relationships among methods for spatial analysis.
Ecography 25 (2002) 558-–577.



\bibitem{delauanay}
Delaunay ~B. Sur la sph\`{e}re vide, 
Izvestia Akademii Nauk SSSR, Otdelenie Matematicheskikh i Estestvennykh Nauk,
7 (1934) 793--800.


\bibitem{gabriel_1969}
Gabriel~K.~R. and Sokal~R.~R. 
A new statistical approach to geographic variation analysis. Systematic Zoology 18 (1969) 259–-270.



\bibitem{jaromczyk} 
Jaromczyk~J.~W. and G.~T.~Toussaint, 
Relative neighborhood graphs and their relatives. Proc. IEEE 80 (1992) 1502--1517.

\bibitem{jombart_2008}
Jombart~T., Devillard~S., Dufour~A.-B., Pontier~D. 
Revelaing cryptic spatial patterns in genetic variability by a new multivariate method. 
Heredity 101 (2008) 92--103.


\bibitem{kirkpatrick}
Kirkpatrick~D.G. and Radke~J.D. 
A framework for computational morphology. In: Toussaint~G.~T., Ed., 
Computational Geometry (North-Holland, 1985) 217-–248.



\bibitem{legendre_1989}
Legendre~P. and Fortin~M.-J.
Spatial pattern and ecological analysis.
Vegetatio 80 (1989) 107--138.


\bibitem{li_2004}
Li~X.-Y. 
Application of computation geometry in wireless networks. 
In: Cheng~X., Huang~X., Du~D.-Z. (Eds.)
Ad Hoc Wireless Networking (Kluwer Academic Publishers, 2004) 197--264.


\bibitem{magwene_2008}
Magwene~P.~W.
Using correlation proximity graphs to study phenotypic integration.
Evolutionary Biology. 35 (2008) 191--198.


\bibitem{matula_1980}
Matula~D.~W. and Sokal~R.~R. 
Properties of Gabriel graphs relevant to geographic variation research and clustering of points in the plane.
Geogr. Anal. 12 (1980) 205–-222.


\bibitem{muhammad_2007}
Muhammad~R.~B.
A distributed graph algorithm for geometric routing in ad hoc wireless networks.
J Networks 2 (2007) 49--57. 


\bibitem{runions_2005}
Runions~A., Fuhrer~M., Lane~B., Federl~P., Rolland-Lagan~A.-G., and Prusinkiewicz~P.
Modeling and visualization of leaf venation patterns. ACM Transactions on Graphics 24 (2005) 702--711.


\bibitem{santi_2005}
Santi~P. Topology Control in Wireless Ad Hoc and Sensor Networks (Wiley, 2005).

\bibitem{sokal_2008}
Sokal~R.~R. and Oden~N.~L.
Spatial autocorrelation in biology 1. Methodology.
Biological Journal of the Linnean Society 10 (2008)
199--228.


\bibitem{song_2004}
Song~W.-Z., Wang~Y., Li~X.-Y.
Localized algorithms for energy efficient topology in wireless ad hoc networks.
In: Proc. MobiHoc 2004 (May 24–-26, 2004, Roppongi, Japan).

\bibitem{sridharan_2010}
Sridharan~M. and Ramasamy~A.~M.~S.
Gabriel graph of geomagnetic Sq variations.
Acta Geophysica (2010) \url{10.2478/s11600-010-0004-y}


\bibitem{toroczkai_2008}
Toroczkai~Z. and Guclu~H. Proximity networks and epidemics.
Physica A 378 (2007) 68. \url{arXiv:physics/0701255v1}


\bibitem{wan_2007}
Wan~P.-J., Yi~C.-W.
On the longest edge of Gabriel Graphs in wireless ad hoc networks.
IEEE Trans. on Parallel and Distributed Systems 18 (2007) 111--125.

\bibitem{watanabe_2005}
Watanabe~D. 
A study on analyzing the road network pattern using proximity graphs.
J  of the City Planning Institute of Japan 40 (2005) 133--138.

\bibitem{watanabe_2008}
Watanabe~D.
Evaluating the configuration and the travel efficiency on proximity graphs
as transportation networks. Forma 23 (2008) 81-–87. 
 
 \end{thebibliography}
\end{document}